\documentclass[
12pt, 
english, 
onehalfspacing, 
nolistspacing, 
liststotoc, 
headsepline, 
]{MastersDoctoralThesis} 
\usepackage{color}
\usepackage{tikz}
\usepackage{amsmath}
\usepackage{amssymb}
\usepackage{bm}
\usepackage{tikz} 
\usepackage{doi}
\usepackage{url}
\usepackage{xcolor}
\usepackage{enumitem}
\usepackage{lscape}
\usepackage{float}
\usepackage{pdfpages}
\usepackage{epstopdf}
\newcounter{numln}
\newlist{numitemise}{itemize}{2}
\setlist[numitemise]{wide}%
\setlist[numitemise, 1]{labelindent=0pt,labelwidth=2em, label=\stepcounter{numln}\makebox[2em]{\thenumln.\hfill}, leftmargin=\dimexpr\labelwidth+\labelsep\relax}%
\setlist[numitemise, 2]{labelindent=\dimexpr -2em-\labelsep\relax, labelwidth=\dimexpr 2em+\labelsep\relax, label=\stepcounter{numln}\makebox[\dimexpr\labelwidth + \labelsep\relax]{\thenumln.\hfill\textbullet}, leftmargin=\dimexpr\leftmargin+2\labelsep\relax}%

\usepackage[utf8]{inputenc} 
\usepackage[T1]{fontenc} 

\usepackage{mathpazo} 

\usepackage[backend=bibtex,style=nature,natbib=true]{biblatex}

\usepackage[autostyle=true]{csquotes} 
\usepackage{setspace}
\usepackage{graphicx}
\usepackage{amsfonts}
\usepackage{amssymb}
\usepackage{amsmath}
\usepackage{amsbsy}
\usepackage{mathrsfs}
\usepackage{latexsym}
\usepackage{subfigure}
\usepackage{wasysym}
\usepackage{bm}
\usepackage{subfigure}
\usepackage{color}
\usepackage{comment}
\usepackage{textgreek}
\usepackage{hyperref}

\thesistitle{Application of the Raychaudhuri Equation in Gravitational Systems} 

\supervisor{ADG} 

\examiner{} 

\degree{Doctor of Philosophy} 

\author{Shibendu Gupta Choudhury} 
\addresses{} 

\subject{Physical Sciences} 
\university{\href{https://www.iiserkol.ac.in/}{IISER Kolkata}} 
\department{\href{https://physics.iiserkol.ac.in/}{Department of Physical Sciences}} 
\def\DateSub{December, 2022}

\AtBeginDocument{
\hypersetup{pdftitle=\ttitle} 
\hypersetup{pdfauthor=\authorname} 
\hypersetup{pdfkeywords=\keywordnames} 
}

\begin{document}
\sloppy
\frontmatter 

\pagestyle{plain} 


\begin{titlepage}
\begin{center}

\textsc{\Large Doctoral Thesis}\\[0.5cm] 
\HRule \\[0.4cm] 
{\huge \bfseries \ttitle\par}\vspace{0.4cm} 
\HRule \\[1.5cm] 

\begin{center}
	\large{By\\
    Shibendu Gupta Choudhury \\
	Roll No.: 14IP003\\[1ex]
	\emph{Supervisor:} Dr. Ananda Dasgupta\\[1ex]
    Department of Physical Sciences\\\vspace{0.0cm}
    Indian Institute of Science Education and Research Kolkata\vspace{0.0cm}}
\end{center}

\vspace{1.5cm}

\centering
	\includegraphics[width=0.3\textwidth]{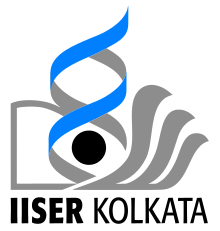}
	
\vspace{1.2cm}

\textit{ A thesis submitted in fulfillment of the requirements for 
the degree of \degreename \ 
in the \deptname \ at \ {\href{https://www.iiserkol.ac.in/}{Indian Institute of 
Science Education and Research Kolkata}} }

\vspace{0.7cm}

{\large \DateSub}\\[4cm] 

\vfill
\end{center}
\end{titlepage}


\begin{declaration}
\addchaptertocentry{\authorshipname} 
\begin{flushright}
Date: \emph{December 23, 2022} \\
\end{flushright}
\noindent
\par I, Mr. \emph{Shibendu Gupta Choudhury} Registration No. \emph{14IP003} dated 
\emph{July 24, 2014}, a student of the {Department of Physical Sciences} of the
Integrated PhD Programme of Indian Institute of Science Education and Research 
Kolkata (IISER Kolkata), hereby declare that this thesis is my own work and, to 
the best of my knowledge, it neither contains materials previously published or 
written by any other person, nor has it been submitted for any degree/diploma or 
any other academic award anywhere before. I have used the originality checking service to prevent inappropriate copying.\\
 
I also declare that all copyrighted material incorporated into this thesis is in 
compliance with the Indian Copyright
Act, 1957 (amended in 2012) and that I have 
received written permission from the copyright owners for my use of their work.\\
 
I hereby grant permission to IISER Kolkata to store the thesis in a database 
which can be accessed by others.

\vspace{2.cm}
\begin{flushleft}
	\emph{Shibendu Gupta Choudhury}\\
	Department of Physical Sciences\\
	Indian Institute of Science Education and Research Kolkata \\
	Mohanpur 741246, West Bengal, India
\end{flushleft}

\vspace{3.0cm}

\end{declaration}

\cleardoublepage

\addcontentsline{toc}{chapter}{Certificate from the Supervisor}
\thispagestyle{plain}
\null\vfil
{\noindent\huge\bfseries Certificate from the Supervisor \par\vspace{10pt}}

\begin{flushright}
	Date: \emph{December 23, 2022} \\
\end{flushright}
\noindent
\par
{\doublespacing This is to certify that the thesis titled \emph{``Application of the Raychaudhuri Equation in Gravitational Systems''} submitted by Mr.
\emph{\authorname} Registration No. \emph{14IP003} dated \emph{24/07/2014}, a
student of the Department of Physical Sciences of the Integrated PhD
Programme of IISER Kolkata, is based upon his own research work under my 
supervision. I also certify, to the best of my knowledge, that neither
the thesis nor any part of it has been submitted for any degree/diploma or any other academic award
anywhere before. In my opinion, the thesis fulfils the requirement for the award of the degree of Doctor
of Philosophy.} \\

\vspace{2.5cm}

\begin{flushleft}
	\emph{Dr. Ananda Dasgupta} \\
	Associate Professor\\
	Department of Physical Sciences \\
	Indian Institute of Science Education and Research Kolkata\\
    Mohanpur 741246, West Bengal, India
\end{flushleft}

\cleardoublepage

\addcontentsline{toc}{chapter}{Acknowledgements}
\thispagestyle{plain}
{\noindent\huge\bfseries Acknowledgements \par\vspace{6pt}}
It is a pleasure to convey my heartfelt gratitude to the people who have made invaluable contributions toward completing my PhD thesis and progress in my life. I consider their presence as blessings in this journey.

First and foremost, I cannot express my gratitude enough to my supervisor, {\it Dr. Ananda Dasgupta}. I would not be in my position today without his guidance, motivation and belief in me. His advice, insights and patience encouraged me throughout my research, in writing my thesis and in many hurdles in life, whether academic or else. His captivating teaching motivates me and helps me to discover physics at the fundamental level. He made me able to think critically and develop some essential research skills. I could not have imagined having a better advisor and mentor for my PhD study.

Along with my advisor, I am incredibly grateful to {\it Prof. Narayan Banerjee (NB)}, who contributed equally to my research. He inscribed the path of my research with his vision and wisdom. His guidance compelled me to make research a passion and ambition. I am thankful to him for always staying by my side in any research-related or personal problem. He taught me how to pursue quality research while having fun during my PhD journey.

I want to thank {\it Dr. Ritesh K. Singh (RKS)} for his encouragement, insightful comment and valuable suggestions, which improved the quality of my thesis. I also thank him for supporting me during some hard times. I am thankful to {\it RKS} \& {\it NB} for monitoring the progress of my research.

My sincere thanks also go to {\it Dr. Golam Mortuza Hossain}, {\it Prof. Ayan Banerjee}, {\it Prof. Chiranjib Mitra}, {\it Dr. Rumi De}, {\it Prof. Rangeet Bhattacharyya}, {\it Dr. Anandamohan Ghosh}, {\it Prof. Nirmalya Ghosh}, {\it Prof. Bhavtosh Bansal}, {\it  Dr. Siddhartha Lal}, {\it Prof. Supratim Sengupta}, {\it Dr. Koushik Dutta} for helping me in various ways. I thank {\it Prof. Sayan Kar} for his generosity in letting me visit the {\it Indian Institute of Technology, Kharagpur} and for many valuable discussions.

I owe my gratitude to {\it Soumya da} for his encouragement and the many enthralling discussions on research work and life. I am blessed to have a collaborator and senior like him.

I must thank {\it Sangita di}, {\it Munna da} \& {\it Ipsita di} in the departmental office for helping me with all the official work.  I thank people from {\it DOAA, DOSA, DORD, CCC} \& {\it Sports sections, Library, SAC} \& {\it Medical unit}. 

My stay at \emph{IISER Kolkata} was a memorable experience, thanks to all my friends, seniors and juniors. I want to thank {\it Shantanu da} \& {\it Subhajit da} for lending helping hands in many needs. I am grateful to {\it Ankan da}, {\it Gopal da}, {\it Chiranjeeb da}, {\it Sachin da}, {\it Avijit da} \& {\it Srijita di} for their valuable suggestions.

Words cannot express how much I am grateful to my friends {\it Tanima}, {\it Prashanti}, {\it Sampurna}, {\it Shreya}, {\it Purba}, {\it Basabendra} \& {\it Sourav}. They are lovely companions who are always with me during my ups and downs and are the closest to me in {\it IISER Kolkata}. I hope our bonding will become stronger as we grow old and experienced.

Special thanks to {\it Dipanjan da} for his help at different times. I thank {\it Debangana, Avijit, Tapas da, Toushik da, Priyanka di, Manna, Sayani, Bishnu, Anjan, Soumya, Debojyoti,  Narayan, Gopal, Debraj, Medha, Siddhartha, Abhirup, Diganta, Swarup, Brataraj} for making this journey memorable. {\it Budhaditya, Poulomi, Saikat, Arkayan, Amulya} \& {\it Chiranjit} deserve a special mention. We all have had quite a lot of cheerful times together. I should also thank friends from my college, schools and other places - {\it Arindam}, who is very special to me, {\it Arghya, Rathin, Sudeep, Subhajit, Atanu, Nupur, Bratati, Lovely, Sulagna, Swarna} and the list goes on. I wish all my friends, seniors and juniors a lot of success.

I want to express my regards and gratitude to my beloved friend, the late {\it Subhadip Roy}, who cared for and protected me during some tough times. A day does not go by when I don’t miss you. 

I am always grateful to have teachers like {\it Dr. Ashim K. Mukherjee, Dr. Parimal Ghosh}, whose teaching, guidance and encouragement motivated me to pursue my dreams. I thank all my teachers at {\it Asansol BB College}, {\it Andal High School} \& {\it Andal E. Rly. High School}.

I will never forget the support and kindness of {\it Swami Mahamedhananda}, whose helping hands took me out of some intimidating situations. He always motivates me to follow my dream and passion.

I am eternally indebted to my parents, {\it Susmita Gupta Choudhury} \& {\it Hiralal Gupta Choudhury}, for their unconditional love and support. I am grateful to my elder sister {\it Aparajita Gupta Choudhury}, her husband, {\it Sagnik Sen} \& her son {\it Richak Sen} for everything they have done for me. My family is my lifeline and the most extensive support in my darkest hours. I want to thank all my relatives who care for and love me.

I sincerely thank the {\it Council of Scientific and Industrial Research, India}, for providing financial support
through the CSIR-NET fellowship (Award No. 09/921(0188)/2017-EMR-I) during the PhD programme. 

Finally, I want to thank the {\it IISER Kolkata campus} for always healing my wounds and keeping me happy.

\cleardoublepage
\dedicatory{Dedicated to my parents\ldots}
\cleardoublepage
\addcontentsline{toc}{chapter}{Abstract}
\thispagestyle{plain}
\null\vfil
{\noindent\huge\bfseries Abstract \par\vspace{12pt}}

The works reported in this thesis primarily address the application of the Raychaudhuri equation in two intriguing problems in gravitational physics. These problems still lack universally accepted explanations. The first problem is related to the existence of spacetime singularities. We aim to find possible escape routes from these problematic singularities at the classical level. The second problem is associated with the late time accelerated expansion of the Universe. In this context, our goal is to find a possible explanation of this phenomenon without assuming the presence of any exotic contribution to the stress-energy tensor.

To investigate the problem of singularities, we begin with the study of gravitational collapse within the scope of General Relativity. We consider two different gravitational collapse scenarios in our study. At first, we examine a self-similar gravitational collapse of a matter distribution containing a fluid and a scalar field. We work under the assumption of spherical symmetry and conformal flatness. We obtain the general focusing condition for this system. This condition is quite helpful in getting insights into the corresponding evolution. We show that the general condition leads to crucial constraints on the metric and matter variables for a few special cases. These constraints distinguish between circumstances where a singularity is inevitable and where it can be avoided. The results obtained from the focusing condition are verified against exact solutions whenever available. Using both approaches, i.e., employing the focusing condition and the exact solutions, we find that non-singular evolution is possible for the system under consideration. We also explore the connection between the Raychaudhuri equation and the critical phenomena in gravitational collapse.

The second system addresses the role of magnetic fields in a gravitational collapse. There is an inherent tendency of magnetic fields to act against gravity due to the repulsion between magnetic field lines. Inhomogeneous models with no restriction on the magnetic field strength to begin with have not been studied significantly in the literature. We investigate such general models of gravitational collapse in our work. Our system consists of a charged fluid distribution collapsing in the presence of a magnetic field under cylindrical symmetry. For such systems, we find the necessary constraints that the magnetic field strength needs to obey to avert collapse.

Next, we explore if it is possible to avoid singularities in a class of modified theories of gravity, namely the scalar-tensor theories of gravity. We study focusing of timelike geodesic congruences, particularly the fate of the timelike convergence condition within this framework. In General Relativity, the strong energy condition implies the timelike convergence condition. The latter is a necessary assumption in the proof of the singularity theorem. But for a scalar-tensor theory, such convergence conditions may not necessarily follow from the energy conditions as the field equations in this theory contain additional terms. We choose the Brans-Dicke theory and Bekenstein's conformally coupled scalar-tensor theory as specific examples within the broad class of non-minimally coupled scalar-tensor theories. Additionally, we consider static, spherically symmetric and spatially homogeneous and isotropic backgrounds for our study. We observe that violation of the convergence condition is possible in both theories. This gives rise to the possibility of avoiding singularities.

Finally, we examine the phenomenon of late time accelerated expansion of the Universe. We will look into this within the purview of the $f(R)$-gravity. In $f(R)$-gravity, repulsive effects can emerge from the geometry itself, and therefore we do not need to introduce any exotic matter to explain this phenomenon. We find a new strategy for reconstructing $f(R)$-gravity models for an accelerated universe employing the Raychaudhuri equation. We investigate two different models -  one of them represents an ever-accelerating universe. In the other one, the evolution mimics the Lambda Cold Dark Matter ($\Lambda$CDM) expansion history. We start by characterizing these evolution histories in terms of the corresponding kinematical quantities, namely the deceleration parameter and the jerk parameter, respectively. For the first example, we find that a combination of power-law terms gives the expression for $f(R)$. In the second example, the form of $f(R)$ involves hypergeometric functions. Viability analysis of these models reveals that the corresponding $f(R)$-gravity models for both examples are unsuitable options.

\cleardoublepage


\doublespacing

\addchaptertocentry{List of Publications}
\thispagestyle{plain}
\null\vfil
{\noindent\huge\bfseries List of Publications \par \vspace{12pt}}

\onehalfspacing

\paragraph*{Publications included in the thesis:}
  \begin{numitemise}
  
        \item {\textbf{Shibendu Gupta Choudhury}, Soumya Chakrabarti, Ananda Dasgupta and Narayan Banerjee, {\it Self Similar Collapse and the Raychaudhuri equation}, \href{https://doi.org/10.1140/epjc/s10052-019-7559-9}{Eur. Phys. J. C {\bf 79} (2019) 1027}, \href{https://doi.org/10.48550/arXiv.1912.05824}{arXiv:1912.05824}. (Chapter 2)}
        
        \item {\textbf{Shibendu Gupta Choudhury}, {\it Role of a magnetic field in the context of inhomogeneous gravitational collapse}, \href{https://doi.org/10.1140/epjp/s13360-022-03205-5}{Eur. Phys. J. Plus {\bf 137} (2022) 971}, \href{https://doi.org/10.48550/arXiv.2203.14877}{arXiv:2203.14877}. (Chapter 3)}

     	\item {\textbf{Shibendu Gupta Choudhury}, Ananda Dasgupta and Narayan Banerjee, {\it Raychaudhuri equation in scalar-tensor theory}, \href{https://doi.org/10.1142/S0219887821501152}{Int. J. Geom. Meth. Mod. Phys. {\bf 18} (2021) 08, 2150115}, \href{https://doi.org/10.48550/arXiv.2103.08869
}{arXiv:2103.088694}. (Chapter 4)}

        \item {\textbf{Shibendu Gupta Choudhury}, Ananda Dasgupta and Narayan Banerjee, {\it Reconstruction of f(R) gravity models for an accelerated universe using Raychaudhuri equation}, \href{https://doi.org/10.1093/mnras/stz731}{Mon. Not. Roy. Astron. Soc. {\bf 485} (2019) 4, 5693}, \href{https://doi.org/10.48550/arXiv.1903.04775}{arXiv:1903.04775}. (Chapter 5)}
  	
\end{numitemise}

\paragraph*{Other publications:}

\begin{numitemise}

   \item{\textbf{Shibendu Gupta Choudhury}, Ananda Dasgupta and Narayan Banerjee, {\it The Raychaudhuri equation for a quantized timelike geodesic congruence}, \href{https://doi.org/10.1140/epjc/s10052-021-09714-4}{Eur. Phys. J. C {\bf 81} (2021) 906}, \href{https://doi.org/10.48550/arXiv.2103.13203}{arXiv:2103.13203}.}

\end{numitemise}

\cleardoublepage


\begin{abbreviations}{ll} 
\textbf{$\Lambda$CDM} &  {\bf L}ambda {\bf C}old {\bf D}ark {\bf M}atter\\[1ex]
\textbf{BBMB} & {\bf B}ocharova--{\bf B}ronnikov--{\bf M}elnikov--{\bf B}ekenstein\\[1ex]
\textbf{BCCSTT} & {\bf B}ekenstein {\bf C}onformally {\bf C}oupled {\bf S}calar--{\bf T}ensor {\bf T}heory (BCCSTT)\\[1ex]
\textbf{BD} & {\bf B}rans--{\bf D}icke \\[1ex]
\textbf{CMBR} & {\bf C}osmic {\bf M}icrowave {\bf B}ackground {\bf R}adiation \\[1ex]
\textbf{CP} & {\bf C}ritical {\bf P}henomena\\[1ex]
\textbf{ESR} & {\bf E}xpansion, {\bf S}hear, {\bf R}otation\\[1ex]
\textbf{FC} & \textbf{F}ocusing \textbf{C}ondition\\[1ex]                                                                                                                                              
\textbf{FRW} & {\bf F}riedmann--{\bf R}obertson--{\bf W}alker\\[1ex]
\textbf{FT} & \textbf{F}ocusing \textbf{T}heorem \\[1ex]
\textbf{GTR} & \textbf{G}eneral \textbf{T}heory of \textbf{R}elativity\\[1ex]
\textbf{GW} & \textbf{G}ravitational \textbf{W}aves\\[1ex]
\textbf{NCC} & \textbf{N}ull \textbf{C}onvergence \textbf{C}ondition\\[1ex]
\textbf{NEC} & \textbf{N}ull \textbf{E}nergy \textbf{C}ondition\\[1ex]
\textbf{NMCSTT} & \textbf{N}on--\textbf{M}inimally  \textbf{C}oupled \textbf{S}calar--\textbf{T}ensor \textbf{T}heories\\[1ex]
\textbf{QFT} & \textbf{Q}uantum  \textbf{F}ield \textbf{T}heory\\[1ex]
\textbf{RE} & \textbf{R}aychaudhuri \textbf{E}quation\\[1ex]
\textbf{SEC} & \textbf{S}trong \textbf{E}nergy \textbf{C}ondition\\[1ex]
\textbf{STR} & \textbf{S}pecial \textbf{T}heory of \textbf{R}elativity\\[1ex]
{\bf TCC} & {\bf T}imelike {\bf C}onvergence {\bf C}ondition\\[1ex]
\textbf{WEC} & \textbf{W}eak \textbf{E}nergy \textbf{C}ondition
\end{abbreviations}

\cleardoublepage

\tableofcontents 
\cleardoublepage
\listoffigures 
\cleardoublepage

\mainmatter 

\pagestyle{thesis} 


\def\rs{r_{s}}
\def\rh{r_H}
\def\rstar{r_{\star}}
\def\rsh{r_{\square}}
\def\roplus{r_{\oplus}}
\def\scriplus{\mathscr{I}^{+}}
\def\scriminus{\mathscr{I}^{-}}

\def\observerminus{\mathbb{O}^{-}}
\def\observerplus{\mathbb{O}^{+}}
\def\kr{\kappa}
\def\ksg{\mathrm{\varkappa}}
\def\rhash{r_{\sharp}}

\def\Phit{\tilde{\Phi}}
\def\omegat{\tilde{\omega}}
\def\mstar{m_{\star}}
\def\mk{\mathbb{K}}

\def\Tr{\mathrm{Tr}\hspace{1pt}}
\def\varphit{\tilde{\varphi}}
\def\pit{\tilde{\pi}}

\def\lstar{l_{\star}}
\def\sstar{s_{\star}}

\chapter{Introduction} 
\label{chapter1}

\section{General theory of relativity: successes and challenges}
Einstein's {\it General Theory of Relativity (GTR)} \cite{Misner:1974qy, Wald:1984rg, Carroll:1997ar, Carroll:2004st, JayantVNarlikar:2010vvf, d1992introducing, hooft2001introduction, book:Schutz, hartle2003gravity} is one of the pioneering concepts of physics. This theory based on {\it Riemannian geometry} was proposed in its final form by Einstein in 1915. Gravity is described in terms of geometry within the framework of GTR. One can sum up Einstein's GTR by the statement  {\it ``matter tells spacetime how to curve and curved spacetime tells matter how to move''} \cite{Misner:1974qy}. This statement means that the matter content of the spacetime dictates the curvature of the spacetime, and this curvature, in turn, determines the motion of test particles. The statement {\it ``matter tells spacetime how to curve''} is written in terms of a set of mathematical equations (in natural units where the speed of light $c=1$) as,
\begin{equation}\label{es}
G_{ab}=R_{ab}-\frac{1}{2}g_{ab}R=8\pi G T_{ab}.
\end{equation}
These justly famous equations are known as \emph{Einstein's field equations}.
Here, $G$ is Newton's gravitational constant; $g_{ab}$ is the metric tensor which determines the geometry of the spacetime. $R_{ab}$ is the Ricci tensor, $R$ is the Ricci scalar and $T_{ab}$ is the energy-momentum tensor. As is usual in the field of gravitation we use a metric with the signature $\left(-, +, +, +\right)$ (see appendix \ref{appenpre} for a brief review).

\subsection{Achievements}
GTR is the most successful theory of gravity to date. Among other things, this theory has successfully explained the perihelion shift of Mercury's orbit. Newtonian gravity could not fully account for this phenomenon. {\it Gravitational time dilation} and redshift, the bending of light etc. are important predictions of GTR. During a solar eclipse in 1919, Eddington demonstrated that deflection of sunlight occurs in accordance with GTR. Einstein's theory of gravity achieved worldwide recognition following this work. In the weak field limit, GTR reproduces the results of Newtonian gravity. The most recent successes of this theory are the detection of {\it Gravitational waves (GW)}\cite{Abbott:2016blz, Abbott:2016nmj} and the image of the \emph{shadow of the supermassive black hole M87*} at the centre of the galaxy Messier 87 \cite{Akiyama:2019cqa, Akiyama:2019brx, Akiyama:2019sww, Akiyama:2019bqs, Akiyama:2019fyp, Akiyama:2019eap}. GW has been detected nearly after a century of its prediction by Einstein. The Kerr black hole solution of GTR explains the shadow of M87* consistently. The discovery of GTR led to the study of the Universe as a gravitational system, giving birth to modern {\it Cosmology}. GTR explains different observational aspects of our Universe quite well. These, and related findings, keep on emphasizing the essential truth behind Einstein's theory a century after he first described it.

\subsection{Problem of singularities}\label{psvbd}
In spite of its huge achievement in describing physical reality, there are a few drawbacks of GTR. The most worrying among them is that GTR predicts the appearance of singularities in spacetimes. Spacetime curvature, energy density and all other physical quantities diverge when a spacetime singularity is approached. Thus, the theory itself breaks down. The first-ever solution of Einstein's field equations, called the {\it Schwarzschild solution}\cite{Misner:1974qy} contains a singularity at the centre of symmetry $r=0$. The cosmological {\it {F}riedmann--{R}obertson--{W}alker (FRW) spacetime}\cite{Misner:1974qy} contains a singularity at the beginning of the Universe. It was initially believed that singularities are artefacts of the symmetry assumptions made in these specific solutions, and will be absent in more general spacetimes. In the year 1965, the works of Penrose\cite{Penrose:1964wq}, and after that, those of Hawking\cite{Hawking:1965mf, Hawking:1966vg} established that under a few physically reasonable assumptions (positivity of energy, causality, the existence of trapped surfaces etc.) singularities of spacetime are inevitable within a general framework. Elegant and detailed discussions regarding the {\it Singularity theorems} can be found in
\cite{Wald:1984rg, Hawking:1973uf, Joshi:1987wg}. A key ingredient of the proof of the singularity theorems was a result published by an Indian scientist, Amal Kumar Raychaudhuri, nearly a decade before this. This result, the RE, forms the central theme of this thesis.


\section{The Raychaudhuri equation: a brief introduction}
The appearance of singularities is a major drawback of any physical theory. Resolution of this issue is necessary for a complete physical description. Therefore, despite being troublesome,
singularities are one of the central research topics in relativity and astrophysics \cite{Joshi:2008zz, Joshi:2012mk}.

To resolve the singularity issue, we must first know how to define them. We shall begin with a precise definition of a spacetime singularity. The notion of {\it geodesic incompleteness}
effectively characterizes a singularity \cite{Hawking:1973uf}. The existence of incomplete timelike and null geodesics, in turn, leads to the appearance of singularities by means of singularity theorems. These theorems are based on the focusing of a \emph{geodesic congruence} within a finite affine parameter value \cite{Wald:1984rg}. It is, therefore, necessary to study the behaviour of congruence of geodesics to characterize a singularity. It is to be noted that the structure and different geometric features of spacetime, as a whole, are reflected through the evolution of geodesic congruences.
There are a few kinematic variables called \emph{expansion}, \emph{shear} and
\emph{rotation}, respectively which describe the evolution of these congruences \cite{book:Poisson, Wald:1984rg}. The behaviour
of these variables is dictated by \emph{Raychaudhuri equation(s) (RE(s))}\cite{Raychaudhuri:1953yv, Ehlers:1993gf}. Focusing of geodesic congruences is an important consequence of
this equation. Thus, the RE is an important tool for dealing with spacetime geometry and singularities.

As mentioned earlier, the RE forms the basis of our present work. We have used this equation to study different gravitational systems. We aim to explore the vast potential for the application of this equation. We shall briefly discuss the history and significance of this equation in the following.

Raychaudhuri started to investigate the subject of singularities in the 1950s. His primary focus was the generic nature of cosmological singularity. He began to search for non-singular solutions during this time. His works in this era include studies regarding {\it cosmological perturbations}, construction of non-static solutions of Einstein's equations etc. In 1955, his paper titled {\it Relativistic Cosmology I}\cite{Raychaudhuri:1953yv} was published. This paper contains a derivation of the equation known today as the RE. The form and derivation of the equation are somewhat different from those discussed today in the standard literature. The approach followed in his subsequent paper in 1957\cite{1957Rayc} bears similarity with that followed nowadays. The original version of this equation considers the geodesic motion in a dust universe. Jordan, Ehlers, Kundt and Sachs introduced the general form of this equation which we see today\cite{Ehlers:1993gf}. In the abovementioned works, the evolution equation for the expansion parameter is referred to as the RE. In \cite{Ehlers:1993gf}, the evolution equations for shear and rotation appeared for the first time. All three evolution equations are sometimes referred to as the REs. The RE for a null geodesic congruence was introduced by Sachs\cite{Sachs:1961zz, Sachs:1962wk}.

 Although the RE was referred to and used in different contexts following the works mentioned above, it came into the limelight after Penrose and Hawking\cite{Penrose:1964wq, Hawking:1965mf, Hawking:1966vg} reported their seminal works on singularity theorems. According to George Ellis, {\it ``it is the fundamental equation of gravitational attraction''}\cite{2007Prama..69...15E}. Writing about \cite{Raychaudhuri:1953yv}, Jurgen Ehlers said, {\it ``The paper reviewed here is truly a landmark in cosmology. It showed how results are in principle observable and local variables can be obtained without imposition of isometries. It can be considered a forerunner of a programme begun by J Kristian and R K Sachs and continued by G Ellis and his coworkers. Moreover, it paved the way to the Penrose-Hawking singularity theorems''}\cite{article}. The importance of the RE was further stressed by J. Earman, {\it ``Further evidence was introduced in 1955 that, in the cosmological context at least, singularities in general relativistic spacetimes are not artifacts of symmetry assumption. The evidence came not from one of the havens of general relativity in America or Europe, but from Calcutta. In 1953 Raychaudhuri produced an analysis of cosmological singularities that was to have a profound effect on later developments but because it ran into difficulties with referees, the paper did not appear until 1955''}\cite{goenner1998expanding}.

\section{The Raychaudhuri equation: significance}
The major consequence of the RE in GTR is the \emph{Focusing Theorem}\cite{book:Poisson, wald1984general}. This theorem states that a collection of geodesics (called a congruence) in spacetime must focus within finite affine parameter value if certain energy conditions hold. As already mentioned, this theorem plays a central role in proving the existence of singularities\cite{Penrose:1964wq, Hawking:1965mf, Hawking:1966vg}. These kinds of generic conclusions can be extracted using the RE without referring to any particular scenario. Thus, this equation enables us to analyze the structure of spacetime from a general point of view, especially when finding exact solutions of field equations is difficult. The RE has been applied to study {\it Gravitational collapse} and explore whether there exist situations in which a collapse can be halted\cite{Germani:2005ar, Tsagas:2006sh, Kouretsis:2010nu, Tsagas:2020lal}. This, in turn, leads to a possibility of avoiding the final crunch resulting in a singularity. A study of the structure and geometric features of different spacetime backgrounds through the application of the RE can be found in \cite{Dasgupta:2008in, Ghosh:2010gq, Dasgupta:2012zf, Shaikh:2014xna}. This equation is also a key equation in Cosmology. The application of this equation in different contexts of Cosmology can be found in  \cite{Tsagas:1997vf, Tsagas:2000zq, Tsagas:1999ft, Barrow:2006ch, Tsagas:2007yx, 5caada012a0341acb9a414c39b49a32a, Spyrou:2008ub, Lyth:1988pj} and references therein. In {\it Relativistic Astrophysics}, application of this equation includes role of voids in light propagation\cite{Sugiura:1999fm}, study of crack formation\cite{Herrera:1992lwz}, nonlinear hydrodynamics and {\it magnetohydrodynamics}\cite{DiPrisco:1994np} etc.
The RE has been employed in the context of thermodynamics of spacetimes in \cite{Jacobson:1995ab, Akbar:2006mq}. This equation plays a crucial role in deriving the laws of {\it black hole mechanics}. For a beautiful discussion regarding this subject, we refer to the book by Poisson\cite{book:Poisson}. The extensive use of the RE to study black hole thermodynamics has been discussed in the brief review \cite{Sarkar:2019xfd}. It should be mentioned that an emerging area of application of the RE is the GW memory effect\cite{Chakraborty:2019yxn}.

It is to be noted that in its general form, the RE follows from the assumption of Riemannian geometry only. It does not need any specific theory of gravity at the outset. This is why the RE has a wide range of applicability. This equation can easily be applied to different modified theories of gravity, e.g. \emph{$f(R)$-theory}\cite{Capozziello:2003gx, Nojiri:2003rz, Nojiri:2003ft,  Carroll:2003wy, Das:2005bn}, \emph{Brans--Dicke (BD) theory}\cite{Brans:1961sx}, {\it Hoyle-Narlikar theory}\cite{1962RSPSA.270..334H, hoyle1963mach}, {\it Gauss-Bonnet theory}\cite{Lovelock:1971yv, Lovelock:1972vz, Lanczos:1938sf, Padmanabhan:2013xyr} etc., where the Riemannian structure of the spacetime is respected. Application of the RE in different modified theories of gravity can be found in \cite{Santos:2007bs, Albareti:2012va, Santos:2016vjg, Burger:2018hpz, Burger:2018hpz, 1964ZA58187R, Banerji:1974ge,  Kung:1995he, 1975PThPh531360R, Page:1987cb, Muller:1985mr, Barrow:1983rx, Kung:1995nh}.


A considerable amount of research is being devoted to studying the application of the RE in quantum settings at current times. In the absence of a well-established quantum theory of gravity, this approach helps us understand whether the problem of singularities can be resolved in the quantum regime. A quantum version of the RE was first reported by Das\cite{Das:2013oda}. In this work, using the concept of {\it Bohmian trajectories}, the author established that geodesic congruences do not focus due to the existence of quantum potentials. The application of the RE to establish singularity avoidance in {\it Loop quantum cosmology} can be found in the references \cite{Bojowald:2001xe, Ashtekar:2008ay, Singh:2009mz, articleli}. Similar applications in other different contexts are there in the literature\cite{Burger:2018hpz, Ali:2014qla, Ali:2015tva, Ashtekar:2005qt, Moti:2019mws, Chakraborty:2019vki, Alsaleh:2017ozf, Choudhury:2021huy}.

Being a geometric result, the RE is found to be useful in many different fields beyond the fields of Gravitation, Cosmology and Astrophysics. This includes study of relativistic membranes\cite{Capovilla:1994yk, Kar:1995pd, Kar:1996rx, Carter:1996wr, Zafiris:1997he} and deformable medias\cite{Dasgupta:2008jt, Dasgupta:2007nr}, different contexts  in {\it Quantum Field Theory (QFT)}\cite{Borgman:2003dm, Kar:2001qm, Alvarez:1998wr, Sahakian:1999bd, Balasubramanian:2000wv, Goldstein:2005rr, Bhattacharjee:2015qaa}, {\it Classical Mechanics}\cite{Shaikh:2013iea} etc.

For a beautiful discussion regarding the history, background, significance and applications of the RE, we refer to the special issue of {\it Pramana – Journal of Physics} - Volume 69, Issue 1, with specific emphasis on \cite{2007Prama..69...15E, article, Dadhich:2007pi, Kar:2006ms}.

We will now discuss the concepts essential for the central part of this thesis. This includes derivation and consequences of the RE. We will mainly follow \cite{book:Poisson, wald1984general} for this discussion.

\section{Timelike congruences and derivation of the Raychaudhuri equation}\label{dre}\sectionmark{Timelike congruences and derivation of the RE}
The definition of a {\it congruence} is as follows. Let us consider a manifold $M$. $O\subset M$ is open. A family of curves in $O$ is called a congruence if precisely one curve in this family passes through each $p \in O$.

In this thesis, we are concerned with timelike congruences. Therefore, we will only discuss this particular class of congruences in the following. We refer the reader to Appendix \ref{appen1} for a discussion on the null case.

\subsection{Congruence of timelike geodesics and the Raychaudhuri equation}\label{tmlkc}
We will first consider the case of timelike geodesic congruences (see appendix \ref{geodesics} for a brief review on geodesics). In this case, one can always choose proper time $\tau$ as the affine parameter of the geodesics. The vector field $u^a$ of tangents to the geodesics is normalized as $u^a u_a = -1$.
Then a purely spatial tensor field can be defined by,
\begin{equation}
B_{ab} = \nabla_b u_a,
\end{equation}
where $\nabla$ denotes the {\it covariant derivative}.
It is easy to prove that,
\begin{equation}\label{eq 9.2.2}
 B_{ab} u^a= B_{ab} u^b =0,
\end{equation}
because $u^a$ satisfies the geodesic equation and is normalized.

\begin{figure}
 \centering
 \frame{\includegraphics[width=0.6\linewidth]{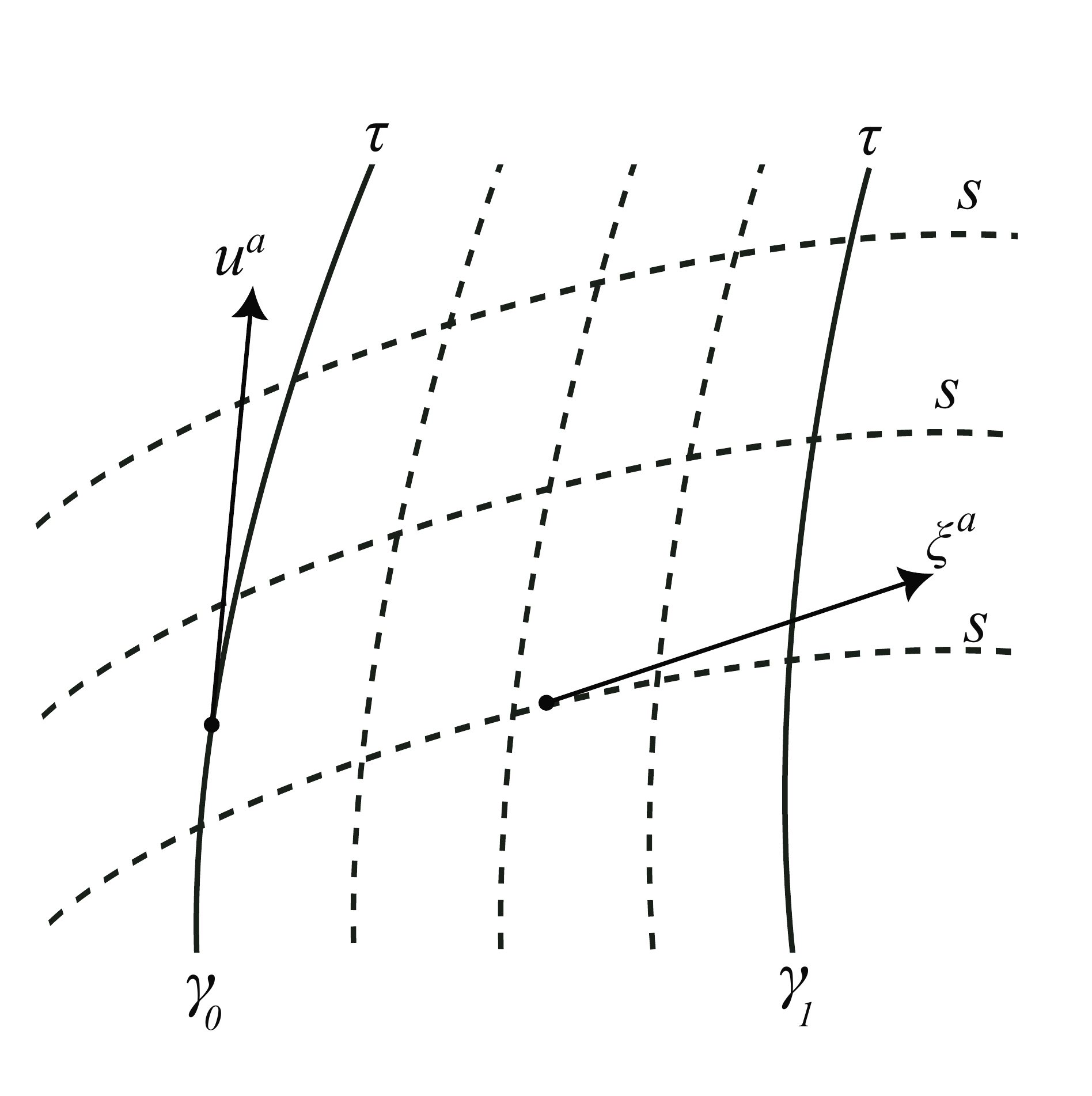}}
 \caption{Deviation vector between two neighboring geodesics.}\label{fig0}
\end{figure}

To understand the physical interpretation of $B_{ab}$, let us consider a one-parameter subfamily $\gamma_s(\tau)$ of geodesics in the congruence, where $s$ is a labeling parameter.
We can represent the geodesics in the congruence collectively with relations $x^a(\tau,s)$.
The orthogonal deviation vector from a reference geodesic $\gamma_0$ in this subfamily is given by $\xi^a$ (see figure \ref{fig0}). Now, using $u^a=\frac{\partial{x^a}}{\partial{\tau}}$ and $\xi^a=\frac{\partial{x^a}}{\partial{s}}$ we have (see appendix \ref{geoddevdis}),
\begin{equation}\label{eq 9.2.4}
 u^b\nabla_b\xi^a=\xi^b\nabla_bu^a={B^a}_b\xi^b.
\end{equation}
Thus ${B^a}_b$ is a measure of the failure of $\xi^a$ to be parallelly transported. An observer sitting on $\gamma_0$ finds geodesics near it to be stretched
and rotated. This is characterized by the linear map ${B^a}_b$.

The \textit{spatial metric} $h_{ab}$ is defined as,
\begin{equation}
 h_{ab}=g_{ab}+u_au_b.
\end{equation}
We can use ${h^a}_b=g^{ac}h_{cb}$ as the projection operator onto the subspace which is the tangent space perpendicular to $u^a$. The \textit{expansion} $\theta$, \textit{shear} $\sigma_{ab}$ and \textit{rotation} (or, \textit{vorticity}) $\omega_{ab}$ of
the congruence are defined as the trace, symmetric traceless and antisymmetric parts of $B_{ab}$ -
\begin{equation}
 \theta=h^{ab}B_{ab},
\end{equation}
\begin{equation}
 \sigma_{ab}=B_{(ab)}-\frac{1}{3}\theta h_{ab},
\end{equation}
\begin{equation}
 \omega_{ab}=B_{[ab]},
\end{equation}
where the round bracket stands for the symmetric part and the square bracket for the anti-symmetric part. Therefore, one can decompose  $B_{ab}$ as,
\begin{equation}
 B_{ab}=\frac{1}{3}\theta h_{ab}+\sigma_{ab}+\omega_{ab}.
\end{equation}
 It follows from Frobenius's theorem\cite{book:Poisson, wald1984general} that a congruence is hypersurface orthogonal if and only if $\omega_{ab}=0$.

\begin{figure}
 \centering
 \frame{\begin{tikzpicture}
 \draw (5,4) .. controls (7,7) and (7,4) .. (15,6);
 \draw (5,3) .. controls (7,7) and (7,4) .. (15,5);
 \draw (5,5) .. controls (7,7) and (7,4) .. (15,7);
 \draw (5,7) .. controls (7,7) and (7,4) .. (15,8);
 \draw (9,5.5) ellipse (0.4cm and 0.8cm);
 \end{tikzpicture}}
 \caption{Area enclosing flow lines within a congruence. \label{fig1}}
\end{figure}
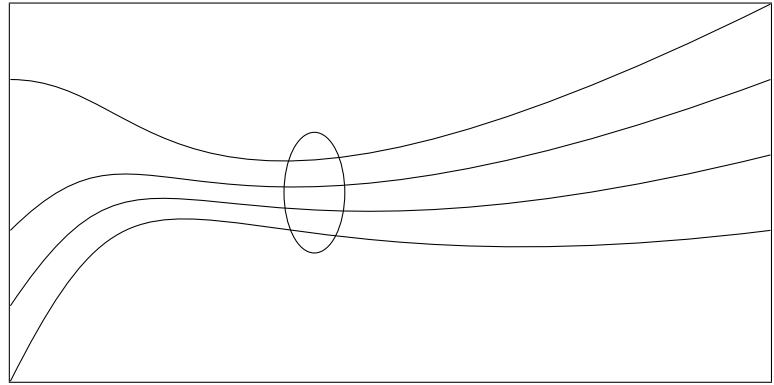

\begin{figure}
 \centering
 \frame{\includegraphics[width=0.45\textwidth]{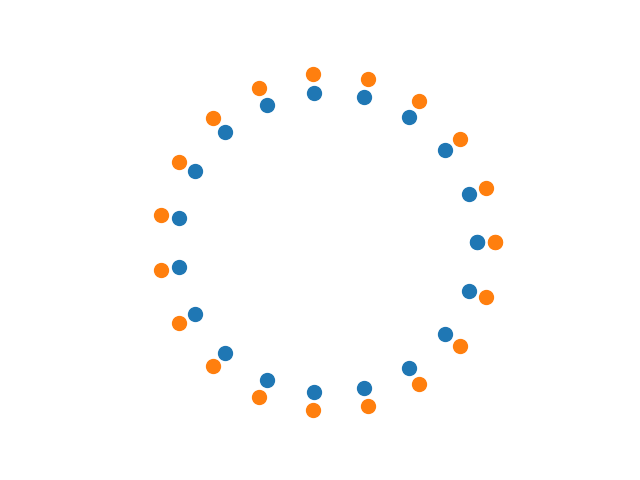} \includegraphics[width=0.45\textwidth]{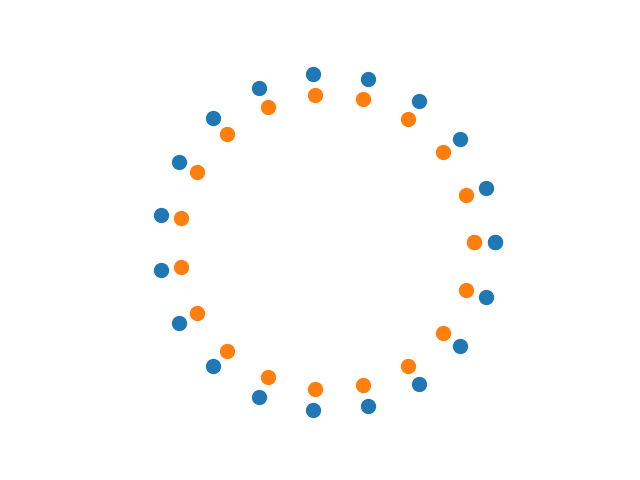}}
 \caption{Pictorial representation of the role of expansion parameter $\theta$ through the change of shape of the cross-sectional area drawn in figure \ref{fig1}. In this, and in figures \ref{fig3} \& \ref{fig4}, blue and orange dots represent initial and final configurations respectively.\label{fig2}}
\end{figure}

\begin{figure}
 \centering
 \frame{\includegraphics[width=0.45\textwidth]{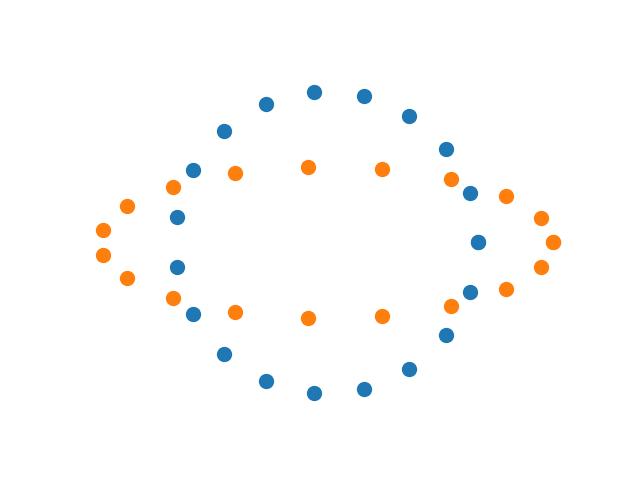} \includegraphics[width=0.45\textwidth]{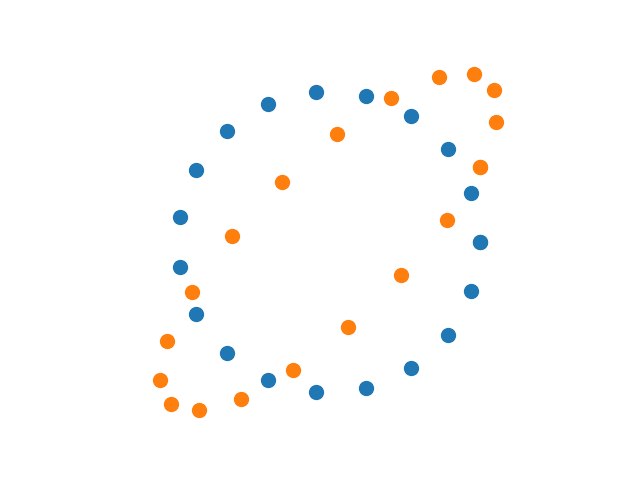}}
 \caption{Pictorial representation of the role of shear $\sigma_{ab}$ through the change of shape of the cross-sectional area drawn in figure \ref{fig1}. \label{fig3}}
\end{figure}

\begin{figure}
 \centering
 \frame{\includegraphics[width=0.6\linewidth]{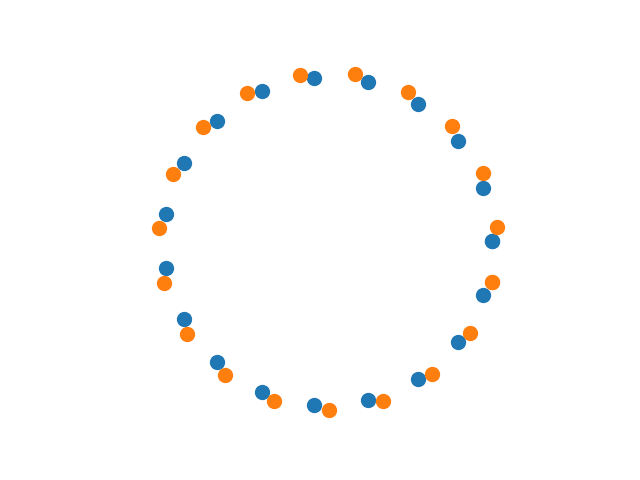}}
 \caption{Pictorial representation of the role of rotation $\omega_{ab}$ through the change of shape of the cross-sectional area drawn in figure \ref{fig1}. \label{fig4}}
\end{figure}

Using equation \eqref{eq 9.2.2}, we have,
$\sigma_{ab}u^b=\omega_{ab}u^b=0$, i.e. $\sigma_{ab}$ and $\omega_{ab}$  are purely spatial.
It can be understood from equation \eqref{eq 9.2.4} that if we take a reference geodesic in the congruence, $\theta$ is a measure of the average expansion of the infinitesimally nearby
surrounding geodesics. $\omega_{ab}$ and $\sigma_{ab}$ measure their rotation and shear respectively. We use a pictorial representation to understand the geometric interpretation of Expansion, Shear and Rotation (ESR) more clearly. Let us consider the congruence as a collection of flow lines (geodesics) as shown in figure \ref{fig1}.
The cross-sectional area enclosing the geodesics may evolve as one moves along the congruence. If the geodesics come closer (or go apart) the area will decrease (or increase) (see figure \ref{fig2}). This behaviour is encapsulated in $\theta$. The shape of the area may also be sheared or twisted (see figures \ref{fig3} and \ref{fig4}). These latter changes are described by $\sigma_{ab}$ and $\omega_{ab}$ respectively.

Let us now derive the evolution equations for the ESR -
\begin{equation}\label{eq 9.2.10}
\begin{split}
 u^c\nabla_cB_{ab} &=u^c\nabla_c \nabla_bu_a=u^c\nabla_b\nabla_c u_a+{R_{cba}}^d u^cu_d \\
 &=\nabla_b\left(u^c\nabla_cu_a\right)-\left(\nabla_bu^c\right)\left(\nabla_cu_a\right)+{R_{cba}}^du^cu_d \\
 &=-{B^c}_b B_{ac}+{R_{cba}}^d u^cu_d,
\end{split}
\end{equation}
where we have used the relation defining the {\it Riemann tensor} ${R^a}_{bcd}$ (see appendix \ref{rieric}),
\begin{equation}
 \left[\nabla_a, \nabla_b\right]u^c=\left(\nabla_a\nabla_b- \nabla_b\nabla_a\right) u^c= -{R^c}_{dab}u^d.
\end{equation}

The trace of equation \eqref{eq 9.2.10} yields,
\begin{equation}\label{rc eq}
 u^c\nabla_c \theta=\frac{d\theta}{d\tau}=-\frac{1}{3}\theta^2-\sigma_{ab}\sigma^{ab}+\omega_{ab}\omega^{ab}-R_{cd} u^cu^d.
\end{equation}
The {\it Ricci tensor} $R_{cd}$ is defined as the contraction (see appendix \ref{ret}),
\begin{equation}
 R_{cd} = {R^a}_{cad}.
\end{equation}

From the symmetric, trace-free part of equation \eqref{eq 9.2.10}, we have,
\begin{equation}\label{rc sig}
 \begin{split}
  u^c\nabla_c\sigma_{ab}&=-\frac{2}{3} \theta\sigma_{ab}-\sigma_{ac}{\sigma^c}_b-\omega_{ac}{\omega^c}_b+\frac{1}{3}h_{ab}\left(\sigma_{cd}\sigma^{cd}-\omega_{cd}\omega^{cd}\right)\\
&+C_{cbad}u^cu^d+\frac{1}{2}\tilde{R}_{ab},
 \end{split}
\end{equation}
where $\tilde{R}_{ab}$ is the trace-free, spatial part of $R_{ab}$,
\begin{equation}
 \tilde{R}_{ab}=h_{ac}h_{bd}R^{cd}-\frac{1}{3}h_{ab}h_{cd}R^{cd},
\end{equation}
and $C_{cbad}$ is the {\it Weyl tensor} defined as,
\begin{equation}
 C_{abcd}=R_{abcd}+\frac{1}{2}\left(R_{ad}g_{bc}-R_{ac}g_{bd}+R_{bc}g_{ad}-R_{bd}g_{ac}\right)+\frac{1}{6}R\left(g_{ac}g_{bd}-g_{ad}g_{bc}\right).
\end{equation}

From the anti-symmetric part of equation \eqref{eq 9.2.10}, we obtain,
\begin{equation}\label{rc om}
 u^c\nabla_c\omega_{ab}=-\frac{2}{3}\theta \omega_{ab}-2{\sigma^c}_{[b}\omega_{a]c}.
\end{equation}
Equation \eqref{rc eq}, which describes the evolution of $\theta$, is known as the RE for timelike geodesic congruences. Sometimes the term {\it REs} is used to refer to all of the three equations \eqref{rc eq}, \eqref{rc sig} and \eqref{rc om}.

\subsection{Energy conditions, convergence conditions and focusing of timelike geodesic congruences}\label{fgc}
We will now discuss the focusing theorem for timelike geodesic congruences (for the null case, refer to Appendix \ref{appen1}). We will confine ourselves to standard GR in this section.
Using Einstein's equations \eqref{es},
We can write the last term in the right-hand side of equation \eqref{rc eq} as,
\begin{equation}\label{eq 9.2.15}
 R_{ab}u^au^b=8\pi\left[T_{ab}-\frac{1}{2}Tg_{ab}\right]u^au^b=8\pi\left[T_{ab}u^au^b+\frac{1}{2}T\right].
\end{equation}
Here,  $T$ is the trace of the energy-momentum tensor. Now, $T_{ab}u^au^b$ is the energy density measured by an observer having four velocity $u^a$. For any physically reasonable
matter distribution, it is generally believed that this energy density is non-negative, i.e.
\begin{equation}\label{wec}
 T_{ab}u^au^b\geq 0,
\end{equation}
for all timelike $u^a$. The condition \eqref{wec} is known as the \textit{Weak Energy Condition (WEC)}. It is also physically reasonable to assume that stresses of matter distribution will not become so large and negative that the right-hand side of the equation \eqref{eq 9.2.15} becomes negative.
It can therefore be assumed that,
\begin{equation}
 T_{ab}u^au^b\ge-\frac{1}{2}T,
\end{equation}
for all unit timelike $u^a$. This assumption is known as the \textit{Strong Energy Condition (SEC)}.

If SEC holds,  $R_{ab}u^au^b$ is positive. The condition
\begin{equation}\label{tcc}
 R_{ab}u^au^b\geq 0,
\end{equation}
 is known as the {\it Timelike Convergence Condition (TCC)}. This condition implies that gravity is attractive.

Let us now examine the RE \eqref{rc eq} with the assumption of the TCC.
For {\it hypersurface orthogonal} congruences, $\omega_{ab}=0$. As $\sigma_{ab}$ is purely spatial, ${\sigma}_{ab}{\sigma}^{ab}\geq0$. So,  all the remaining terms in the right-hand side of \eqref{rc eq} are non-positive. Thus,  $\frac{d\theta}{d\tau}\leq 0$. This implies that
the expansion must decrease as the congruence evolves. The divergence of
an initially diverging $(\theta_0 > 0)$ congruence will be slowed down in the future,
while the convergence of an initially converging $(\theta_0 < 0)$ congruence will be accelerated in the future. Under the conditions mentioned above, the RE \eqref{rc eq} gives us,
\begin{equation}\label{tfc}
 \frac{d\theta}{d\tau}\leq -\frac{1}{3}\theta^2,
\end{equation}
which yields
\begin{equation}
 \theta^{-1}(\tau)\geq{\theta_0}^{-1}+\frac{\tau}{3},
\end{equation}
where $\theta_0\equiv\theta(0)$. Therefore, for an initially converging ($\theta_0<0$) timelike geodesic congruence,  $\theta(\tau) \rightarrow -\infty $ within a proper time $\tau \leq 3 /|{\theta_0}|$.
To summarize, focusing of geodesics in an initially converging congruence occurs within a finite proper time when the SEC holds. This conclusion is known as the \emph{Focusing Theorem} (FT).
The congruence will therefore form a \textit{caustic}, a point at which all geodesics focus.
A caustic is also understood as a singularity of the congruence.
We should note that a congruence singularity may not be an actual curvature singularity being present in the structure of spacetime. We need a few additional requirements to show that congruence singularities lead to geodesic incompleteness. This result follows from the Penrose-Hawking singularity theorem\cite{Hawking1970TheSO}. We shall discuss this in section \ref{phst}.

\subsection{Raychaudhuri's equation for non-geodesic motion}
Non-geodesic motions will come into the picture if, for example, non-gravitational forces are present in spacetime. In such cases, acceleration $a^b=u^c\nabla_c u^b$ will be non-zero and the RE will be modified\cite{Ehlers:1993gf, Kar:2006ms}.
For non-geodesic timelike motion, equation \eqref{eq 9.2.10} yields,
\begin{equation}\label{eq 9.2.10ng}
 u^c\nabla_cB_{ab} =\nabla_b\left(u^c\nabla_cu_a\right)-{B^c}_b B_{ac}+{R_{cba}}^d u^cu_d.
\end{equation}
Taking the trace of the above equation, we get the corresponding RE as,
\begin{equation}\label{rc eq ng}
 u^c\nabla_c \theta=\frac{d\theta}{d\tau}=\nabla_b a^b-\frac{1}{3}\theta^2-\sigma_{ab}\sigma^{ab}+\omega_{ab}\omega^{ab}-R_{cd} u^cu^d.
\end{equation}\\

This thesis addresses two essential themes. These are related to two important issues in gravitational physics which are still in need of a complete resolution. The first one is the issue of singularities mentioned in section \ref{psvbd}. The other one corresponds to the late time accelerated expansion of the Universe. We shall use the RE as the common tool to study both of these cases.

\section{Singularities and the Raychaudhuri equation}\label{phst}
In this section, we will briefly mention the ingenious Penrose-Hawking singularity theorem The Penrose-Hawking singularity theorem is stated as \cite{Hawking1970TheSO}-
{\it If the convergence condition (timelike \eqref{tcc} or null \eqref{ncc}) is satisfied, there are no closed
future pointing timelike curves, a generic condition on the curvature holds and if there is one of the
following:
\begin{itemize}
 \item a compact spacelike (and achronal) hypersurface,
 \item a closed trapped surface,
 \item a point with re-converging light cone,
\end{itemize}
then the spacetime is causal geodesically incomplete.}

Geodesic incompleteness implies spacetime incompleteness. This novel characterization of singularities is a remarkable idea which was proposed by Penrose and later established by Hawking\cite{Hawking:1973uf} and Geroch\cite{Geroch:1968ut}.

Although all the assumptions mentioned in the singularity theorem are physically reasonable, the central one among them is  the convergence condition for causal geodesics. This condition leads to focusing of geodesics as discussed in the previous section. Geodesic focusing is essential for the proof of any singularity theorem. We will not discuss all of the assumptions in the singularity theorem and its proof here. For such a discussion we refer to \cite{Wald:1984rg, Hawking:1973uf, Joshi:1987wg}. A short but very compact and beautiful discussion regarding this theorem and the underlying assumptions has been reported recently by Senovilla\cite{Senovilla:2021pdg}. Our main focus is on the subject of the convergence condition (equation \eqref{tcc}, and more generally on the Focusing Condition (FC) (equation \eqref{tfc}) for the timelike case.  Examination of these conditions is the first step when one wishes to resolve the issue of singularities.\\

The focus of the first part of this thesis is on the issue of singularities. Our aim is to study the possibility of avoiding singularities in GTR and modified theories of gravity through the application of the RE. In GTR, we have to consider the presence of non-gravitational agents, pressure gradients, scalar fields etc. for this, since they can give rise to agents that oppose gravity. We will study two such scenarios in this thesis. The first one is a self-similar gravitational collapse of a distribution containing a fluid and a scalar field (chapter \ref{chapter2}). In the second, we will examine the collapse of a charged fluid distribution in the presence of a magnetic field (chapter \ref{chapter3}).

We have seen in section \ref{fgc} that the SEC implies the TCC in GTR. In modified theories of gravity, the field equations differ from those of GTR. Therefore, even if the matter distribution satisfies the physical requirement of the SEC, other contributions exist which take part in determining the fate of the TCC. In this context, modified theories like $f(R)$-gravity and Gauss-Bonnet gravity have already been studied in the literature\cite{Burger:2018hpz}. We will consider another example in this thesis, namely the {\it Non--Minimally  Coupled Scalar--Tensor Theories (NMCSTT)}\cite{Fujii:2003pa} to explore whether a violation of the TCC is possible in this class of theories (chapter \ref{chapter4}).

\section{Accelerated expansion of the Universe and the Raychaudhuri equation}\label{accex}\sectionmark{Accelerated expansion of the Universe and the RE}
Observations indicate that our Universe is expanding with an acceleration at late times\cite{SupernovaCosmologyProject:1998vns, SupernovaSearchTeam:1998fmf}. This is the most worrying puzzle in cosmology\cite{Rubin:2016iqe, Haridasu:2017lma}. The driving force behind this acceleration is known as {\it ``Dark Energy''}. There is no satisfactory and universally accepted explanation of this phenomenon.

The theoretical structure used to analyze the observational data in cosmology is GTR and the cosmological principle. The latter assumes the Universe to be spatially homogeneous and isotropic. The Universe is then described by the {\it FRW metric} characterized by a scale factor $a$. This factor determines the expansion or contraction of the spatial geometry of the Universe. The evolution of the Universe is thus dictated by the time variation of the scale factor $a(t)$. Therefore, the second time derivative of $a$ will provide the acceleration or deceleration of the Universe. The RE applied to this case connects the second time derivative of the scale factor with the energy density $\rho$ and the pressure $p$ of the matter content of the Universe,
\begin{equation}\label{REacc}
 \frac{\ddot{a}}{a}=-\frac{4\pi G}{3}\left(\rho+3p\right).
\end{equation}
There are four known types of known observational matter: ordinary matter (comprised of baryons and leptons), radiation (photons), neutrinos and dark matter. These four types of matter can describe a significant part of the history of the Universe. For all of these, we have $(\rho+3p)>0$ which implies $\ddot{a}<0$. Thus, these types of matter cannot account for accelerated expansion.

We can see that negative pressure, such that $p< -\frac{1}{3}\rho$, can indeed ensure acceleration. Thus, if we invoke the existence of an exotic fluid with a negative equation of state parameter and also assume that it is dominant at the late stages of the evolution of the Universe, this issue can easily be solved. This fluid is known as dark energy. The simplest and most well-known choice for this fluid was given by Einstein himself: the cosmological constant. It corresponds to a constant vacuum energy density. A non-zero value of the cosmological constant explains the observational data\cite{Padmanabhan:2002ji}. However, its theoretically predicted value from QFT is nowhere near to its observationally required value\cite{Weinberg:1988cp}.

Several dark energy models have been proposed to explain this accelerated expansion of the Universe. Quintessence models, involving scalar fields with a potential, explain the observations quite well after a fine-tuning of parameters. But, none of these models has strong support from theoretical physics. For a comprehensive discussion regarding different dark energy models, we refer to \cite{amendola_tsujikawa_2010, Brax:2017idh}.

An alternative way to explain this accelerated expansion of the Universe is to modify the theory of gravity. Modified theories like $f(R)$-gravity (where $R$ is the Ricci scalar) and NMCSTT are the prime choices that are considered in this context\cite{Bertolami:1999dp, Banerjee:2000mj, Capozziello:2003gx, Nojiri:2003rz, Nojiri:2003ft, Carroll:2003wy, Das:2005bn}. We will consider the $f(R)$-theory of gravity to analyze the problem of late time acceleration (chapter \ref{chapter5}). The idea is to reconstruct $f(R)$-gravity models (i.e. viable forms of $f(R)$) which can explain this accelerated expansion. We will use the RE for this purpose.\\

We will now briefly review the subjects of gravitational collapse, $f(R)$-theory of gravity and NMCSTT.

\section{Gravitational collapse and spacetime singularity}
The phenomenon of shrinking of a massive astronomical body due to the effect of its own gravitational pull is known as gravitational collapse. It is generally believed that structure formation in the Universe occurs due to gravitational collapse. For example, stars are created by the gradual collapse of interstellar clouds. When the nuclear fuel inside a star is exhausted, the following scenarios can occur. It may eventually become a {\it white dwarf} in which the electron degeneracy pressure balances gravitational pull inside the star or a {\it neutron star} in which the balancing agent is the neutron degeneracy pressure. Apart from this, according to GTR, when the mass of the star crosses the Tolman-Oppenheimer-Volkoff limit\cite{PhysRev.55.374}, nothing can balance the gravitational pull. Then the star undergoes a phase of continuous gravitational collapse where curvatures and densities grow without bound. One can find a detailed discussion on the relevance of gravitational collapse in physics, and the related open problems in the references \cite{Joshi:1987wg, Joshi:2008zz}.

\subsection{End state of a gravitational collapse: Black holes or Naked singularities}
A central question in this context is what is the final fate of this collapse. To answer this question, one must solve Einstein's equations \eqref{es} under dynamical collapse scenarios for different kinds of matter distribution. This is a formidable task due to the notoriously non-linear nature of Einstein's equations. Therefore, simplifying assumptions which retain the essential physics are required to solve these equations. A study of gravitational collapse employing the field equations of GTR was initiated by Oppenheimer and Snyder\cite{Oppenheimer:1939ue} and independently by Datt\cite{datt1938klasse}.
They studied a spherically symmetric collapse of an idealized star with a homogeneous density of dust inside. The end state of such a collapse is a {\it black hole} which traps all matter, including light, inside an {\it event horizon}. When the horizon forms before the formation of a singularity, this horizon of the eventual black hole envelops the collapsed region containing the singularity. The situation is troublesome inside the event horizon due to the presence of the singularity. However, this is not a problem for the outside Universe. This is because no communication is possible between the regions inside and outside the event horizon.

A general enough collapse may not always lead to a black hole as an end state. Another possibility is that the horizon forms either after the formation of the singularity or does not form at all. As a result, the final singularity may become exposed. This visible singularity can causally interact with the rest of the Universe through emissions of matter and electromagnetic waves. This is known as a {\it naked singularity}.

The main objection against the existence of naked singularities is the issue of asymptotic predictability\cite{Hawking:1973uf}. Suppose we are provided with regular initial data on a spacelike hypersurface. In that case, we should, in principle, be able to predict the evolution (in past or future) of spacetime for all times through the Einstein equations. As the classical theory fails near a singularity, the presence of a naked singularity ruins this predictability. This is because the singularity causally interacts with the Universe. As we do not know the physics near a singularity, we cannot predict what may emerge from it.

The first significant attempt to address these issues was by Penrose\cite{1969NCimR...1..252P}. He conjectured that during gravitational collapse formation of a singularity is necessarily accompanied with the formation of an event horizon. The horizon hides the singularity from the outside Universe. This conjecture is known as the \emph{Cosmic censorship hypothesis}. If censorship holds, predictability in spacetime is ensured. However, the censorship hypothesis lacks any direct proof which is universally applicable. Counterexamples to censorship do exist, which question the generality of the censorship conjecture. One can find these examples in  \cite{Yodzis:1973gha, Eardley:1978tr, Christodoulou:1984mz, Newman:1985gt, Joshi:1987wg, Joshi:1992vr, Joshi:1993zg, Waugh:1989qcs, Ori:1987hg, Ori:1989ps, Giambo:2003fd, Iguchi:1998qn, Shapiro:1991zza} and references therein.

Though the concept of black holes takes care
of the issue of predictability, it suffers from major paradoxes. When some matter enters a black hole, it must be lost into the spacetime singularity where every known law breaks down. This creates instability as far as the classical description is concerned. A major paradox in the current context is the {\it information paradox}\cite{Hawking:1976ra, Hawking:1975vcx}. This points out that the presence of a black hole violates the fundamental principles of quantum theory by destroying any information which goes inside. For elegant discussions regarding the information paradox, we refer to \cite{Mathur:2009hf, Almheiri:2012rt, Raju:2020smc} and references therein.

Therefore the real problem stems from the very existence of singularity and not really from the fact whether it is hidden inside an event horizon or naked. It is generally believed that quantum effects necessarily come into the picture in the strong gravity region. These quantum effects may eliminate the problematic singularities. This is a promising area of research of current times\cite{rovelli_2004}. There exist different candidate theories of quantum gravity in this regard\cite{Thiemann:2007zz, Ashtekar:1986, Ashtekar:1987gu,
Thiemann:2002nj, Ashtekar:2004eh, Rovelli1998, Thiemann:2006cf, Smolin:2004sx,zwiebach2004first,
polchinski1998string, Polchinski:1998rr, Donoghue:1997hx, Donoghue:1995cz, Kothawala:2013maa, Kothawala:2014tya, Stargen:2015hwa, Awad:2014bta, Trugenberger:2015xma, Garattini:2015aca, Oriti:2016ueo, Hofmann:2015xga}.  However, all of them have their own problems\cite{Ajello:2012fb, Roszkowski:2014wqa, Nicolai:2005mc}. Therefore, exploring the possibility of avoiding singularities even at the classical level is worthwhile. In this thesis, we shall explore possible ways of resolution of the issue of singularity within the classical regime.  \\

We must know the corresponding metric to get a complete picture of the dynamics of gravitational collapse. As mentioned earlier, finding the metric by solving Einstein's equations for sufficiently general systems is very difficult. The RE provides an alternative approach towards exploring these situations\cite{Germani:2005ar, Tsagas:2006sh, Kouretsis:2010nu, Tsagas:2020lal}. In this approach, the collection of flow lines of the particles inside the collapsing distribution is treated as a congruence. The motivation behind using the RE is that important general conclusions about the dynamics can often be obtained without having to solve the field equations.

\subsection{Critical phenomena in gravitational collapse}\label{critph}
{\it Critical phenomena (CP)} in gravitational collapse is one of the most important discoveries in recent times. Choptuik first discovered it in the numerical investigation of the evolution of a spacetime consisting of a minimally coupled massless scalar field\cite{Choptuik:1992jv}. CP can be described as follows. Let us consider a parameter $\beta$ characterizing the solutions of a collapsing distribution.
When this parameter crosses a critical value, the solution changes from one corresponding to a black hole to one which describes an ultimately flat spacetime resulting from a dispersal.
In the study by Choptuik, the solution for the critical value of $\beta$ corresponds to an asymptotically flat spacetime consisting of a null curvature singularity at $r=0$. This is a conceptually rich finding and has created several avenues for further examination. 



After Choptuik's pioneering work, existence of CP has been found in various collapse models with different kinds of symmetry and matter distributions\cite{Hawley:2000tv, Olabarrieta:2001wy, Olabarrieta:2007di, Ventrella:2003fu, Noble:2007vf, Radice:2010rw, Kellermann:2010rt, Noble:2015anf, Choptuik:2003ac}. These phenomena are now established as a common feature in gravitational collapse. The critical collapse problem was studied analytically by Brady\cite{Brady:1994xfa} to find a theoretical explanation of the numerical results reported by Choptuik. Analytical examinations of more general models are reported in \cite{deOliveira:1995cn, Frolov:1997uu, Soda:1996nq, Roberts:1989sk}.  We refer to the review by Gundlach and Martin-Garcia for a comprehensive discussion on CP in gravitational collapse\cite{Gundlach:2007gc}.

 We will now discuss an analytical example to explain the CP. This example will help to understand the possible connection between CP and FC which we will discuss in chapter \ref{chapter2} in the context of a self-similar gravitational collapse.
 
\subsubsection{An analytical example}
We will consider an example discussed by Roberts in \cite{Roberts:1989sk}.
Brady\cite{Brady:1994xfa} used this example to illustrate the CP in massless scalar field collapse analytically. The energy-momentum tensor for the system is given by,
\begin{equation}
     T^\phi_{ab}=\partial_a\phi\partial_b\phi-g_{ab}\Bigg[\frac{1}{2}g^{cd}\partial_c\phi\partial_d\phi\Bigg],
     \end{equation}
The solutions for the metric (in the double null coordinates $u$ and $v$) and the scalar field obtained from the field equations are given by,
\begin{equation}\label{robsol}
 \mathrm{d}s^2=-2\mathrm{d}u\mathrm{d}v+\left(\alpha v^2+u^2-uv\right) \mathrm{d}\Omega^2,
\end{equation}
and
\begin{equation}
 \phi=
 \begin{cases}
       \pm  \frac{1}{2}\ln{\left|\frac{2\alpha v-u\left(1+\sqrt{1-4\alpha}\right)}{2\alpha v-u\left(1-\sqrt{1-4\alpha}\right)}\right|} & \alpha\neq 0 \\
       \pm \frac{1}{2}\ln{\left|1-\frac{v}{u}\right|} & \alpha=0.
      \end{cases}
\end{equation}
respectively. Here $\phi$ is assumed to be zero for $v<0$.

These solutions can be classified into three distinct classes depending on the value of the parameter $\alpha$: subcritical ($0<\alpha<\frac{1}{4}$), critical ($\alpha=0$) and supercritical ($\alpha<0$). We will now discuss these cases one by one. In the following, it is assumed that the influx of the scalar field is turned on at $v=0$.

\begin{itemize}
 \item {\bf Subcritical ($0<\alpha<\frac{1}{4}$)} - In this case, the scalar field collapses, interacts gravitationally and then disperses. This leaves behind a flat spacetime. Here the gravitational interaction is not strong enough to form a black hole.

 \item {\bf Critical ($\alpha=0$)} - This case represents a scenario where the scalar field collapse proceeds toward a null, scalar-curvature singularity at $r=0$.

 \item {\bf supercritical ($\alpha<0$)} - This case corresponds to black hole formation, where an apparent horizon surrounds the singularity at $r=0$.
\end{itemize}
 It is evident from the above discussion that the critical value of the parameter $\alpha$, characterizing the scalar field collapse, is $\alpha=0$.

\section{Modified theories of gravity}\label{modg}
Though GTR is a widely accepted theory of gravity, alternative relativistic theories of gravity continue to be studied for a variety of reasons. Alternative theories have been introduced primarily to propose a solution for the accelerated expansion of the Universe and other cosmological problems. These theories may also serve as an attempt at a semi-classical description of gravitational interactions where the quantum effects are characterized via an effective action. There are several varieties of modified theories. The most discussed are those that stay within the framework of Riemannian geometry and incorporate effects of higher-order curvature invariants and/or coupling with scalar fields. These corrections have a strong motivation from quantum gravity theories. One can find a useful discussion regarding this in the references \cite{Buchbinder:1992rb, Birrell:1982ix, Vilkovisky:1992pb, Gasperini:1991ak}.

These theories find most of their applications in the field of cosmology.
Observational cosmology indicates that there are two phases of cosmic acceleration in the evolution history of the Universe. The first one is known as {\it inflation}. It is generally believed to have occurred before the radiation-dominated phase. Inflation is required to solve some fundamental issues of big bang cosmology, e.g. the flatness problem, the horizon problem and observed features of the {\it Cosmic {M}icrowave {B}ackground {R}adiation (CMBR)}. For a comprehensive discussion regarding inflationary phenomenology we refer to \cite{Starobinsky:1980te, Kazanas:1980tx, Guth:1980zm, Sato:1980yn, Liddle:2000cg, Lyth:1998xn, Bassett:2005xm, COBE:1992syq}. Another phase is the late time acceleration of the Universe. This begins in the late matter-dominated era and has already been discussed in section \ref{accex}. Within the regime of GTR, standard matter cannot explain these two accelerating phases and components that can provide an effective negative pressure are needed to explain such phenomena.

Scalar fields subject to a potential which varies slowly are often considered to explain inflation. Scalar fields can also serve as a choice for explaining dark energy. NMCSTT are a class of modified theories where a scalar field is non-minimally coupled with the curvature sector. Another way to describe these accelerated phases is to modify the action of the theory such that an effective contribution to the energy-momentum tensor arises from the geometry itself\cite{Clifton:2011jh, delaCruzDombriz:2010xy}. The simplest example of such a theory is $f(R)$-gravity\cite{Sotiriou:2008rp}.

We should note that besides explaining physical and observational phenomena, a relativistic theory of gravity should pass some fundamental tests for its viability. These include explaining the physics within the solar system, the phenomena of stable structure formation etc. In the context of modified theories, these topics have been summarized in the references \cite{Faraoni:2008mf, Clifton:2011jh, Sotiriou:2007yd, Sotiriou:2008rp}. GTR does pretty well in this aspect, but this is still an important issue for modified theories.

We shall briefly discuss $f(R)$-gravity and NMCSTT in the following.

\subsection{$f(R)$-gravity}
The action for GTR is given by
\begin{equation}\label{gtrac}
 S=\frac{1}{2}\int \mathrm{d}^4 x \sqrt{-g} R+S_m,
\end{equation}
where $R$ is the Ricci scalar, $g$ is the determinant of the spacetime metric $g_{ab}$, $S_m$ is the action for the matter distribution present, and we have set $8\pi G=1$. This action, known as the {\it Einstein-Hilbert action}, is the simplest invariant action for gravity which contain up to second-order derivatives of metric (see appendix \ref{ehac} for a brief discussion).
The stress-energy tensor is defined as,
\begin{equation}
 T_{ab}=-\frac{2}{\sqrt{-g}}\frac{\delta S_m}{\delta g^{ab}}.
\end{equation}
When we replace the Ricci scalar $R$ in the action \eqref{gtrac} by a general analytic function of $R$, i.e. $f(R)$, we have the invariant action corresponding to {\it $f(R)$-gravity}.
Thus, the action for $f(R)$-gravity with matter fields is given by,
\begin{equation}\label{actfr}
 S=\frac{1}{2}\int \mathrm{d}^4 x \sqrt{-g} f(R)+S_m.
\end{equation}
$f(R)$-gravity theories can explain phases of cosmic acceleration without any exotic matter component for specific choices of the function $f$. A beautiful discussion regarding the utility of this theory in explaining cosmic acceleration consistent with the standard cosmological model can be found in the review of Capozziello, De Laurentis and Faraoni\cite{Capozziello:2009nq}.

\subsubsection{$f(R)$-gravity field equations}
If we vary the action \eqref{actfr} with respect to the metric tensor $g_{ab}$, the field equations are obtained as,
\begin{equation}\label{fieldeqfr}
 f^{\prime}R_{ab} -\frac{f}{2} g_{ab}-(\nabla_a\nabla_b-g_{ab}\square)f^\prime=T_{ab},
\end{equation}
where $f^\prime(R)=\dfrac{\mathrm{d}f(R)}{\mathrm{d}R}$. This equation can be recast in terms of the Einstein tensor as,
\begin{equation}\label{frenf}
 G_{ab}=\frac{1}{f^\prime(R)}(T_{ab}+T_{ab}^{\text{eff}}),
\end{equation}
where
\begin{equation}
 T_{ab}^{\text{eff}}=\left[\frac{f-Rf^\prime}{2}g_{ab}+(\nabla_a \nabla_b-g_{ab}\square)f^\prime\right].
\end{equation}
Here $\square=g^{ab}\nabla_a \nabla_b$
is the d$'$Alembertian operator. The term $T_{ab}^{\text{eff}}$ is purely geometrical in origin and is known as the \emph{effective stress-energy tensor}. Equation \eqref{frenf} reflects the existence of a non-minimal coupling through the presence of $f^\prime$ in the denominator. As a result, the effective gravitational constant becomes a variable in this theory.

\subsubsection{$f(R)$-gravity and phases of accelerated expansion}
Here we shall briefly discuss why $f(R)$-models are important in the context of cosmic acceleration. In $f(R)$-theory, we have an effective energy-momentum tensor along with the standard matter energy-momentum tensor. As the effective energy-momentum tensor is purely geometrical, it need not always obey the SEC. Therefore, the curvature can give rise to an effective repulsive contribution which can drive the accelerated expansion of the Universe.

Many $f(R)$ models have been discussed in the literature. Starobinsky was the first to propose an $f(R)$ model for inflation\cite{Starobinsky:1980te}, given by $f(R)=R+\alpha R^2(\alpha>0)$. The term quadratic in the Ricci scalar dominates at the early stages of the Universe when the curvature is large and drives inflation. The Starobinsky model was found to be consistent with  observations\cite{Clifton:2011jh, Sotiriou:2008rp, Starobinsky:1980te, Paul2009AcceleratingUI}. Following the same line of thought, we may say that presence of negative powers of $R$ in $f(R)$ can explain the late time acceleration, which is believed to be driven by dark energy. Paul, Debnath and Ghose\cite{Paul2009AcceleratingUI} proposed an $f(R)$ model where $f(R)=R-\frac{\alpha}{R^n}(\alpha>0, n>0)$ for this purpose. However, this kind of models leads to instabilities\cite{Clifton:2011jh, Paul2009AcceleratingUI}. In addition, such models have a few other drawbacks: the absence of a matter-dominated epoch, difficulties in passing local gravity tests, etc. Nojiri and Odintsov\cite{Nojiri:2003ft, Nojiri:2003ni} later pointed out that if further terms like $R^2$ or $\ln R$ are included in the action proposed in \cite{Paul2009AcceleratingUI}, it is possible to avoid these difficulties.

Ideally, an $f(R)$-model should be able to explain the inflationary phase as well as the late time accelerated expansion phase. At the same time, it should incorporate a matter-dominated era within these two phases. Therefore, models like $f(R)=R+\alpha R^m+\frac{\beta}{R^n}$
are generally considered. Here the constants $\alpha$, $\beta$ represent the coupling and $m,n>0$. We need $1\leq m\leq 2$ for a power law inflation\cite{Paul2009AcceleratingUI}.

It is to be noted that there do not exist any definite criteria which can select a particular model so that observation at all scales can be explained. Observational constraints on a few $f(R)$-models have been discussed in the review by Copeland, Sami and Tsujikawa\cite{Copeland:2006wr}. Das, Banerjee and Dadhich\cite{Das:2005bn} discussed that $f(R)$-gravity can explain a smooth transition from decelerated to accelerated expansion phase, observed to have occurred in the recent past. In addition to observational aspects, another critical issue is the viability of $f(R)$-models. This serves as a central criterion in choosing physical $f(R)$-models.

\subsubsection{Viability criteria for $f(R)$-gravity models}\label{vcfr}
Some $f(R)$ models suffer from a pathology known as non-linear instability. The formation of stable structures is hardly possible in the presence of such instabilities. Therefore, the explanation of the existence of relativistic stars becomes difficult. This occurs because singularity develops at higher curvatures. An example is the model - $f(R)=R-\frac{\mu^4}{R}$, which suffers from an instability known as the Dolgov and Kawasaki instability\cite{Dolgov:2003px}. This instability manifests within a very short timescale. Nojiri and Odintsov\cite{Nojiri:2003ft, Nojiri:2003ni} also confirmed this conclusion and proved that this pathology can be removed by adding a $R^2$ contribution to the $f(R)$. Later Baghram, Farhang and Rahvar\cite{Baghram:2007df} found the same kind of instability for a specific form of $f(R)$. Faraoni\cite{Faraoni:2008mf, Faraoni:2006sy} proposed certain viability criteria needed to avoid this instability for general $f(R)$-models. These criteria are used to rule out candidates which are not theoretically viable. Satisfying these criteria is now established as an important check for the viability of $f(R)$-models.

The stability analysis uses a parametrization that characterizes the deviation from GTR. For this, the form of $f(R)$ is assumed to be,
\begin{equation}
 f(R)=R+\epsilon \chi(R),
\end{equation}
where the constant $\epsilon$ takes a small positive value. We will only state the conclusions of this stability analysis. For a detailed calculation and discussion, we refer to the reviews by Sotiriou and Faraoni\cite{Faraoni:2008mf, Sotiriou:2008rp}.

First, we need $f'(R)>0$ for viability. A particular $f(R)$-theory is stable if and only if $\frac{\mathrm{d}^2 f}{\mathrm{d}R^2}\geq 0$ and unstable otherwise. Seifert\cite{Seifert:2007fr} confirmed this conclusion using a generalized variational approach. He showed that in theories with  $\frac{\mathrm{d}^2 f}{\mathrm{d}R^2}< 0$, stars become unstable for any kind of matter distribution. Sawicki and Hu\cite{Sawicki:2007tf} studied cosmological perturbations in the context of $f(R)$-gravity and reached the same stability condition as above.

Higher-order theories of gravity are often plagued by the appearance of ghosts in the form of derivatives of order higher than two - often regarded as undesirable in gravitational physics. In the corresponding quantum theory, ghosts are characterized by massive states of negative norm, which cause an apparent lack of unitarity. $f(R)$-gravity theories have no ghosts when the conditions $f^\prime>0$ and  $f^{\prime\prime}\geq 0$ are satisfied. These conditions ensure that the associated tensor and scalar fields of $f(R)$-gravity are not ghosts.

We can understand the interpretation of the viability criteria in a simple way by following the arguments of Faraoni\cite{Faraoni:2007yn}.
It follows from equation \eqref{frenf} that the effective gravitational coupling, $G_\text{eff}$ contains $\frac{1}{f^\prime(R)}$.
So, $f'(R)>0$ ensures a positive gravitational coupling. Now,
\begin{equation}
 \frac{\mathrm{d}G_\text{eff}}{\mathrm{d}R}\sim -\left(\frac{\mathrm{d}f}{\mathrm{d}R}\right)^{-2}\frac{\mathrm{d^2 f}}{\mathrm{d}R^2}.
\end{equation}
Thus, when $\frac{\mathrm{d}^2 f}{\mathrm{d}R^2}< 0$, $\frac{\mathrm{d}G_\text{eff}}{\mathrm{d}R}>0$. The effective gravitational coupling, therefore, increases with the curvature when $f''<0$. As a result, a positive feedback mechanism emerges. Large curvature makes gravity stronger, which again causes the curvature to increase. This drives the system into instability. A tiny curvature grows without bounds, and no stable ground state then exists. On the other hand, when  $\frac{\mathrm{d}^2 f}{\mathrm{d}R^2}\geq 0$, there is a negative feedback which prevents the instability.

To describe the cosmological dynamics properly, a model should incorporate an inflationary phase followed by a radiation-dominated phase. After that, a matter-dominated phase is necessary to explain structure formation in the Universe. A late time accelerated phase should follow after this. Moreover, the transition between these phases should be smooth. Some $f(R)$-gravity theories fail to describe this adequately. For example, some $f(R)$-models show inconsistency in describing exit from the radiation-dominated phase. These issues of cosmological viability and their possible cures are extensively discussed in the references \cite{Faraoni:2008mf, Amendola:2006we, Capozziello:2006dj}. A study of structure formation consistent with the standard cosmological model reveals that $f(R)$-corrections do not affect the structure formation modes when compared with GTR\cite{Song:2006ej}. The condition which ensures the stability of perturbations is again given by $\frac{\mathrm{d}^2 f}{\mathrm{d}R^2}\geq 0$.

Explaining local gravity phenomena is one of the most important requirements for the physical viability of any theory. GTR is the most successful theory in describing local physics as verified by experiments\cite{will2018theory}. Therefore, GTR should be recovered from any viable $f(R)$-model in some specific limit corresponding to the scales of local physics. However, this is still a matter of debate. Although the results in the weak field limit for these extended theories do not exactly coincide with those for GTR, the latter are generalized in some sense. For instance, in $R^2$-theory, as reported by Stelle\cite{Stelle:1977ry}, a few corrections to the Newtonian potential arise. Later, Sanders\cite{Sanders:2002pf} claimed that these corrections can explain the flat rotation curves of galaxies. For a brief and compact discussion regarding the Newtonian limit of extended theories of gravity, we refer to the work of Capozziello\cite{Capozziello:2004sm}. Clifton and Barrow\cite{Clifton:2005aj} considered a model like $f(R)=R^{1+\delta}$. They explored the weak field limit and cosmological properties of this model. Compatibility with local astronomical tests (e.g. perihelion shift) demands a tiny value of $\delta$ so that the model is very close to GTR. An important point to note here is that a viable $f(R)$-model needs a {\it chameleon mechanism} for passing tests in the weak field limit\cite{Sotiriou:2008rp, Chiba:2006jp, Olmo:2006eh}. The chameleon effect\cite{Khoury:2003rn, Khoury:2003aq} can naively be understood as follows: a theory behaves differently at local and cosmological scales such that it can cause the acceleration of the Universe while keeping local physics intact. So, there exists a screening mechanism at the local scale. For an elaborate discussion on the chameleon mechanism in the context of $f(R)$-gravity we refer to \cite{Cembranos:2005fi, Faulkner:2006ub, Navarro:2006mw, Starobinsky:2007hu}.

\subsection{Scalar-tensor theory}\label{bdbeken}
Another important class of modified gravity theories are scalar-tensor theories, originally introduced by Jordan\cite{Jordan:1949zz} and Brans and Dicke\cite{Brans:1961sx}. Their work was further generalized by Nordvedt\cite{osti_4090467} and Wagoner\cite{Wagoner:1970vr}. It is worthwhile to mention that $f(R)$-gravity has been shown to be conformally equivalent to scalar-tensor theories such as the BD theory\cite{Sotiriou:2008rp}.
A general class of NMCSTT is represented by the action
\begin{equation}\label{nmcstta}
  S=\frac{1}{2}\int \sqrt{-g} \mathrm{d}^4 x \left( f(\phi) R-\frac{\omega(\phi)}{\phi}\partial_a \phi \partial^a \phi\right)+S_m.
\end{equation}
Here the scalar field $\phi$ is non-minimally coupled with the Ricci scalar $R$. This, in turn, leads to the variation of the gravitational coupling.
BD theory is the prototype of this class. Different scalar-tensor theories follow from the general action \eqref{nmcstta} for different choices of the functions $f(\phi)$ and $\omega(\phi)$. We refer to \cite{Faraoni:2004pi, Fujii:2003pa, Quiros:2019ktw} for detailed discussions about scalar-tensor theories.

NMCSTT have a strong motivation from the quantum gravity side, specifically from {\it string theory}. The origin of the scalar degree of freedom can be explained by string theory. However, the original idea of Brans and Dicke was to propose a theory consistent with the well-known Mach principle while using the simplest possible choice of a scalar field. NMCSTT is an excellent candidate for describing inflation and late time acceleration of the Universe. For an elegant discussion regarding the abovementioned topics, we refer to \cite{Brans:2014, Faraoni:2004pi} and references therein.

\subsubsection{Field equations}
The field equations for the class of NMCSTT given by the action \eqref{nmcstta} can be found by varying the action with respect to the metric $g_{ab}$ and the scalar field $\phi$. These are given by,
\begin{equation}\label{fsce1}
f\left(R_{ab}-\frac{1}{2}g_{ab}R\right)+g_{ab}\square f-\nabla_b \nabla_a f-\frac{\omega}{\phi}\partial_a\phi \partial_b\phi+\frac{1}{2}g_{ab}\frac{\omega}{\phi}\partial_a\phi \partial^a\phi=T_{ab},
\end{equation}
and
\begin{equation}\label{fsce2}
 R \frac{df}{d\phi}+\frac{2\omega}{\phi}\square\phi+\left(\frac{1}{\phi}\frac{d\omega}{d\phi}-\frac{\omega}{\phi^2}\right)\partial_a\phi \partial^a\phi=0,
\end{equation}
respectively. Here the effective gravitational coupling is $G_\mathrm{eff}=f(\phi)^{-1}$. Hence, $f(\phi)$ is assumed to be positive to ensure a positive coupling.

We will consider two important candidate theories in this class of NMCSTT to carry out our investigations. These are the BD theory\cite{Brans:1961sx} and the {\it Bekenstein conformally coupled scalar-tensor theory (BCCSTT)}\cite{Bekenstein:1974sf}.

\subsection{Brans-Dicke theory}\label{bdtheory}
BD theory is one of the most discussed generalizations of GTR. This theory follows from the action \eqref{nmcstta} for the choice $f(\phi)=\phi$ and $\omega=\text{constant}$. A large value of $\omega$ is necessary to explain observations within the framework of BD theory\cite{reasenberg1979viking, will2018theory,Bertotti:2003rm}. BD theory was claimed to reduce to GTR in the limit $\omega\rightarrow \infty$\cite{weinberg1972gravitation}. Later it has been established that the trace of the energy-momentum tensor of the matter distribution actually limits this conclusion\cite{Banerjee:1996iy, Faraoni:1999yp}. Nevertheless, many problems in cosmology can be resolved within the purview of the BD theory. It can explain the graceful exit problem in the context of inflation\cite{Mathiazhagan:1984vi, La:1989za} and the late time accelerated expansion of the Universe without the need for any dark energy\cite{Banerjee:2000mj}.

We will now discuss exact solutions in BD theory under two specific symmetry assumptions. These correspond to the static spherical symmetric case and the spatially homogeneous and isotropic case. The first one provides insights about black hole (or naked) singularities, while the second helps analyze the big bang type singularity in cosmology.

\subsubsection{Exact Solutions: static spherically symmetric vacuum solutions} Finding static spherically symmetric solutions is a natural first step in any proposed theory of gravity. This provides analogues of the Schwarzschild solution in GTR. Four families of such vacuum solutions were reported by Brans\cite{Brans:1962zz} shortly after the introduction of the BD theory. These are called {\it Brans class I, II, III and IV solutions}, respectively. Their explicit forms can be found in the reference \cite{Brans:1962zz}. Later it was shown that class III and class IV solutions are equivalent\cite{Bhadra:2001my}. The same conclusion holds for class I and II solutions\cite{Bhadra:2005mc}.

Among these, the class I solution is the most talked about example in the literature. It is valid for all values of $\omega$ in its original form. The Brans class I solution using isotropic coordinates can be written as,
\begin{equation}
\begin{split}
\mathrm{d}s^{2}=& -e^{\nu_{o}}\left( \frac{1-\frac{B}{r}}{1+\frac{B}{r}} \right)^{\frac{2}{\lambda}} \mathrm{d}t^{2} \\ &+ e^{\mu_{o}}\left( 1 + \frac{B}{r}\right) ^{4} \left( \frac{1-\frac{B}{r}}{1+\frac{B}{r}} \right)
^{\frac{2(\lambda -C -1)}{\lambda}} \left( \mathrm{d} r^{2} +r^{2} \mathrm{d}\theta ^{2} +r^{2} sin^{2} \theta \mathrm{d}\varphi ^{2} \right),
\end{split}
\end{equation}
\begin{equation}
\phi = \phi_{0} \left( \frac{1-\frac{B}{r}}{1+\frac{B}{r}} \right)^{\frac{C}{\lambda}},
\end{equation}
with the constraint
\begin{equation}
\lambda^2 =  (C+1)^{2} - C(1-\frac{\omega C}{2}).
\end{equation}
Here $\mu_{o}, \nu_{o}, B, C $ are arbitrary constants. This solution has been examined in the context of explaining solar system physics and local tests in the weak field limit in the seminal paper by Brans and Dicke\cite{Brans:1961sx}. The possible utility of the solution in the context of gravitational collapse has been studied in \cite{Janis:1968zz}. In general, this solution corresponds to a naked singularity\cite{Janis:1968zz}. However, while studying a three-parameter family of static and spherically symmetric vacuum solutions of BD theory, Campanelli and Lousto\cite{Campanelli:1993sm} found that the class I solution may represent a black hole. Campanelli and Lousto have discussed the astrophysical applications of these solutions and have pointed out the solutions that agree with solar system observations. The implications of these solutions in the context of the {\it no-hair conjecture}\cite{Ruffini:1971bza} for black holes have also been discussed in \cite{Campanelli:1993sm}. However, in contradiction to this work, Faraoni, Hammad and Belknap-Keet\cite{Faraoni:2016ozb} have argued that Brans solutions correspond to {\it wormholes} or naked singularities for different parameter choices but cannot describe a black hole.

Other classes of solutions have not evoked much interest primarily because they require negative $\omega$ to be valid\cite{Brans:1962zz}. This causes a violation of the effective WEC. However, the extra scalar degree of freedom may be allowed to violate WEC and there exist physical situations where violation of energy conditions can occur (we refer to \cite{Barcelo:2000zf, Klinkhammer:1991ki, Ford:1992ts, Ford:1993bw, Hawking:1975vcx, doi:10.1098/rspa.1976.0045, doi:10.1098/rspa.1977.0130} and references therein for such examples). In addition, solar system tests do not place any constraint on the sign of $\omega$\cite{will2018theory, Turyshev:2003ut}. Therefore, these solutions cannot be eliminated straight away. These solutions and their applications have been discussed in references \cite{Bhadra:2001fx, Agnese:1995kd, Nandi:1997mx, Anchordoqui:1997yb, He:2002bb, Nandi:2004ha, Bronnikov:1998hm, Nandi:2000gt}. Applications include describing wormholes\cite{Bhadra:2001fx, Agnese:1995kd, Nandi:1997mx, Anchordoqui:1997yb, He:2002bb, Nandi:2004ha}, generating solutions in  string theory\cite{Nandi:2004ha} etc. In some situations, the class IV solution corresponds to an object known as a {\it cold black hole}\cite{Bronnikov:1998hm, Nandi:2000gt}. The {\it tidal forces} for such objects do not diverge at the horizon, which is a better behaviour compared to the class I case\cite{Nandi:2000gt}. For comparisons and useful discussions regarding vacuum static spherically symmetric solutions of the BD field equations, we refer to \cite{Brans:1962zz, Bhadra:2005mc, Faraoni:2016ozb}.

\subsubsection{Exact Solutions: spatially homogeneous and isotropic cosmological solutions}
There are several exact cosmological solutions in BD theory. Most of them involve a power law evolution for the scale factor $a(t)$. For a spatially flat universe, a few examples are - The {\it O'Hanlon and Tupper solution}\cite{OHanlon:1972ysn} which is a vacuum solution valid for $\omega>-\frac{3}{2}$, $\omega\neq 0, -\frac{4}{3}$; The {\it BD dust solution}\cite{Brans:1961sx} which describes a universe dominated by pressureless matter and is valid for $\omega\neq -\frac{4}{3}$; The {\it Nariai solution}\cite{10.1143/PTP.40.49, gurevich1973problem} which corresponds to a perfect fluid solution. The latter is the most general solution with a power law behaviour of the scale factor. Further generalizations can be found in the references \cite{gurevich1973problem, Morganstern:1971am}. Perfect fluid solutions were also discussed in other references, e.g. \cite{lorenz1984exact, Morganstern:1971dz}. Solutions involving the cosmological constant as the only source of matter can be found in the references \cite{La:1989za, Mathiazhagan:1984vi, Romero:1992xx}. There also exist a few exact solutions representing FRW universes with closed and open spatial sections. These solutions were reported in \cite{Barrow:1993nt, Dehnen:1971zz, Levin:1993wr, Liddle:2000cg, lorenz1984exact, Mimoso:1994wn, Morganstern:1971dz}. For a detailed discussion on exact cosmological solutions in BD theory, their properties and applications in describing phenomena in the early and late stages of the Universe, we refer to \cite{Faraoni:2004pi} and references therein.

\subsection{Bekenstein's conformally coupled scalar-tensor theory}\label{bkccstt}
BCCSTT\cite{Bekenstein:1974sf} is another important example of a NMCSTT. In this theory, coupling with a {\it conformal scalar field}\cite{Penrose:1965am} is used. A motivation for studying conformal scalar fields follows from their appealing properties observed in QFT\cite{Callan:1970ze, Parker:1973kx}. For having a self-consistent cosmological description in a semi-classical gravity, where the scalar field is quantized, but the gravity is still classical, conformal coupling comes out to be a mandatory choice\cite{Brout:1977ix, Brout:1979bd, gunzig1987self}. This choice is also implemented when the Einstein equivalence principle is employed\cite{Sonego:1993fw, Grib:1995xm, Grib:1995xp}. We refer to \cite{Gunzig:2000yj} for a brief discussion on why this theory is an important candidate within the class of NMCSTT. This theory follows from the action \eqref{nmcstta} for the choice, $f(\phi)=1-\frac{\phi^2}{6}$ and $\omega(\phi)=\phi$.  The theory has attracted more attention because it yields a black hole solution, which is the first ever known counterexample to the no-hair conjecture\cite{Herdeiro:2015waa}. This theory has been investigated significantly in the context of this conjecture\cite{Herdeiro:2015waa, bocharova1970exact, Bekenstein:1975ts, Sudarsky:1997te, Bronnikov:1978mx, Bekenstein:1996pn, doi:10.1063/1.529253, Winstanley:2002jt, Banerjee:2013yua}. The intriguing nature of this theory motivates several studies in many different contexts which can be found in \cite{Martinez:2002ru, Martinez:2005di, Anabalon:2009qt, Padilla:2013jza, deHaro:2006ymc, Dotti:2007cp, Oliva:2011np, Cisterna:2021xxq, Caceres:2020myr, Gunzig:2000yj, Hosotani:1985at, Madsen:1988ph, Bertolami:1987wj, Ford:1987de, Kiefer:1989km, Fakir:1990zi, Futamase:1989hb, Lucchin:1985ip, Barroso:1991aj, Futamase:1987ua, Fakir:1990eg, Abreu:1994fd, Fernandes:2021dsb} and references therein. In this thesis, we will focus on the exact solutions of this conformally coupled theory.

\subsubsection{Exact solutions: static spherically symmetric case}
Bekenstein first proposed a technique to generate solutions for the Einstein-conformal scalar system. For an elaborate discussion on this, we refer to \cite{Bekenstein:1974sf}. This technique uses static vacuum solutions in GTR as the seed and yields two types of static solutions for the Einstein-conformal scalar system. These are called the {\it type A} and the {\it type B} solutions. These two types of solutions are given by,
\begin{equation}
 \mathrm{d}s^2= \frac{1}{4}\left[w(r)^\beta\pm w(r)^{-\beta}\right]^{2}\left(-w(r)^{2\alpha} \mathrm{d}t^2+ w(r)^{-2\alpha}h_{ij}\mathrm{d}{x^i}\mathrm{d}{x^j}\right),
\end{equation}
\begin{equation}\label{conphi}
\phi = \sqrt{6} \frac{1 \mp w^{2\beta}}{1 \pm w^{2\beta}}.
\end {equation}
where upper and lower signs correspond to the type A and type B solutions, respectively. Here
\begin{equation}
\mathrm{d}s^2=-w(r)^{2} \mathrm{d}t^2+ w(r)^{-2}h_{ij}\mathrm{d}{x^i}\mathrm{d}{x^j}
\end{equation}
is the seed solution where $i$ and $j$ denote only the spatial indices. The constant $\alpha=\left(1-3\beta^2\right)^\frac{1}{2}$ and $\beta$ is a free parameter within the range $\left[-\dfrac{1}{\sqrt{3}},\dfrac{1}{\sqrt{3}}\right]$. When the parameter $\beta$ vanishes, the type A solution returns the original seed solution, whereas the type B solution becomes undefined.
Bekenstein used the Schwarzschild solution as the seed solution and obtained the corresponding static spherically symmetric solutions of the Einstein-conformal scalar system in \cite{Bekenstein:1974sf}. The type A solution, in this case, can represent a black hole with scalar hair or a naked singularity, depending on the choice of $\beta$. The black hole solution is known as the {\it Bocharova–Bronnikov–Melnikov–Bekenstein (BBMB) black hole}\cite{Bekenstein:1996pn}. As mentioned earlier, this solution is the first ever known counterexample to the no-hair conjecture. Bekenstein also showed that violation of the energy conditions can occur for a specific range of values of $\beta$. The corresponding solutions are important in the context of our work as a violation of the energy conditions leads to a possibility of avoiding singularities. Similar to the type A solution, the type B solution also represents black holes or naked singularities according to the choice of $\beta$. However, the black hole solutions of type A and type B coincide. Static charged solutions are also discussed in the reference \cite{Bekenstein:1974sf}. For a detailed discussion regarding the properties of these solutions mentioned above, we refer to \cite{Bekenstein:1974sf}. Static spherically symmetric solutions of Bekenstein's theory with a cosmological constant can be found in \cite{Martinez:2002ru}. Solutions of this kind for a more general case, which also includes a Maxwell field, were reported in \cite{Dotti:2007cp, Martinez:2005di, Anabalon:2009qt}.

\subsubsection{Exact solutions: spatially homogeneous and isotropic case}
The first cosmological solutions of this conformally coupled theory were obtained again by Bekenstein and reported in \cite{Bekenstein:1974sf}. The author solved the field equations under the assumption of the FRW background with radiation as the source of the energy-momentum. Solutions corresponding to all the three spatial geometries (spatially flat, open and closed) can be found in \cite{Bekenstein:1974sf}. Solutions for the spatially flat and closed cases always approach a singularity within a finite time, either in the past or in the future. For the spatially open universe, three different solutions exist. Only one of them never approaches a singularity. This is an important example of a classical cosmological model with a bounce. Another exact solution for the spatially flat geometry was obtained in \cite{Demianski:1991zv}. However, the universe corresponding to this solution does not contain any other matter field. Later Abreu, Crawford and Mimoso\cite{Abreu:1994fd} extended the solution generating technique of Bekenstein\cite{Bekenstein:1974sf}. They studied self-interacting models with conformal scalar fields and obtained new solutions describing FRW geometries with zero or non-zero spatial curvature. This is an important technical application of the work carried out by Bekenstein. For a flat spatial geometry, a few exact solutions, having some interesting characteristics, were reported in \cite{Gunzig:2000yj}.

\section{A brief outline of the thesis}
This thesis aims to explore the application of the RE in two important yet unresolved topics. Chapters \ref{chapter2}, \ref{chapter3}, \ref{chapter4} constitute the first part of this thesis. This part is concerned with the problem of singularities in spacetime. Another part is devoted to the problem of the late time accelerated expansion of the Universe. This is discussed in chapter \ref{chapter5}.\\

In chapters \ref{chapter2} and \ref{chapter3}, we investigate the subject of gravitational collapse in two different systems within the purview of GTR. We aim to check whether the formation of singularities can be avoided in these models.

We study a self-similar gravitational collapse of a spherically symmetric distribution consisting of a scalar field and fluid in chapter \ref{chapter2}. The spacetime is assumed to be {conformally flat} to carry out the investigation. We start from a general fluid and a scalar field and find the necessary expressions. After that, we consider different special cases and discuss the corresponding conclusions.

Chapter \ref{chapter3} examines the collapse of a charged fluid distribution in the presence of a magnetic field. We assume the system to be cylindrically symmetric, with the magnetic field lines oriented along the axis of symmetry. We consider two specific cases: a separable case and a self-similar one. We aim to find conditions on the magnetic field strength for which the final formation of the singularity can be avoided.\\

In chapter \ref{chapter4}, we consider a modified theory of gravity, namely NMCSTT. We investigate whether a violation of the TCC is possible in NMCSTT. Such a violation leads to the possibility of avoiding singularities. For this investigation, we pick up two candidate theories, the BD theory and the BCCSTT. We work under two types of symmetry assumptions - the static spherically symmetric case and the spatially homogeneous and isotropic case. \\

To find an explanation for the late time acceleration of the Universe, we consider $f(R)$-gravity theories in chapter \ref{chapter5}. The goal is to reconstruct viable models of $f(R)$-gravity that can account for this acceleration. We start from different ansatz for the kinematical quantities (e.g. quantities like the \emph{deceleration parameter}, the \emph{jerk parameter} etc.), which represent different types of accelerated expansion. Using such ansatz, we express the RE as a differential equation for $f(R)$. The solution of this equation will give us the corresponding functional form of $f(R)$.\\

In the final chapter (chapter \ref{chapter6}), we summarize the contents and findings of the works presented in this thesis. We also mention some concluding remarks and possible future avenues of exploration in this chapter.

\cleardoublepage
\chapter{Analysis of a self-similar gravitational collapse} 
\label{chapter2}
In this chapter, we investigate the role of the Raychaudhuri Equation (RE) in the context of gravitational collapse and associated Critical Phenomena (CP). For this purpose, we study the evolution of a matter distribution containing a fluid and a minimally coupled scalar field in a conformally flat geometry. A conformally flat geometry has a vanishing Weyl tensor. This kind of spacetimes containing a radiating fluid was first considered by Som and Santos\cite{SOM198189}. Later more general cases having a conformally flat geometry have been studied by several authors in the context of radiating and/or shear-free stars\cite{PhysRevD.25.2518, modak1984cosmological, PhysRevD.40.670, schafer2000gravitational, Ivanov:2012hy}. Gravitational collapse of a shear-free fluid with heat flux was explored by Herrera {et. al.}\cite{Herrera:2004ys} under the assumption of a conformally flat geometry. The study of collapsing configurations under this specific geometry has received significant interest in recent times\cite{Hamid:2014kza, Chakrabarti:2016xbk, Banerjee:2017njk}.

We carry out our work with an additional assumption that the spacetime admits a \emph{homothetic Killing vector}, i.e., the evolution is self-similar. This simplifies the problem so that we can attempt an analytical treatment.

\section{Conformal symmetry and self-similarity}
Understanding the underlying symmetry is very important in the context of analyzing different properties of spacetimes and finding exact solutions of the gravitational field equations. A system has \emph{conformal symmetry} if it is invariant under {\it conformal transformation}. In conformal transformation, the angle between two curves does not change. Therefore, conformal transformation preserves the structure of the null cone. However, the lengths of these curves change by a scale factor which depends on the spacetime points.

A customary way to characterize spacetime symmetries is to utilize the concept of \emph{Killing vectors}\cite{Misner:1974qy, Wald:1984rg}. A vector field $X$ is known as a Killing vector field if the Lie derivative of the metric tensor, $g_{ab}$ with respect to $X$ vanishes, i.e.,
\begin{equation}\label{ke}
 \mathcal{L}_{X}g_{ab}=0.
\end{equation}
where $\mathcal{L}_X$ denotes the Lie derivative along the direction of $X$. Equation \eqref{ke} implies that the spacetime structure does not change along the direction of the Killing vector. Therefore, the existence of a Killing vector indicates the presence of a symmetry and hence, the existence of a conserved quantity. For example, if we have a timelike Killing vector in a spacetime, the spacetime is stationary, and the energy is conserved.

When the spacetime admits a vector field $X$ such that the right hand of the equation \eqref{ke} becomes proportional to the metric tensor instead of vanishing,
\begin{equation}\label{homo}
 \mathcal{L}_{X}g_{ab}=2\Phi g_{ab},
\end{equation}
the spacetime is said to have a conformal symmetry. Then $X$ is called a \emph{conformal Killing vector}. Here $\Phi$ is a function of spacetime points and is known as the {\it conformal factor}.

A homothetic Killing vector is a special case of a conformal Killing vector. The former corresponds to the case when $\Phi$ in equation \eqref{homo} is a constant. When a homothetic Killing vector exists in a spacetime, the latter is known as a self-similar spacetime of the first kind\cite{Moopanar:2013rla, Maartens_1986, Moopanar:2010edo}. This is sometimes referred to as \emph{continuous self-similarity}. Self-similarity implies a scale-invariance and plays an essential role in the context of CP in gravitational collapse (see section \ref{critph}). For example, the critical solution for a spherically symmetric perfect fluid distribution shows continuous self-similarity, whereas that for a scalar field under the same symmetry shows discrete self-similarity\cite{Gundlach:2007gc}. Numerical investigations indicate that CP often involves \emph{discrete self-similarity}. Because it is difficult to pursue an analytical study of a problem having discrete self-similarity, continuous self-similarity is often assumed\cite{Brady:1994xfa}. Spherically symmetric dynamical spacetimes containing a perfect fluid were studied by Cahill and Taub\cite{Cahill:1970ew} under the assumption of continuous self-similarity. The authors showed that for these spacetimes, the metric coefficients can be written as functions of a single variable, $z=\frac{t}{r}$ by a suitable coordinate transformation.

Self-similarity implies that the spacetime distribution is similar to itself during the evolution. Though the assumption of self-similarity puts a stringent constraint on the geometry, it is not unphysical. In the references \cite{Cahill:1970ew, Joshi:1987wg}, the authors have discussed the importance of self-similar spacetimes in cosmology. The relevance of the self-similarity feature in the context of the CP has been discussed briefly in Chapter \ref{chapter1} (see section \ref{critph}). For a beautiful discussion regarding the significance of self-similarity in gravitational physics, we refer to the recent review by Carr and Coley\cite{Carr:1998at}. \\

A study of gravitational collapse of a fluid distribution having heat flow under the assumptions of self-similarity and conformal flatness was reported by Chan, Silva and Rocha\cite{Chan:2002bn}. Investigations of self-similar evolution of fluids in more general spacetimes can be found in \cite{Banerjee:2017njk, Brandt:2001nb, Brandt:2006ai, Chan:2014zra, brandt2003} and references therein. Our motivation and approach in this thesis differ from the works mentioned above. Instead of adopting a metric-based approach, where we have to solve the field equations explicitly to find exact solutions, we use an approach based on the RE to study the dynamics.

\section{The system}
We explore the evolution of a matter distribution consisting of a fluid and a scalar field in a conformally flat spacetime. The spacetime metric is given by,
\begin{equation}
\label{metriccnf}
ds^2=\frac{1}{{A^2 (r,t)}}\Bigg[-dt^2+dr^2+r^2d\Omega^2\Bigg],
\end{equation}
where $\frac{1}{A^2 (r,t)}$ is the conformal factor governing the evolution of the 2-sphere.
The energy-momentum tensor for the system can be written as,
\begin{equation}\label{EMT}
T_{ab}=T^\phi_{ab}+T^{\mathrm{fluid}}_{ab},
\end{equation}
where \begin{equation}
     T^\phi_{ab}=\partial_a\phi\partial_b\phi-g_{ab}\Bigg[\frac{1}{2}g^{cd}\partial_c\phi\partial_d\phi+
V(\phi)\Bigg],
    \end{equation}
and \begin{equation}
\begin{split}
       T^{\mathrm{fluid}}_{ab}=\left(\rho+p_t\right)u_a u_b+p_t g_{ab}
+\left(p_r-p_t\right)\chi_a \chi_b +q\left(u_a \chi_b+ u_b \chi_a\right).
\end{split}
      \end{equation}
$\rho$, $p_t$,
$p_r$ and $q$ are the energy density, tangential pressure, radial pressure and radial heat flux of the fluid, respectively;
$u^a=A\delta^a_0$ is the velocity of the fluid
and $\chi^a=A\delta^a_1$ is a unit radial vector.

 As mentioned earlier, we consider our system to be self-similar. Therefore, we write\cite{Cahill:1970ew},
 \begin{equation}\label{ssv}
\begin{split}
 A(r,t) =r B(z),\hspace{0.2cm} \rho(r,t)=\rho(z), \hspace{0.2cm} p_r(r,t)=p_r(z),\\ p_t(r,t)=p_t(z), \hspace{0.2cm}
q(r,t)=q(z), \hspace{0.2cm}  \phi(r,t)=\phi(z),
\end{split}
\end{equation}
 where $z = \dfrac{t}{r}$. This choice makes $z$ the only independent variable in the field equations.

 \subsection{Field equations}
 In the units $8\pi G=1$, the Einstein field equations \eqref{es} lead to the following non-trivial equations,
 \begin{equation}
\label{fe1}
3\dot{A}^2-3A'^2+2AA''+\frac{4}{r}AA'={\rho}+\frac{1}{2}A^2\dot{\phi}^2+\frac{1}{2}A^2\phi'^2+V(\phi),
\end{equation}

\begin{equation}
\label{fe2}
2\ddot{A}A-3\dot{A}^2+3A'^2-\frac{4}{r}AA'=p_r+\frac{1}{2}{\phi'}^2A^2+\frac{1}{2}A^2\dot{\phi}^2-V(\phi),
\end{equation}

\begin{equation}
\begin{split}
\label{fe3}
2\ddot{A}A-3\dot{A}^2+3A'^2-\frac{2}{r}AA'-2AA''= p_t+\frac{1}{2}A^2\dot{\phi}^2 - \frac{1}{2}{\phi'}^2A^2-V(\phi),
\end{split}
\end{equation}
and
\begin{equation}
\label{fe4}
\frac{2\dot{A}'}{A}=\dot{\phi}\phi'-\frac{q}{A^2},
\end{equation}
where dot and prime in the above expressions denote differentiation with respect to $t$ and $r$, respectively.

When expressed in terms of the similarity variable $z$, equation \eqref{fe4} takes the form,
\begin{equation}
\label{fe4s}
\begin{split}
\frac{1}{B}\frac{d^2B}{dz^2}=\frac{1}{2}\left(\frac{d\phi}{dz}\right)^2+\frac{q}{2zB^2}.
\end{split}
\end{equation}
This equation implies that when $\phi$ is constant and the heat flux vanishes, $B$ is proportional to $z$. Similarly, when written in terms of $z$, equations \eqref{fe1}-\eqref{fe3} are given by,
\begin{equation}
\label{fe1s}
\begin{split}
\frac{1}{B}\frac{d^2B}{dz^2}-\frac{3}{B^2}\left(\frac{dB}{dz}\right)^2+\frac{2z}{z^2-1}\frac{1}{B}\frac{dB}{dz}+\frac{1}{z^2-1}\\
=\frac{\rho}{(z^2-1)B^2}-\frac{q(1+z^2)}{2z(z^2-1)B^2}+\frac{V(\phi)}{(z^2-1)B^2},
\end{split}
\end{equation}

\begin{equation}
\label{fe2s}
\begin{split}
\frac{1}{B}\frac{d^2B}{dz^2}-\frac{3}{B^2}\left(\frac{dB}{dz}\right)^2+\frac{2z}{z^2-1}\frac{1}{B}\frac{dB}{dz}+\frac{1}{z^2-1}\\
=-\frac{p_r}{(z^2-1)B^2}+\frac{q(1+z^2)}{2z(z^2-1)B^2}+\frac{V(\phi)}{(z^2-1)B^2},
\end{split}
\end{equation}

\begin{equation}
\label{fe3s}
\begin{split}
\frac{1}{B}\frac{d^2B}{dz^2}-\frac{3}{B^2}\left(\frac{dB}{dz}\right)^2+\frac{4z}{z^2-1}\frac{1}{B}\frac{dB}{dz}-\frac{1}{z^2-1}\\
=-\frac{p_t}{(z^2-1)B^2}-\frac{q}{2zB^2}+\frac{V(\phi)}{(z^2-1)B^2},
\end{split}
\end{equation}
To obtain the above expressions, we have used equation \eqref{fe4s} in the intermediate steps. This helps in replacing the terms which contain derivatives of the scalar field. One can write from equations \eqref{fe1s} and \eqref{fe2s},
\begin{equation}
 \label{rhopr}
 \rho=-p_r+\frac{q(1+z^2)}{z}.
\end{equation}
On the other hand, equations \eqref{fe2s} and \eqref{fe3s} give,
\begin{equation}\label{eqfu}
 \frac{1}{B}\frac{dB}{dz}=\frac{1}{z}+\frac{p_r-p_t}{2zB^2}-\frac{q}{2B^2}.
\end{equation}

\subsection{Conservation equations}
To simplify the calculation, we assume that the respective energy-momentum tensors of the fluid and the scalar field are conserved independently. This means no interaction exists between the fluid and the scalar field. This assumption, while simplifying the problem, retains essential physics. The separate conservation equation corresponding to the scalar field leads to the Klein-Gordon equation,
\begin{equation}
\label{wave}
\nabla_a T^{ab}_\phi=0\implies\square\phi-\frac{dV}{d\phi} = 0.
\end{equation}
This is a consequence of the Bianchi identity $\nabla_a G^{ab}=0.$
Using the form of the metric (\ref{metriccnf}), this equation (\ref{wave}) takes the following form,
\begin{equation}
\label{wave2}
\ddot{\phi}-\phi''-2\frac{\dot{A}}{A}\dot{\phi}-2\frac{\phi'}{r}+2\frac{\phi'A'}{A}+\frac{1}{A^2}\frac{dV}{d\phi}=0.
\end{equation}
In terms of $z$, this equation is given by,
\begin{equation}\label{wave3}
\frac{d^2\phi}{dz^2}-2\frac{d\phi}{dz}\Bigg[\frac{1}{B}\frac{dB}{dz}+\frac{z}{1-z^2}\Bigg]+\frac{1}{B^2(1-z^2)}\frac{dV}{d\phi} = 0.
\end{equation}

The conservation equation for the fluid,
\begin{equation}\label{consfl1}
 \nabla_a T_\mathrm{fluid}^{ab}=0,
\end{equation}
yields two non-trivial equations,
\begin{equation}
 \label{rho}
\dot{\rho}+q^\prime=\frac{3\dot{A}}{A}(\rho+p_t)+\frac{\dot{A}}{A}(p_r-p_t)+\frac{4qA^\prime}{A}-\frac{2q}{r} ,
\end{equation}
and
\begin{equation}\label{prad}
 {p_r}^\prime+\dot{q}=\left(\frac{3A^\prime}{A}-\frac{2}{r}\right)(p_r-p_t)+\frac{A^\prime}{A}(\rho+p_t)+\frac{4q\dot{A}}{A}.
\end{equation}
Using equation \eqref{rhopr} and the assumption of self-similarity, we find that the two equations \eqref{rho} and \eqref{prad} both lead to the same equation,
\begin{equation}\label{consfl2}
 \frac{d\rho}{dz}-z\frac{dq}{dz}=\frac{2}{B}\frac{dB}{dz}(\rho+p_t)+\frac{(1-3z^2)q}{zB}\frac{dB}{dz}+2q.
\end{equation}

\section{The Raychaudhuri equation and focusing condition}
The RE for a timelike congruence (equation \eqref{rc eq ng}) and the corresponding FC (see equation \eqref{tfc}) are given by,
\begin{equation}
\label{raych-eq}
 \frac{d\theta}{d\tau}=-\frac{1}{3}\theta^2+\nabla_{d}a^{d}-\sigma_{ab}\sigma^{ab}
 +\omega_{ab}\omega^{ab}-R_{ab}u^a u^b,
 \end{equation}
 and
 \begin{equation}\label{fcc2}
 \frac{d\theta}{d\tau}+\frac{1}{3}\theta^2\leq 0\implies \nabla_{d}a^{d}-\sigma_{ab}\sigma^{ab}
 +\omega_{ab}\omega^{ab}-R_{ab}u^a u^b\leq 0,
\end{equation}
respectively.
To apply the RE in a gravitational collapse scenario, we have to define a congruence. We consider the collection of worldlines of the fluid particles inside the collapsing system as the congruence in question. Thus the velocity of the congruence is $u^a=A\delta^a_0$.
 The shear and rotation of the congruence will vanish as we are considering a conformally flat spacetime. But the acceleration of the congruence is non-zero in this case. Therefore, the FC leads to,
\begin{equation}\label{abcond}
 R_{ab}u^a u^b\geq \nabla_{d}a^{d}.
\end{equation}
 In this expression, the left-hand side signifies the effect of gravity, whereas the right-hand side is the divergence of acceleration. If the latter works against gravity and dominates, focusing can be prevented. When the divergence of acceleration becomes sufficiently high, the evolution may end up in a dispersal.

In terms of the metric coefficient $A(r,t)$, the condition \eqref{abcond} for focusing can be written as,
\begin{equation}\label{abcond2}
\frac{\partial}{\partial t}\left(\frac{\dot{A}}{A}\right)\geq 0.
\end{equation}
Imposing the condition of self-similarity (i.e. equation \eqref{ssv}), the above equation \eqref{abcond2} yields,
\begin{equation}
 \label{rccond}
\frac{d}{dz}\left(\frac{1}{B}\frac{dB}{dz}\right)\geq 0.
\end{equation}

\subsection{Focusing condition and dynamics of the spacetime}
Finding exact solutions for the system under consideration is difficult without further simplification. The RE-based approach is useful in this context. One can obtain generic conditions regarding the dynamics of the spacetime from the FC. This may provide important information about the evolution.

From the condition of focusing \eqref{rccond}, we have,
\begin{equation}\label{cond2}
 \frac{d}{dz}\left(\frac{p_r-p_t}{2zB^2}-\frac{q}{2B^2}\right)\geq \frac{1}{z^2},
\end{equation}
where equation \eqref{eqfu} is used.

If the conformal factor or the variables of the energy-momentum sector satisfies generic conditions like \eqref{rccond} and \eqref{cond2} respectively, the formation of singularities will occur. Thus, these conditions play a crucial role in understanding the dynamics of the system. To illustrate this, we will now consider a simple yet important example. If we have a perfect fluid with isotropic pressure and no radial heat flux, the left-hand side of the expression in \eqref{cond2} vanishes. So, the condition \eqref{cond2} will be satisfied only when $z\rightarrow \infty$. This signifies either the formation of a central singularity at $r=0$ or an inconsequential singularity as $t\rightarrow\infty$. Therefore, avoiding a central singularity or the formation of a singularity in a finite time is possible only when the fluid has anisotropic pressure and/or heat flux.

Let us discuss a few additional examples. FC \eqref{rccond} can be expressed into different forms (including \eqref{cond2}) using the field equations. For instance, using equations \eqref{fe4s} and \eqref{eqfu}, it can be recast into,
\begin{equation}\label{cond3}
 \frac{1}{2}\left(\frac{d\phi}{dz}\right)^2+\frac{q}{2zB^2}\geq \left(\frac{1}{z}+\frac{p_r-p_t}{2zB^2}-\frac{q}{2B^2}\right)^2.
\end{equation}
When the scalar field is not present or a constant (equivalent to a cosmological constant) and the heat flux is zero, this condition will be satisfied if,
\begin{equation}
 \frac{1}{z}+\frac{p_r-p_t}{2zB^2}=0.
\end{equation}
This equation implies, $\frac{dB}{dz}=0$ by virtue of the equation \eqref{eqfu}. This corresponds to a non-evolving spacetime. So, a violation of FC must occur if we want a non-trivial evolution of the spacetime in this case (i.e. when $\phi=\text{constant}$ and $q=0$). Then this evolution will be non-singular. However, when either $q$, or $\dfrac{d\phi}{dz}$, or both are non-zero, we need further information to comment on the fate of the condition \eqref{cond3}. In this case, both formation and avoidance of a singularity are possible. The outcome depends on the relative strengths of different matter variables. For example, when the rate of change of the scalar field is dominant in the expression \eqref{cond3}, a singularity is inevitable.

As the final example, we consider the case where $\dfrac{d\phi}{dz}=0$ and $p_r=p_t$. Condition \eqref{cond3} then yields,
\begin{equation}
 \frac{q^2}{4 B^4}-\frac{3q}{2zB^2}+\frac{1}{z^2}\leq 0,
\end{equation}
which implies
\begin{equation}\label{constraint}
 \frac{(3-\sqrt{5})B^2}{z}\leq q \leq \frac{(3+\sqrt{5})B^2}{z}.
\end{equation}
Thus the formation of a singularity will occur only when the heat flux obeys this constraint.

\subsection{Exact solutions and the Raychaudhuri equation}
We have analyzed our system without referring to any exact solution so far. However, it is worthwhile to check the consistency of the above conclusions with those following the exact solutions whenever available. For some special cases, it is possible to find out exact solutions. A few of them are listed in Appendix \ref{appen}. An exact solution, if available, gives us an explicit idea about the evolution. This includes whether there is a collapse or an expansion and whether the outcome is the formation of a singularity or a complete dispersal. We have found that for all the cases listed in Appendix \ref{appen}, the conclusions agree with those obtained from the FC. We discuss a non-trivial example among them
in the following.

We consider the example of the case where both a scalar field and a fluid are present. In addition, we assume that $p_r=\beta p_t$ and $q=0$, where $\beta$ is a constant. For details, we refer to Appendix \ref{fifth}. The solution for the conformal factor is given by,
\begin{equation}\label{Bsolprop}
 B=\left(Fz^{2\beta}-k\right)^\frac{1}{2\beta},
\end{equation}
where $F$ and $k$ are constants.
This implies,
\begin{equation}
 A=\left(Ft^{2\beta}-kr^{2\beta}\right)^\frac{1}{2\beta}.
\end{equation}
A singularity occurs when $A\rightarrow\infty$, i.e. the volume of the two-sphere vanishes. Therefore, when $\beta>0$, a zero proper volume singularity may occur as $t\rightarrow\infty$ and/or $r\rightarrow \infty$ which can be excluded from the discussion. In this case, a dispersal (volume going to infinity) takes place at $z=\left(\dfrac{k}{F}\right)^{\frac{1}{2\beta}}$. For $\beta<0$, a singularity forms at $z=\left(\dfrac{k}{F}\right)^{\frac{1}{2\beta}}$ whereas a dispersal occurs at $t=0$. Dispersal may also occur at $r=0$, but this is not physically significant.\\

Let us now look at the FC in this setting. Using the solution for $B$ as given by equation \eqref{Bsolprop}, we have,
\begin{equation}\label{rcprop}
 \frac{d}{dz}\left(\frac{1}{B}\frac{dB}{dz}\right)=-\frac{2F^2\beta z^{(4\beta-2)}}{\left(F z^{2\beta}-k\right)^2}
 +\frac{F(2\beta-1) z^{(2\beta-2)}}{\left(F z^{2\beta}-k\right)}.
\end{equation}
In the limit $z\rightarrow\left(\dfrac{k}{F}\right)^{\frac{1}{2\beta}}$, the first term in the right-hand side of the expression \eqref{rcprop} dominates which means that the condition \eqref{rccond} is satisfied only when $\beta<0$. So, the formation of a singularity is inevitable when $\beta<0$. On the other hand, when $\beta>0$, $\frac{d}{dz}\left(\frac{1}{B}\frac{dB}{dz}\right)$ is negative in this region, which makes a dispersal possible. One can also note from the expression \eqref{rcprop} that formation of a singularity as $z\rightarrow \infty$ is a possibility when $\beta>0$. On the other hand, for $\beta<0$, a dispersal is possible as $z\rightarrow 0$.

\subsection{The Raychaudhuri equation and critical phenomena}
Examining all the exact solutions in Appendix \ref{appen} we have found that collapse and dispersal depend on the sign of the quantity $\dfrac{d}{dz}\left(\dfrac{1}{B}\dfrac{dB}{dz}\right)$. This quantity being positive (negative) corresponds to collapse (dispersal). So, if there is a transition from collapse to dispersal or vice versa, it will be accompanied with a change of sign of $\dfrac{d}{dz}\left(\dfrac{1}{B}\dfrac{dB}{dz}\right)$. This suggests a relation between FC and CP (see section \ref{critph} for a discussion on CP). We will consider the example discussed in the previous section to illustrate this point.

In this example, we consider the limit $z\rightarrow\left(\dfrac{k}{F}\right)^{\frac{1}{2\beta}}$. A singularity forms at $z=\left(\dfrac{k}{F}\right)^{\frac{1}{2\beta}}$ when $\beta$ is negative, whereas a dispersal takes place at this point when $\beta$ is positive. Thus $\beta$ can be treated as a critical parameter. It has been discussed in the previous section that in the limit $z\rightarrow\left(\dfrac{k}{F}\right)^{\frac{1}{2\beta}}$, $\dfrac{d}{dz}\left(\dfrac{1}{B}\dfrac{dB}{dz}\right)$ is negative or positive when $\beta>0$ or $\beta<0$ respectively. This implies that the critical parameter $\beta$  determines focusing (collapse) or dispersal of the matter distribution, and its critical value is zero.

\section{Summary and discussion}
In this chapter, we have studied a self-similar gravitational collapse of a matter distribution in a conformally flat spherically symmetric spacetime using the FC. It should be emphasized that though our system has a stringent symmetry requirement, the matter distribution is quite general and the evolution is still inhomogeneous. The matter distribution includes anisotropic fluid pressure, heat flux and a minimally coupled scalar field.

We have demonstrated that general conditions regarding the evolution of spacetime can be obtained from the Focusing Condition (FC). This leads to important conclusions which are useful in the absence of exact solutions. We have discussed the consistency of these conclusions with those from the metric-based approach by investigating exact solutions for a few simplified special cases. We have described one particular example in detail for explicit illustration.
Another important finding of this study is that the condition developed from the RE helps to understand the CP from a general point of view. With the help of the aforementioned example, we have demonstrated the connection between the FC and the CP. We have found the corresponding critical parameter and its critical value for this particular example. It is intriguing that this parameter is not related to the scalar field at all but is rather determined only by the fluid.

In the next chapter, we will make use of the concepts discussed in this chapter to study the role of a magnetic field in a gravitational collapse. We aim to see whether the presence of non-gravitational agents (which is a magnetic field for our case) can halt a collapse and thereby prevent the final singularity formation.

\cleardoublepage
\chapter{Inhomogeneous gravitational collapse in the presence of a magnetic field}
\label{chapter3}
\chaptermark{Gravitational collapse incorporating magnetic fields}

The presence of non-gravitational agents, such as electromagnetic fields, is expected to play a significant role in deciding the fate of a gravitational collapse. Several models of gravitational collapse have been explored to investigate whether forces arising from such agents can halt the collapse and prevent the formation of singularities. One can find these in \cite{novikov, Germani:2005ar, Tsagas:2006sh, Kouretsis:2010nu, Tsagas:2020lal, de1967gravitational, rayc12, Ardavan:1977up, PhysRevD.44.2278, 1995MNRAS, deFelice:1999qp, Ray:2003gt, Ghezzi:2005iy, Krasinski:2006sb} and references therein. Among these, the choice of magnetic fields is pretty well-motivated as has been known for several decades\cite{1969tsraconf443T}. There is an inherent ability of magnetic fields to work against gravity. This idea first emerged from the works by Melvin\cite{Melvin:1963qx, Melvin:1965zza}. He proposed a cylindrically symmetric non-singular static solution of the Einstein field equations. The corresponding spacetime contains a magnetic field concentrated near the axis of symmetry. The pressure arising from the magnetic field balances the gravitational pull in {\it Melvin spacetime}. This solution was found to be absolutely stable against radial perturbations. It does not collapse to a singularity or encounters a dispersal\cite{Melvin:1965zza, PhysRev.139.B244}. Later studies by different authors revealed that magnetic fields do offer contributions resisting gravitational contraction\cite{Germani:2005ar, Tsagas:2006sh, Tsagas:2020lal, Ardavan:1977up, Tsagas:2000zq}. Repulsion between magnetic force lines is generally believed to be responsible for this phenomenon. Magnetic fields give rise to a \emph{magneto-curvature stress} that acts against any agent trying to deform these force lines\cite{Tsagas:2006sh, Tsagas:2020lal, Tsagas:2000zq, Tsagas:1999tu}.

Although several previous and recent studies indicate that magnetic fields may avert the final formation of a singularity due to gravitational collapse, there exists very little by the way of concrete conclusions. In this chapter, we examine the role of magnetic fields in an inhomogeneous gravitational collapse for further investigation in this direction. The motivation for this study comes from the ideas put forward in a few recent works\cite{Tsagas:2006sh, Kouretsis:2010nu, Tsagas:2020lal}. In these contributions, the authors have reported a condition which is pivotal in deciding the fate of a gravitational collapse in the presence of magnetic fields. The authors employed the RE to obtain this condition. A simple interpretation of this condition is that the formation of singularities can be avoided if the repulsive effect arising due to the presence of magnetic field balances or overpowers the gravitational pull. For further insights, one needs to test the validity of this condition in realistic models of gravitational collapse. Recently, Mavrogiannis and Tsagas\cite{Mavrogiannis:2021nrl} considered magnetized collapse in a perturbed Bianchi I background, which is almost homogeneous. Their purpose was to check if this condition holds at some stage during the evolution. Here we will study the dynamics of a cylindrically symmetric inhomogeneous spacetime. The spacetime contains a charged fluid obeying the Strong Energy Condition (SEC) and a magnetic field directed along the axis of symmetry. In this case, we shall look for the possibilities of averting a collapse and the final singularity formation by working at the ideal-Magnetohydrodynamics (ideal--MHD) limit. Before that, however, we will briefly revisit the homogeneous case to highlight the utility of the approach adopted here.

 \section{A brief review of the homogeneous non-static case}
 We begin with a brief review of the homogeneous non-static case incorporating a fluid and a magnetic field. This example explains the utility of applying the RE-based approach quite easily. The spacetime is anisotropic as the magnetic field is assumed to be oriented along a specified direction. We assume that the magnetic field is directed along the $z$-axis in the rest frame of the fluid and choose the {\it Bianchi I spacetime} for our study,
 \begin{equation}\label{metricih}
 \mathrm{d}s^2= -\mathrm{d}t^2+e^{2\lambda(t)}\left(\mathrm{d}x^2+\mathrm{d}y^2\right)+e^{2\gamma(t)}\mathrm{d}z^2.
\end{equation}

 We consider a congruence comoving with the fluid. The velocity of the congruence is $u^a=\delta^a_0$. So, the congruence has $a^b=0$ and $\omega_{ab}=0$. The term $\left(T_{ab}u^a u^b+\frac{1}{2}T\right)$ can be decomposed as,
 \begin{equation}\label{fmem}
 \left(T_{ab}u^a u^b+\frac{1}{2}T\right)=\left(T_{ab}u^a u^b+\frac{1}{2}T\right)_\mathrm{f}+\left(T_{ab}u^a u^b \right)_\mathrm{B}.
\end{equation}
The subscripts f and B signify the corresponding contributions offered by the fluid and the magnetic field, respectively. In the above expression, we have used the information that the energy-momentum tensor of the magnetic field is traceless. The energy density of the magnetic field with respect to an observer comoving with the fluid is given by the quantity $\left(T_{ab}u^a u^b \right)_\mathrm{B}$. So, we always have $\left(T_{ab}u^a u^b \right)_\mathrm{B}\geq 0$. Therefore, if the fluid obeys the SEC $\left(\left[T_{ab}u^a u^b+\frac{1}{2}T\right]_\mathrm{f}\geq 0\right)$, the Timelike Convergence Condition (TCC) $\left(R_{ab}u^a u^b\geq 0\right)$ follows. This implies that focusing within a finite proper time is inevitable. So, singularities cannot be avoided in this model. Thorne\cite{1967ApJT} showed that all possible solutions for this model lead to singularities. He solved the field equations explicitly to reach this conclusion. We have reached the same generic conclusion just by using the RE. Solving the field equations is not always necessary when one uses this approach to investigate whether a singularity is present. We have also seen this in the previous chapter.

The acceleration $a^\mu$ vanishes for a homogeneous distribution. Non-zero acceleration is the only agent that can resist the attractive effect of gravity for hypersurface orthogonal congruences. So, it is expected that the conclusions obtained for this particular example should stay valid for general homogeneous models. Here it should be mentioned that the study by  Mavrogiannis and Tsagas\cite{Mavrogiannis:2021nrl} treated perturbation with respect to the homogeneous background as the origin of a non-vanishing acceleration. Perturbation also enables one to study collapse with closed spatial sections. The authors showed that the initial conditions of the problem determine the precise fate of this collapse.

\section{Inhomogeneous models of gravitational collapse}
In this section, we study inhomogeneous collapsing models of a charged fluid distribution. The distribution collapses under the effect of a magnetic field. The collapse of inhomogeneous magnetized systems has not been studied significantly in the literature. Germani and Tsagas\cite{Germani:2005ar} investigated the collapse of a weakly magnetized dust in an inhomogeneous background using a RE-based approach. The authors assumed that the magnetic field is relatively weak, and a spherically symmetric Tolman-Bondi metric represents the system. Their investigation revealed that the magnetic field severely distorts the spherical symmetry as the collapse proceeds.

In the present study, we consider the system to be cylindrically symmetric, where the direction of the magnetic field is along the axis of symmetry. For this analysis, we use the ideal--MHD approximation. In the ideal--MHD limit, the electric field vanishes in the frame of the fluid. Moreover, the magnetic field lines are frozen in the fluid\cite{priest_forbes_2000, parker1979cosmical, mestel2003stellar}. There exist several works regarding gravitational collapse of cylindrically symmetric distributions (see for example \cite{Apostolatos:1992qqj, Shapiro:1992heg, Echeverria:1993wf, david, Letelier:1994pli, Nolan:2002zd, Wang:2003vf, Nolan:2004ur, Harada:2008rx} and references therein). However, magnetized collapse of an inhomogeneous fluid distribution under cylindrical symmetry is yet to be studied in sufficient detail. We write the metric representing our system in the Einstein-Rosen form as\cite{david, stephani_kramer_maccallum_hoenselaers_herlt_2003} -
\begin{equation}\label{metric}
 \mathrm{d}s^2=e^{2\mu(r,t)-2\nu(r,t)}\left(-dt^2+dr^2\right)+e^{2\nu(r,t)}dz^2\\ +R^2(r,t)e^{-2\nu(r,t)}d\phi^2.
 \end{equation}
 The magnetic field, having strength $B(r,t)$ (not to be confused with $B(z)$ defined in \eqref{ssv}), is oriented along the $z$-axis in the rest frame of the fluid. The energy-momentum tensor for the matter distribution is,
 \begin{equation}
 T_{ab}=\left(T_{ab}\right)_\mathrm{f}+\left(T_{ab}\right)_\mathrm{B},
\end{equation}
where
\begin{equation}
 \left(T_{ab}\right)_\mathrm{f}
=\left(\rho_\mathrm{f}+p_\mathrm{tf}\right)u_a u_b+p_\mathrm{tf} g_{ab}
+\left(p_\mathrm{rf}-p_\mathrm{tf}\right)\chi_a \chi_b+q_\mathrm{f}\left(u_a \chi_b+ u_b \chi_a\right),
\end{equation}
and\cite{Barrow:2006ch}
\begin{equation}
 \left(T_{ab}\right)_\mathrm{B}=\left(\rho_\mathrm{B}+p_\mathrm{B}\right)u_a u_b+p_\mathrm{B} g_{ab}+\pi_{ab}.
\end{equation}
$\rho_\mathrm{f}$, $p_\mathrm{rf}$, $p_\mathrm{tf}$ and $q_\mathrm{f}$ respectively denote the energy density, radial pressure, tangential pressure and radial heat flux of the fluid. $\rho_\mathrm{B}$, $p_\mathrm{B}$ and $\pi_{ab}$ stand for the energy density, isotropic and anisotropic pressure of the magnetic field. These are defined as\cite{Tsagas:2020lal},
 $\rho_\mathrm{B}=\frac{B^2}{2}$, $p_\mathrm{B}=\frac{B^2}{6}$ and $\pi_{ab}=\frac{B^2}{3}h_{ab}- B_{a}B_{b}$ respectively. Here, $B^a=B n^a$ and $n^a$ is the unit vector along the $z$-axis. $\chi^a$ is a radial unit vector, and $u^a$ is the velocity vector of the fluid. These three vectors are given by,
 \begin{equation}
 u^a=e^{\nu-\mu}\delta^a_0, \hspace{0.2cm} \chi^a = e^{\nu-\mu}\delta^a_1, \hspace{0.2cm} n^a=e^{-\nu}\delta^a_2.
\end{equation}
Here we have $u^a u_a=-1$, $u^a\chi_a=0$, $\chi^a n_a=0$ and $u^a n_a=0$.

\subsection{Field equations}
The non-trivial components of the Einstein field equations (where $\kappa=8\pi G$),
\begin{equation}
 G_{ab}=\kappa T_{ab},
\end{equation}
for our model are
\begin{eqnarray}
 && G_{00}=\kappa T_{00}\implies \nonumber \\ &&
 \frac{\dot{R}}{R}\dot{\mu}+\frac{R^\prime}{R}\mu^{\prime}-\left(\dot{\nu}^2+{\nu^\prime}^2\right)-\frac{R^{\prime\prime}}{R}=\kappa e^{2\mu-2\nu}\left(\rho_\mathrm{f}+\frac{B^2}{2}\right),\label{fec31} \\[2ex]
&& G_{01}=\kappa T_{01}\implies
  \dot{\mu}\frac{R^\prime}{R}+\frac{\dot{R}}{R}\mu^\prime-2\dot{\nu}\nu^\prime-\frac{\dot{R}^\prime}{R}=-\kappa  e^{2\mu-2\nu} q_\mathrm{f}, \label{fec32} \\[2ex]
 &&G_{11}=\kappa T_{11}\implies \nonumber \\ &&
 \frac{\dot{R}}{R}\dot{\mu}-\frac{\ddot{R}}{R}+\frac{R^\prime}{R}\mu^{\prime}-\left(\dot{\nu}^2+{\nu^\prime}^2\right)=\kappa e^{2\mu-2\nu}\left(p_\mathrm{rf}+\frac{B^2}{2}\right), \label{fec33}\\[2ex]
 && G_{22}=\kappa T_{22}\implies \nonumber \\
 && 2\frac{\dot{R}}{R}\dot{\nu}-\frac{\ddot{R}}{R}-2\frac{R^\prime}{R}\nu^{\prime}+\frac{R^{\prime\prime}}{R}-\left(\dot{\nu}^2-{\nu^\prime}^2+\ddot{\mu}-2\ddot{\nu}-\mu^{\prime\prime}+2\nu^{\prime\prime}\right)\nonumber \\
&& =\kappa e^{2\mu-2\nu}\left(p_\mathrm{tf}-\frac{B^2}{2}\right),\label{fec34} \\[2ex]
&& G_{33}=\kappa T_{33}\implies
  \mu^{\prime\prime}+{\nu^\prime}^2-\ddot{\mu}-\dot{\nu}^2=\kappa e^{2\mu-2\nu}\left(p_\mathrm{tf}+\frac{B^2}{2}\right), \label{fec35}
 \end{eqnarray}
where dot and prime stand for derivatives with respect to $t$ and $r$, respectively.

\subsection{Focusing condition}
The congruence containing the worldlines of the fluid particles (having velocity $u^a$) has a zero vorticity but non-zero acceleration and shear. In this case, the FC,
\begin{equation}
 \frac{d\theta}{d\tau}+\frac{1}{3}\theta^2\leq 0 \implies R_{ab} u^a u^b+\sigma_{ab}\sigma^{ab}-\nabla_b a^b \geq 0,
\end{equation}
where the RE \eqref{rc eq ng} has been used to obtain the final expression. The divergence of acceleration term is effectively related to the magneto-curvature stress, which opposes gravity\cite{Tsagas:2006sh, Tsagas:2020lal}. This term, therefore, resists focusing and the ultimate collapse to a singularity. Our aim is to examine if this term can balance or even overpower the combined contribution from the curvature ($R_{ab} u^a u^b$) and the shear ($\sigma_{ab}\sigma^{ab}$) at some stage during the evolution. The FC will then be violated at this stage. Collapse may be averted in such a scenario.

Extracting useful information is very difficult in a completely general case. Thus, we will limit ourselves to special situations to see how the magnetic field actually affects the evolution of an inhomogeneous distribution. We will consider two representative situations. In the first one, the metric coefficients are assumed to be separable into radial and time dependent parts. In the second, the evolution is assumed to be self-similar.

\subsection{Case I: separable case }
In the first case, we use the \emph{ansatz},
\begin{equation}\label{anz1}
 \mu=k \nu, \hspace{0.2cm} \nu(r,t)=\nu_1(t)+\nu_2(r) \hspace{0.2cm} \mathrm{and} \hspace{0.2cm} R^2(r,t)=r^2 e^{2m\nu_1(t)},
\end{equation}
where $k, m$ are constants.
In this setting, we can express the geometric quantities contributing to the FC in terms of the matter variables as,
\begin{equation}\label{rmns}
 R_{ab} u^a u^b=\frac{\rho_\mathrm{f}+p_\mathrm{rf}+2 p_\mathrm{tf}}{2}+\frac{B^2}{2},
\end{equation}
\begin{equation}\label{divaccs}
 \nabla_b a^b=\kappa \left[\frac{k-1}{2}B^2+(k-1)\left(\frac{1}{m}-\frac{1}{2}\right)\left(\rho_\mathrm{f}-p_\mathrm{rf}\right)\right],
\end{equation}
 and
\begin{equation}\label{sigs}
 \sigma_{ab}\sigma^{ab}=\frac{2}{3}e^{2(1-k)\nu}\left(m^2+k^2-m(k+2)-2k+4\right)\dot{\nu}_1^2,
\end{equation}
where
\begin{equation}\label{nu1dots}
 \dot{\nu}_1^2 =\kappa e^{2(k-1)\nu}\left[\frac{2-k}{4(mk-1)}B^2+\frac{k+2}{4(mk-1)}\left(\rho_\mathrm{f}-p_\mathrm{rf}\right)\right. +\left. \frac{p_\mathrm{rf}+p_\mathrm{tf}}{2(mk-1)}\right].
 \end{equation}
 We have used the field equations \eqref{fec31}-\eqref{fec35} to obtain these expressions.
 The ansatz (given by equation \eqref{anz1}) helps us to carry out an analytical investigation which leads to useful information about the resulting evolution. Despite the simplifying assumptions, this does not correspond to a completely unphysical scenario. We have kept the matter distribution quite general and have not imposed any restrictions on the strength of the magnetic field at the outset.

 Now, violation of the FC implies $R_{ab}u^a u^b+\sigma_{ab}\sigma^{ab}-\nabla_b a^b<0$. Here, the constants $k$ and $m$ in \eqref{anz1} play a decisive role in determining the structure of the spacetime. If values of $k$ and $m$ are such that $\nabla_b a^b$ dominates at some stage during the evolution, the collapse will halt. An expansion may even occur in that case. Otherwise, the collapse will lead to the formation of a singularity.

 We use $m=2$ for illustration. For this specific case, $\nabla_b a^b$ is completely determined by the magnetic field strength,
 \begin{equation}
 \nabla_b a^b= \frac{\kappa(k-1)}{2}B^2.
\end{equation}
Thus, we need $k>1$ to get a repulsive contribution from the magnetic field.
In this case we have,
\begin{equation}\label{fcs}
\begin{split}
 R_{ab} u^a u^b+\sigma_{ab}\sigma^{ab}-\nabla_b a^b =  \kappa\left[\frac{(k+1)^2(2-k)}{6(2k-1)}B^2\right. \\ \left. +\frac{(k+1)(k^2-3k+5)}{6(2k-1)}(\rho_\mathrm{f}-p_\mathrm{rf}) +\frac{4k-1}{2(2k-1)}(p_\mathrm{rf}+p_\mathrm{tf})\right].
 \end{split}
\end{equation}
When the magnetic field is absent, we should have $R_{ab} u^a u^b+\sigma_{ab}\sigma^{ab}\geq 0$ under the assumption of the SEC. Therefore, to avoid focusing in the presence of a magnetic field, the first term on the right-hand side of the expression \eqref{fcs} must be negative. In addition, its magnitude should be greater than the magnitude of the other two terms combined (we should note that interaction does exist between the magnetic field and the fluid, but we expect that the terms related to the fluid variables in \eqref{fcs} will have a positive contribution). The above criteria are satisfied when $k>2$ (as we already have $k>1$ for $\nabla_b a^b$ to be positive) and,
\begin{equation}\label{icond1}
 B^2> \frac{k(k-3)+5}{(k+1)(k-2)}\left(\rho_\mathrm{f}-p_\mathrm{rf}\right)+\frac{3(4k-1)}{(k+1)^2(k-2)}\left(p_\mathrm{rf}+p_\mathrm{tf}\right).
\end{equation}
Here one might be tempted to infer that the FC will be violated if the magnetic field strength becomes sufficiently high. However, the magnetic field cannot be arbitrarily large. An upper bound on $B$ follows from the requirement $\dot{\nu}_1^2>0$, which implies (using equation \eqref{nu1dots}),
\begin{equation}
B^2< \frac{k+2}{k-2}\left(\rho_\mathrm{f}-p_\mathrm{rf}\right)+\frac{2\left(p_\mathrm{rf}+p_\mathrm{tf}\right)}{(k-2)}.
\end{equation}

Thus the constraint on $B$ for which focusing will be prevented is,
\begin{equation}\label{ineq1}
\begin{split}
B_1^2<B^2<B_2^2,
 \end{split}
\end{equation}
where
\begin{equation}
 B_1^2=\frac{k(k-3)+5}{(k+1)(k-2)}\left(\rho_\mathrm{f}-p_\mathrm{rf}\right)+\frac{3(4k-1)}{(k+1)^2(k-2)}\left(p_\mathrm{rf}+p_\mathrm{tf}\right),
\end{equation}
and
\begin{equation}
 B_2^2=\frac{k+2}{k-2}\left(\rho_\mathrm{f}-p_\mathrm{rf}\right)+\frac{2\left(p_\mathrm{rf}+p_\mathrm{tf}\right)}{(k-2)}.
\end{equation}
Conventional fluids are found to obey the conditions, $\rho_\mathrm{f}\geq p_\mathrm{rf}$ and $p_\mathrm{rf}, p_\mathrm{tf}\geq 0$. Hence, for $k>2$, $B_1^2$ and $B_2^2$ are always positive which is needed for consistency. Also, it is easy to verify that $B_1^2$ is manifestly greater than $B_2^2$ for $k>2+\sqrt{\frac{3}{2}}$.

At this stage we should note that when $k=2$ and $m=2$, the metric \eqref{metric} takes the form,
\begin{equation}\label{metricconf}
\begin{split}
 \mathrm{d}s^2= e^{2\nu_1(t)}\left[e^{2\nu_2(r)}\left(-dt^2+dr^2+dz^2\right) \right. \left. +r^2 e^{-2\nu_2(r)}d\phi^2\right].
 \end{split}
 \end{equation}
 The model then represents a simple conformal gravitational collapse. Models of conformal collapse have been studied in a few recent contributions\cite{Chakrabarti:2021gqa, Bini:2022xzk}. These models are often considered in the context of analytical investigation of gravitational collapse because of their mathematical simplicity. If we choose $e^{2\nu_2} = (1+r^2)^2$, the conformally related metric within the square bracket in equation \eqref{metricconf} becomes the well-known Melvin static metric. Recently, Bini and Mashhoon\cite{Bini:2022xzk} studied a cylindrically symmetric gravitational collapse where the seed solution is the Melvin solution. The metric given by equation \eqref{metric} with $e^{2\nu_2} = (1+r^2)^2$ and $e^{2\nu_1}=(a+bt)$  represents the corresponding spacetime. This is a very special case of the model discussed in the current section. Here, we can see from equation \eqref{fcs} that avoiding a singularity is not possible when $k=m=2$, if one considers conventional fluids.

\subsection{Case II: self-similar case}
In the second example, the spacetime is assumed to have continuous self-similarity. There exists a significant amount of study regarding cylindrically symmetric self-similar spacetimes in the literature. A numerical study by Nakao et al.\cite{Nakao:2009jp} showed that
the asymptotic nature of the gravitational field outside a collapsing hollow dust cylinder is self-similar. Cylindrically symmetric collapse models under the assumption of self-similarity have been examined in \cite{Harada:2008rx, Wang:2003vf, Sharif:2005rg, Nolan2007CylindricallySM, Condron:2013rha}. Sharif and Aziz\cite{Sharif:2005rf} discussed the behaviour of the kinematical quantities (e.g. Expansion, Shear and Rotation (ESR)) and thereby analyzed the singularity feature for cylindrically symmetric solutions having self-similarity.

When the spacetime has a continuous self-similarity, the metric \eqref{metric} can be expressed as\cite{Nolan2007CylindricallySM},
\begin{equation}
  \mathrm{d}s^2 = e^{2\mu(\xi)-2\nu(\xi)}\left(-dt^2+dr^2\right)+ r^2 \left[e^{2\nu(\xi)}dz^2+R^2(\xi)e^{-2\nu(\xi)}d\phi^2\right].
  \end{equation}
 This can be done using a redefinition of the metric coefficients and the coordinates. Here $\xi$ denotes the similarity variable defined as, $\xi=\frac{t}{r}$.

  In the present case, we consider the following ansatz,
  \begin{equation}
 \mu=k\nu \hspace{0.2cm} \mathrm{and} \hspace{0.2cm} R=e^{m \nu}.
\end{equation}
 This leads to a simplification similar to that in the previous case. Using $u^a=e^{\nu-\mu}\delta^a_0$,  $\chi^a = e^{\nu-\mu}\delta^a_1$ and $ n^a=\frac{e^{-\nu}}{r}\delta^a_2$, the Einstein field equations yields,
 \begin{equation}\label{ricss}
 R_{\alpha\beta} u^\alpha u^\beta=\frac{\rho_\mathrm{f}+p_\mathrm{rf}+2 p_\mathrm{tf}}{2}+\frac{B^2}{2},
\end{equation}
\begin{equation}\label{divaccss}
 \nabla_\alpha a^\alpha=  \frac{\kappa \xi^2\left(k-1\right)}{\left(\xi^2-1\right)\left(2-m\right)}B^2,
\end{equation}
and
\begin{equation}\label{sigss}
 \sigma_{\alpha\beta}\sigma^{\alpha\beta} =\frac{2}{3}\frac{e^{2(1-k)\nu}}{r^2}\left(m^2+k^2-m(k+2)-2k+4\right)\tilde{\nu}^2,
 \end{equation}
where
\begin{equation}
\begin{split}
 \tilde{\nu}^2\equiv \left(\frac{\mathrm{d}\nu}{\mathrm{d}\xi}\right)^2=\frac{\kappa r^2 e^{2(k-1)\nu}}{\xi^2-1}\left[\frac{4-4k-m}{2(m-2)(km-1)}B^2\right. \\ +\frac{k-1}{m(km-1)}(\rho_\mathrm{f}-p_\mathrm{rf}+2)-\left. \frac{p_\mathrm{tf}}{km-1}\right].
 \end{split}
\end{equation}

For this example, if $m=2$, consistency demands $B=0$. So, we exclude this particular case in the following discussion.
Here also the conditions
$R_{\alpha\beta} u^\alpha u^\beta+\sigma_{\alpha\beta}\sigma^{\alpha\beta}-\nabla_\alpha a^\alpha< 0$ and $\tilde{\nu}^2>0$ will lead to a constraint on the magnetic field strength for which focusing of the particle worldlines can be avoided.
But, unlike the previous example, this depends explicitly on the spacetime point (as $\nabla_\alpha a^\alpha$ and $\tilde{\nu}^2$ have explicit dependence on $\xi$).

Here we shall focus on two specific limits. At first we consider $\xi\rightarrow \infty$. This gives an idea about whether the collapse leads to a central singularity (i.e. we are concerned about the limit $r\rightarrow 0$, the limit $t\rightarrow \infty$ is inconsequential).
In this limit, we have $R_{\alpha\beta} u^\alpha u^\beta$, $\nabla_\alpha a^\alpha\gg \sigma_{\alpha\beta}\sigma^{\alpha\beta}$ which is evident from the expressions \eqref{ricss}, \eqref{divaccss} and \eqref{sigss}. So the effect due to shear can be neglected. Therefore, the formation of a central singularity as a result of the collapse can be avoided if we have,
\begin{equation}
 R_{\alpha\beta} u^\alpha u^\beta-\nabla_\alpha a^\alpha<0\implies \left[\frac{\left(k-1\right)}{\left(2-m\right)}-\frac{1}{2}\right]B^2>\frac{\left(\rho_\mathrm{f}+p_\mathrm{rf}+2p_\mathrm{tf}\right)}{2}.
 \end{equation}
 From this we see that the constants $k$ and $m$ must satisfy, $\frac{\left(k-1\right)}{\left(2-m\right)}>\frac{1}{2}$.

 We consider $t\rightarrow 0$ as another limit. This corresponds to $\xi\rightarrow \infty$. In this case, one can see $\nabla_\alpha a^\alpha\ll R_{\alpha\beta} u^\alpha u^\beta, \sigma_{\alpha\beta}\sigma^{\alpha\beta}$. The FC, therefore, holds in this limit. This gives rise to two different possibilities. A singularity may occur at $t=0$. Alternatively, the choice of initial conditions of the problem should be such that the evolution must start from a contracting phase. As the collapse proceeds, a singularity will form if the FC continues to hold. Alternatively, the divergence of acceleration may start to dominate before the evolution hits a singularity. In that scenario, the collapse will halt, and an expansion may even occur. This is possible because of the
presence of the magnetic field. During the collapse, the deformation of the magnetic field lines increases. This, in turn, leads to an increasing repulsive effect coming from the magnetic field. This fact has been discussed quite intuitively and in detail by  Tsagas and Mavrogiannis\cite{Tsagas:2020lal}.

\section{Summary and discussion}
In this chapter, we have explored how the presence of a magnetic field affects the dynamics of a  gravitational collapse. Examining the ability of magnetic fields to halt a collapse and avert the final formation of a singularity has been an essential topic of research for a long time. We have used the same approach as in the previous chapter, i.e. employing the RE to get insights into the dynamics. We have considered cylindrically symmetric inhomogeneous models where a charged fluid distribution collapses in the presence of a magnetic field. We have carried out the analysis within the ideal--MHD limit. The examples considered here are very simple. There exists only one independent metric coefficient in these examples. Despite this, they are useful enough in providing vital information about the role of a magnetic field in gravitational collapse. Using these models, we have observed that the magnetic field does give rise to repulsive stresses. We have obtained constraints on the magnetic field strength for which focusing can be prevented.

Findings of the current study strengthen the claims reported in the works \cite{Tsagas:2020lal, Mavrogiannis:2021nrl}. The current study acquires importance since it leads to important conclusions regarding magnetized inhomogeneous collapsing systems in spite of the fact that no exact solution for such systems seems to be available in the literature.

It should be made clear that obtaining solutions of the field equations is necessary to reach a definite conclusion in this context. This will fix the allowed values of $k$ and $m$. This will also tell us if the constraints obtained on the strength of the magnetic field for averting a collapse are obeyed. Detailed studies of this kind go beyond our current focus. Our primary intent here is to use the RE to investigate how magnetic fields influence the dynamics of a gravitational collapse involving charged matter distributions. The RE enables us to conduct such an investigation without solving the field equations explicitly.

\cleardoublepage
\chapter[Timelike convergence condition in scalar-tensor theories of gravity]{Timelike convergence condition in scalar-tensor theories of gravity}
\chaptermark{TCC in scalar-tensor theories of gravity}
\label{chapter4}
In the last two chapters, we have studied gravitational collapse to look into the issue of singularities. These studies have been carried out within the framework of General Theory of Relativity (GTR). In this chapter, we consider a modified theory of gravity, namely Non-Minimally Coupled Scalar--Tensor Theory (NMCSTT), to study whether the resolution of singularities is possible in this class of scalar-tensor theories. As mentioned in chapter \ref{chapter1}, the Raychaudhuri Equation (RE) can be applied in any modified theory respecting the Riemannian structure. Therefore, one can use the RE in the form \eqref{rc eq} for a timelike geodesic congruence in the framework of NMCSTT.

The field equations in NMCSTT contain additional terms which depend on the scalar degree of freedom. Therefore, even if we continue to assume that the energy conditions hold, these additional terms may give rise to repulsive effects. This leads to a possibility of violation of the convergence conditions. As a result, focusing of geodesics and formation of singularities may be avoided in these theories.
Here our aim is to use the RE to investigate whether the Timelike Convergence Condition (TCC) can be violated in NMCSTT.
We will begin by considering a general NMCSTT given by the action \eqref{nmcstta}.

\section{Timelike convergence condition in non-minimally coupled scalar-tensor theories}
From the field equations \eqref{fsce1} in NMCSTT we have,
\begin{equation}\label{maineq}
R_{ab}u^a u^b=\frac{1}{2f}\left[2(T_{ab}u^a u^b+\frac{1}{2}T)-\square f +2 u^a u^b\nabla_a \nabla_b f +\frac{2\omega}{\phi}u^a u^b\partial_a\phi \partial_b\phi \right].
\end{equation}
We need $f>0$ to ensure a positive gravitational coupling. We can see from equation \eqref{maineq} that additional terms are present in this expression which depend on the scalar field. Therefore, the assumption of the Strong Energy Condition (SEC) $\left(T_{ab}u^a u^b+\frac{1}{2}T\geq 0\right)$ does not necessarily lead to the TCC $\left(R_{ab}u^a u^b\geq 0\right)$ within this framework. This is an essential difference between this class of theories and GTR.

One can deal with a large number of scalar-tensor theories using the framework which we will discuss here. However,
we shall specialize to the two important candidate theories discussed in section \ref{bdbeken}, namely the Brans-Dicke (BD) theory and the Bekenstein Conformally Coupled Scalar--Tensor theory (BCCSTT) in the following sections. Continuing with the assumption of the SEC, we will examine whether the quantity  $R_{ab}u^a u^b$ can be negative in these theories and thus lead to a violation of the TCC. Even if we choose a specific theory, it is difficult to comment on the specific nature of the additional terms present in equation \eqref{maineq}. Therefore, we have to use some exact solutions to reach definite conclusions. We will consider examples from static spherically symmetric and spatially homogeneous and isotropic cases. As mentioned in section \ref{bdbeken}, these choices help analyze the two most common types of singularities: black hole (or naked) singularities and big bang type singularities. At this point, we should emphasize that we shall deal with isotropic cases both for the stationary and cosmological scenarios. Using these simplest examples, we aim to demonstrate how we can obtain valuable results from equation \eqref{maineq}. Assuming other symmetries (e.g. axial symmetry) may lead to different conclusions. However, it is expected that the difference there will be caused by the terms related to the shear ($\sigma_{ab}\sigma^{ab}$) or the rotation ($\omega_{ab}\omega^{ab}$) in the RE, instead of the specific characteristics of the NMCSTT.

\section{Static spherically symmetric case}
At first, we consider static spherically symmetric spacetimes to examine the fate of the TCC in the two sub-classes of NMCSTT mentioned earlier.
\subsection{The Brans-Dicke theory}
As already mentioned in section \ref{bdtheory}, the BD action is given by the equation \eqref{nmcstta} with $f(\phi)=\phi$ and $\omega=\text{constant}$. If we use isotropic  coordinates, a static spherically symmetric metric can be represented as,
\begin{equation}\label{bdmetric}
\mathrm{d}s^2= -A^2(r)\mathrm{d}t^2+B^2(r)(\mathrm{d}r^2+r^2\mathrm{d}\theta^2+r^2\sin^2\theta \mathrm{d}\varphi^2).
\end{equation}
The velocity vector for a radially incoming geodesic in this spacetime is given by,
\begin{equation}
 u^a=\left(\frac{1}{A^2},-\frac{\sqrt{1-A^2}}{AB},0,0\right).
\end{equation}
Putting this in equation \eqref{maineq} we have,
\begin{equation}\label{bdstatic}
\begin{split}
 R_{ab}u^a u^b=\frac{1}{2\phi}\left[2\left(T_{ab}u^a u^b+\frac{1}{2}T\right)+\frac{2-3A^2}{A^2}\square\phi+
 \frac{2\omega}{\phi}\frac{1-A^2}{A^2B^2}(\phi^\prime)^2\right.\\ \left.-\frac{2\phi^\prime}{A^2 B^2}\left({2(1-A^2)}\frac{B^\prime}{B}+(2-A^2)\frac{A^\prime}{A}+
 \frac{2}{r}{(1-A^2)}\right)\right].
 \end{split}
\end{equation}
Combining equation \eqref{fsce2}  with equation \eqref{bdstatic} we have,
\begin{equation}\label{rcbdsss}
\begin{split}
 R_{ab}u^a u^b=P_1+P_2+P_3+P_4,
 \end{split}
\end{equation}
where \begin{equation}\label{p1234}
\begin{split}
 P_1 &=\frac{1}{\phi}\left(T_{ab}u^a u^b+\frac{1}{2}T\right),\\
 P_2 &=\frac{3A^2-2}{A^2}\frac{R}{4\omega},\\ P_3 &=\frac{(\phi^\prime)^2}{2 A^2B^2\phi^2}\left((1+2\omega)-\left(\frac{3}{2}+2\omega\right)A^2\right),\\P_4 &= -\frac{\phi^\prime}{A^2 B^2\phi}\left({2(1-A^2)}\frac{B^\prime}{B}+(2-A^2)\frac{A^\prime}{A}+\frac{2}{r}{(1-A^2)}\right).
 \end{split}
\end{equation}
The right-hand side of the equation \eqref{rcbdsss} has been split up into four terms as indicated for convenience of discussion. These terms are discussed one by one in the following.

Here as $f(\phi)=\phi>0$ (to ensure a positive gravitational coupling), the SEC implies $P_1\geq 0$. There is a switch of sign of $P_2$ at $A^2=\frac{2}{3}$ and that of $P_3$ at $A^2=\frac{2+4\omega}{3+4\omega}$. Therefore $R_{ab}u^a u^b$ may contain terms with negative contribution. The TCC will be violated when these terms dominate.

To examine the situation explicitly, we pick up Brans class I solution\cite{Brans:1962zz} as an example. The significance of this solution has been discussed briefly in section \ref{bdtheory}. If we consider a suitable weak field limit and  assume the condition of asymptotic flatness, the metric representing this solution is given by the equation \eqref{bdmetric} with\cite{Brans:1962zz},
\begin{eqnarray}
A^2&=&\left(\frac{1-\frac{D}{r}}{1+\frac{D}{r}}\right)^{\frac{2}{\lambda}}, \\
B^2&=&\left(1+\frac{D}{r}\right)^4\left(\frac{1-\frac{D}{r}}{1+\frac{D}{r}}\right)^{\frac{2(\lambda-C-1)}{\lambda}}.
\end{eqnarray}
The solution for the scalar field is,
\begin{equation}
 \phi=\frac{1}{\lambda^2 G_0}\left(\frac{1-\frac{D}{r}}{1+\frac{D}{r}}\right)^{\frac{C}{\lambda}}.
\end{equation}
Here
 \begin{eqnarray}
   \lambda=\sqrt{\frac{2\omega+3}{2\omega+4}}, \hspace{0.2cm} C=-\frac{1}{2+\omega}, \hspace{0.2cm} D=\frac{MG_0\lambda}{2}.
 \end{eqnarray}
$M$ and $G_0$ stand for the mass of the distribution causing the spacetime curvature and the Newtonian gravitational coupling constant, respectively. This solution is valid for $\omega>-\frac{3}{2}$ and $\omega<-2$.

As the Brans class I solution is a vacuum solution, $P_1$ vanishes. The other terms in the expression of $R_{ab} u^a u^b$ (equation \eqref{rcbdsss}) are given by,
\begin{eqnarray}
   P_2 &=&\left(\frac{3A^2-2}{A^2}\right)\frac{D^2 C^2}{r^4\lambda^2 \left(1-\frac{D^2}{r^2}\right)^4 }\left(\frac{1-\frac{D}{r}}{1+\frac{D}{r}}\right)^{\frac{2(C+1)}{\lambda}},\label{p2ex}\\
 P_3 &=& \frac{(\phi^\prime)^2}{2A^2B^2\phi^2}\left((1+2\omega)-\left(\frac{3}{2}+2\omega\right)A^2\right), \label{p3ex}\\ \label{P4eq}
 P_4 &=& \frac{1}{A^2B^2}\left(\frac{8(1-A^2)D^2 C(C r-D \lambda +r)}{\lambda ^2 r^5 \left(1-\frac{D^2}{r^2}\right)^2}\right. -\frac{4(2-A^2) D^2 C}{\lambda ^2 r^4 \left(1-\frac{D^2}{r^2}\right)^2}\\ \nonumber && \hspace{1.5cm} \left.-\frac{4{(1-A^2)} D C}{r^3\lambda\left(1 -\frac{D^2}  {r^2}\right)}\right).
 \end{eqnarray}
 It can be verified from equation \eqref{p2ex}  that $P_2$ is negative when $A^2<\frac{2}{3}$. Similarly, $P_3$ is negative when $\frac{2+4\omega}{3+4\omega}<A^2<1$. Fixing the sign of $P_4$ takes a bit more effort. For $C<0$ (i.e. $\omega>-2$), the first term in equation \eqref{P4eq} is negative when $r(1+C)>{D\lambda}$, whereas the other two terms are always positive. For
$C>0$ (i.e. $\omega<-2$), the first term in equation \eqref{P4eq}  is negative when $r(1+C)<{D\lambda}$. The other two terms always remain negative in this case.

\begin{figure}[h]
\centering
\frame{\includegraphics[width=0.7\linewidth]{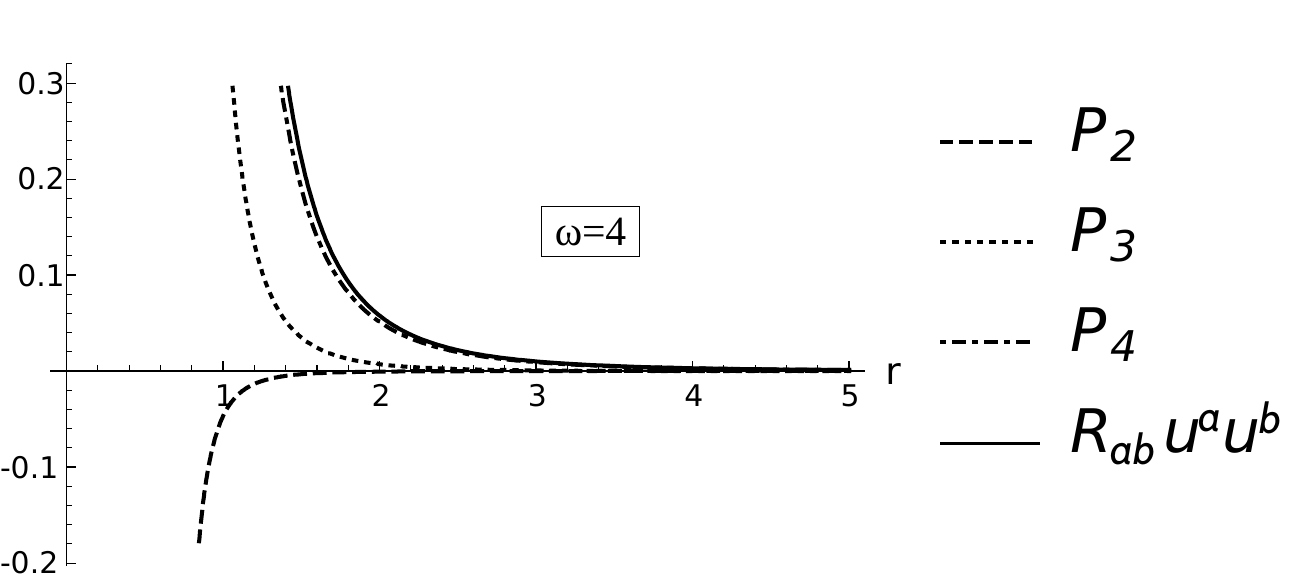}}\\[2ex]
 \frame{\includegraphics[width=0.7\linewidth]{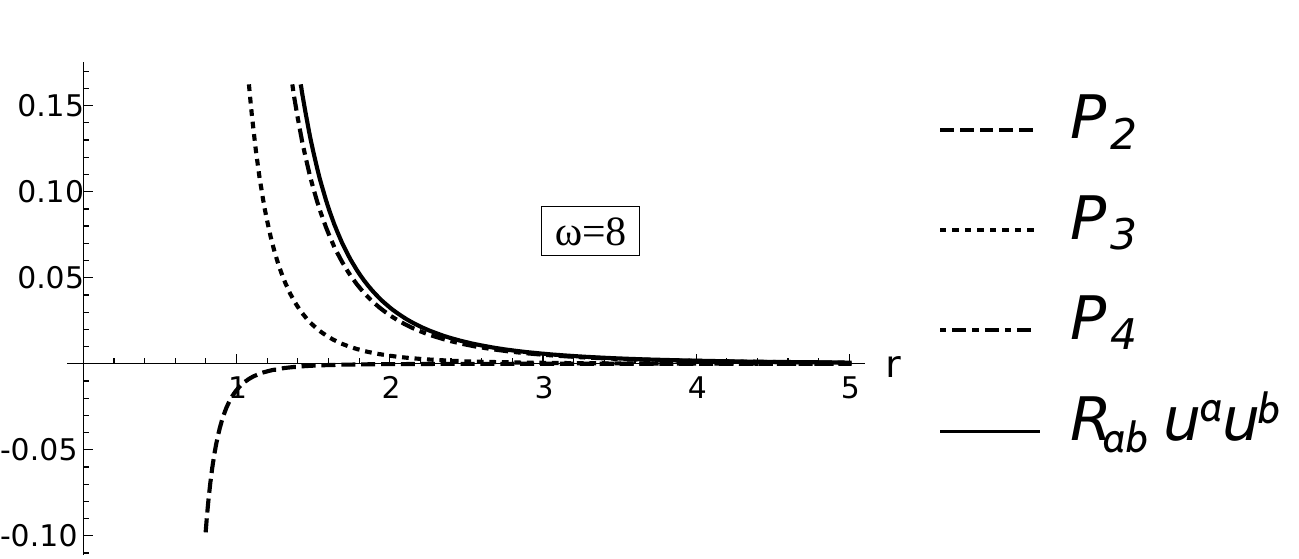}}\\[2ex]
 \frame{\includegraphics[width=0.7\linewidth]{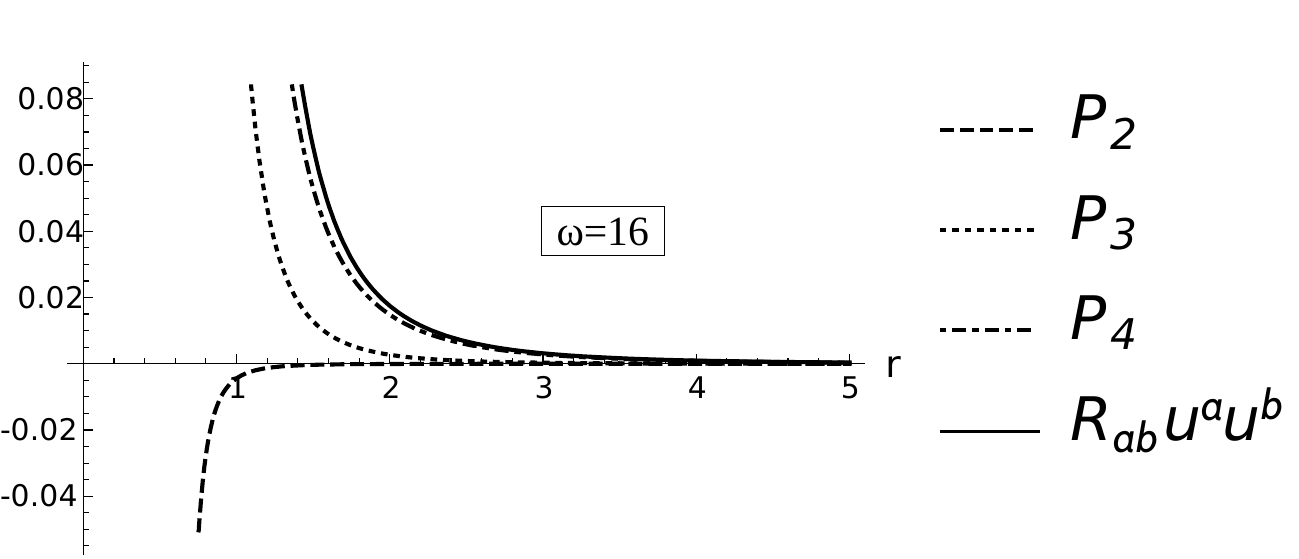}}
\caption{{Radial variation of different terms present in the expression \eqref{rcbdsss} of $R_{ab}u^a u^b$ for different positive values of $\omega$. We have chosen
$G_0=1$, $M=1$.}
 \label{figbd1}}
\end{figure}

\begin{figure}[h]
\centering
\frame{\includegraphics[width=0.7\linewidth]{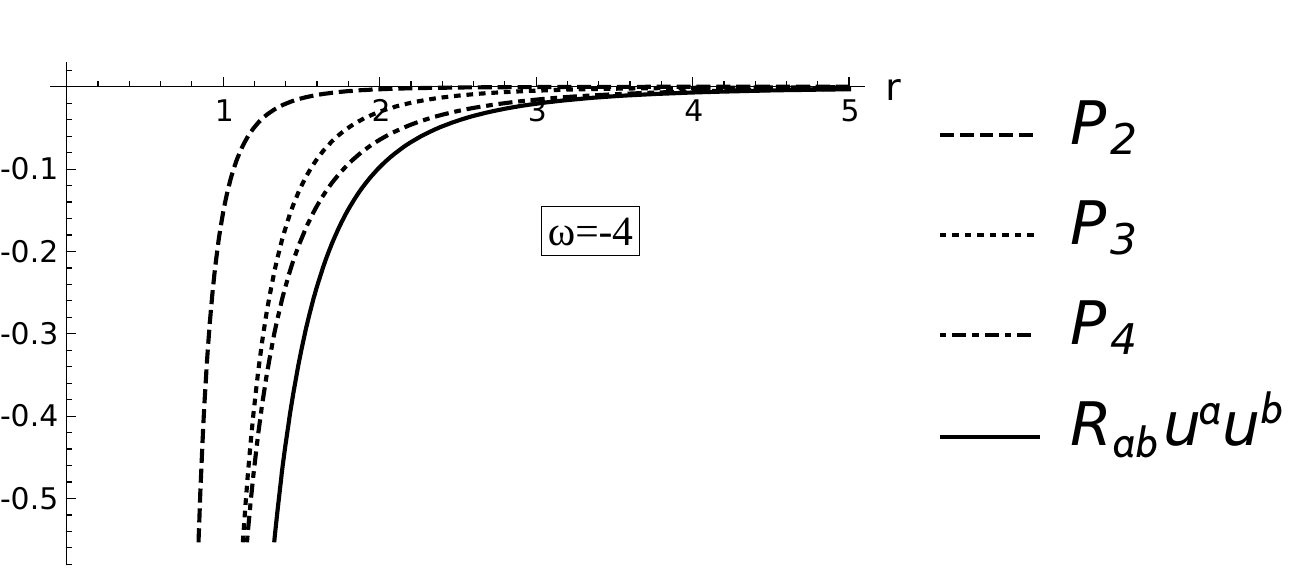}}\\[2ex]
 \frame{\includegraphics[width=0.7\linewidth]{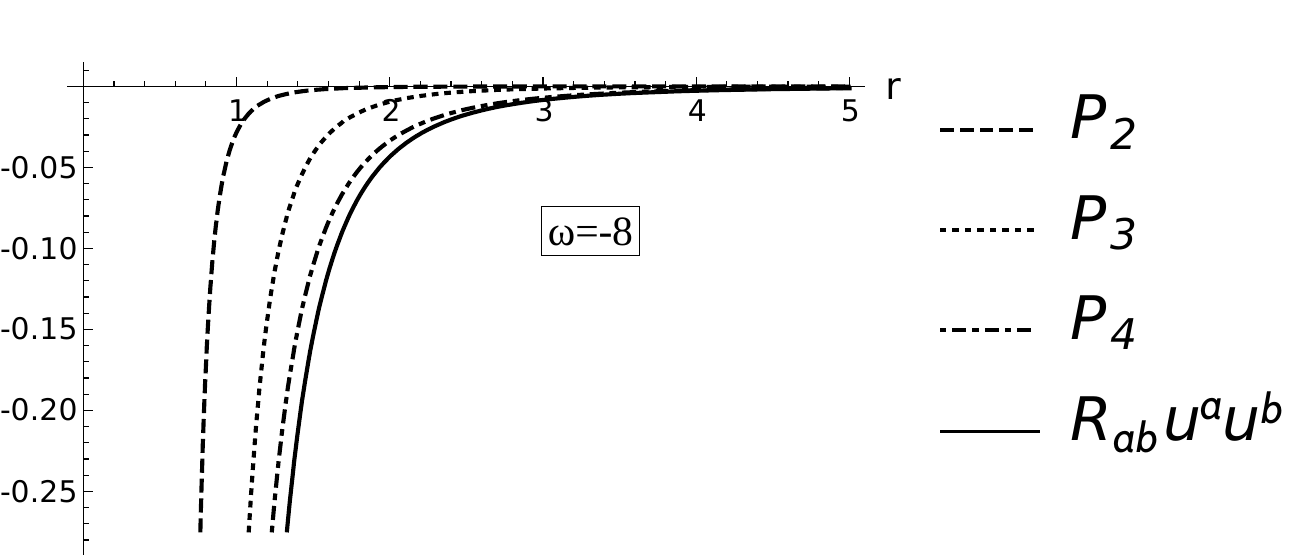}}\\[2ex]
 \frame{\includegraphics[width=0.7\linewidth]{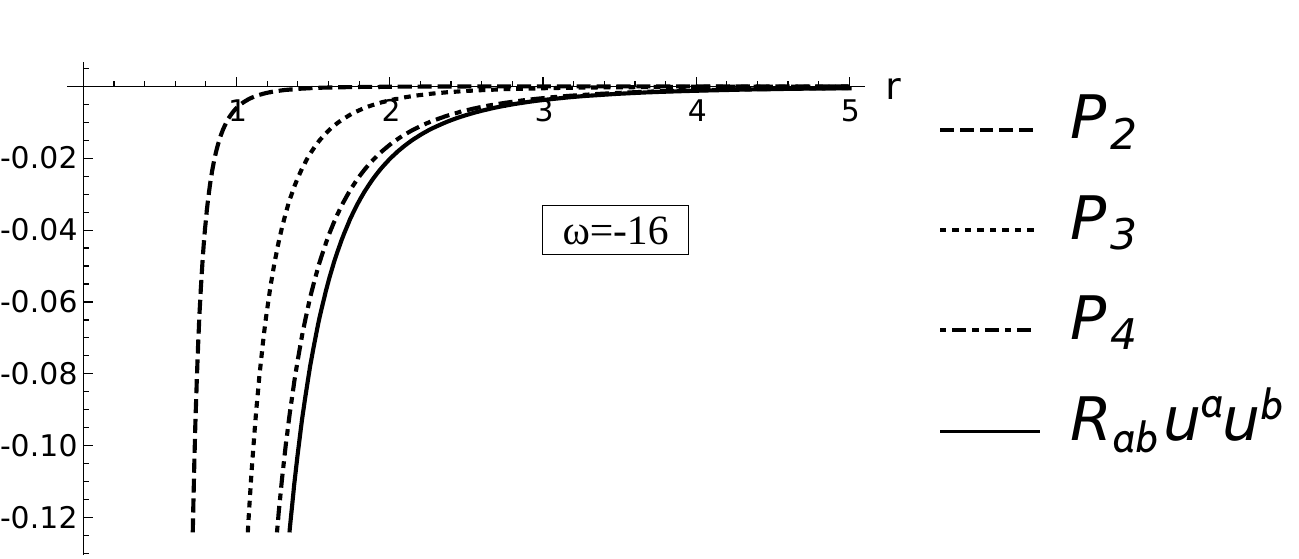}}
\caption{{Radial variation of different terms present in the expression \eqref{rcbdsss} of $R_{ab}u^a u^b$ for different negative values of $\omega$. We have chosen
$G_0=1$, $M=1$.}
 \label{figbd2}}
\end{figure}

We will now depict the behavior of the quantities $P_2$,  $P_3$, $P_4$ and $R_{ab}u^a u^b$ as they vary with $r$. This will give a definite idea regarding whether the TCC can be violated. The variation of the aforementioned quantities is demonstrated through figures \ref{figbd1} and \ref{figbd2} for different positive and negative values of $\omega$. The qualitative behaviour does not appear to depend on the magnitude of $\omega$  as long as the sign of $\omega$ is fixed. This fact is evident from the two sets of plots - figure \ref{figbd1} (plotted for three different positive values of $\omega$) and figure \ref{figbd2} (plotted for three different negative values of $\omega$).

 These plots suggest that the TCC can be violated only for allowed negative values of $\omega$ (figure \ref{figbd2}). $R_{ab}u^a u^b$ always remains positive for positive values of $\omega$ (figure \ref{figbd1}). It approaches zero in the large $r$ limit for both the cases. This is a consequence of assuming asymptotic flatness. Therefore, avoiding a singularity in this model is possible only when $\omega$ is negative.

\subsection{Bekenstein's scalar-tensor theory}
The action for this theory is,
\begin{equation}
 S=\int \sqrt{-g}\mathrm{d}^4x\left[R-\partial_a\phi \partial^a\phi-\frac{1}{6}R\phi^2\right].
\end{equation}
In our work we consider the type A solution proposed in \cite{Bekenstein:1974sf} (see section \ref{bkccstt}). This is given by,
\begin{equation}\label{bks}
 \mathrm{d}s^2= - A^{2}(r) \mathrm{d}t^2+ B^{2}(r) \mathrm{d}r^2+ S^{2}(r) \left(r^2\mathrm{d}\theta^2+r^2\sin^2\theta \mathrm{d}\varphi^2\right),
\end{equation}
where,
\begin{equation}\label{abcp}
 \begin{split}
  A^2(r)=\frac{1}{4}\left[w(r)^\beta+ w(r)^{-\beta}\right]^{2}w(r)^{2\alpha},\\
  B^2(r)=\frac{1}{4}\left[w(r)^\beta+ w(r)^{-\beta}\right]^{2}w(r)^{-2\alpha},\\
  S^2(r)=\frac{1}{4}\left[w(r)^\beta+ w(r)^{-\beta}\right]^{2}w(r)^{-2\alpha+2}.
 \end{split}
\end{equation}
Here $\alpha=\sqrt{1-3\beta^2}$ and $\beta$ lies in the range $-\frac{1}{\sqrt{3}}\leq\beta\leq \frac{1}{\sqrt{3}}$. The expression for the scalar field is,
\begin{equation}\label{conphi}
\phi = \sqrt{6} \frac{1 - w^{2\beta}}{1 + w^{2\beta}}.
\end {equation}
If we put
\begin{equation}\label{wr}
 w=\sqrt{1-\frac{2M}{r}},
\end{equation}
and $\beta=0$, the type A solution reduces to the Schwarzchild solution. For $\beta=\pm \frac{1}{2}$,  this solution represents a black hole with scalar hair. This is the well-known Bocharova--Bronnikov--Melnikov--Bekenstein (BBMB) black hole (refer to section \ref{bkccstt}). It was shown by Xanthopoulos and Zannias\cite{doi:10.1063/1.529253} that the BBMB black hole is the unique static, asymptotically flat solution for the Einstein conformal-scalar system. Another important fact is that the BBMB black hole having a negative scalar charge can mimic a wormhole or an Einstein-Rosen bridge (see \cite{Chowdhury:2018pre} and references therein).
For all other allowed values of $\beta$, the type I solution (with $\omega$ given by equation \eqref{wr})
corresponds to a spacetime which contains a naked singularity at $r=2M$. This is a point singularity when $\frac{1}{2}<|\beta|\leq\frac{1}{\sqrt{3}}$ and a 2-surface  otherwise.

Let us now consider a radially incoming timelike geodesic in this spacetime. The velocity vector for this geodesic is,
\begin{equation}
 u^a=\left(\frac{1}{A^2},-\frac{\sqrt{1-A^2}}{AB},0,0\right).
\end{equation}
Using the above information (and $f(\phi)=1-\frac{\phi^2}{6}$, $\omega(\phi)=\phi$), from equation \eqref{maineq} we have,
\begin{equation}\label{rmunubb}
 \begin{split}
  R_{ab}u^a u^b=Q_1+Q_2+Q_3,
 \end{split}
\end{equation}
with
\begin{equation}
\begin{split}
Q_1 &=\frac{\phi\square \phi}{\left(1-\frac{\phi^2}{6}\right)}\left(\frac{3A^2-2}{6A^2}\right),\\
Q_2 &=\frac{\phi\phi^\prime}{3A^2B^2{\left(1-\frac{\phi^2}{6}\right)}}\left[(2-A^2)\frac{A^\prime}{A}+2(1-A^2)\frac{S^\prime}{S}+\frac{2}{r}(1-A^2)\right],\\
Q_3 &= \frac{(\phi^\prime)^2}{{\left(1-\frac{\phi^2}{6}\right)}}\frac{4-3A^2}{6A^2 B^2}.
\end{split}
\end{equation}
The energy-momentum tensor corresponding to the conformally coupled scalar field has zero trace. So, it leads to a vanishing Ricci scalar by virtue of the field equations. Then equation \eqref{fsce2} yields,
\begin{equation}
 \square \phi=0 \implies Q_1=0.
\end{equation}
As $A^2<1$ and $f(\phi)>0$, we have $Q_3>0$. However, it is difficult to make such general statements about the nature of $Q_2$ and, therefore, that of $R_{ab}u^a u^b$. We need explicit forms of $A$, $B$, $S$ and $\phi$ for this purpose.

\begin{figure}[h]
\centering
 \frame{\includegraphics[width=0.7\linewidth]{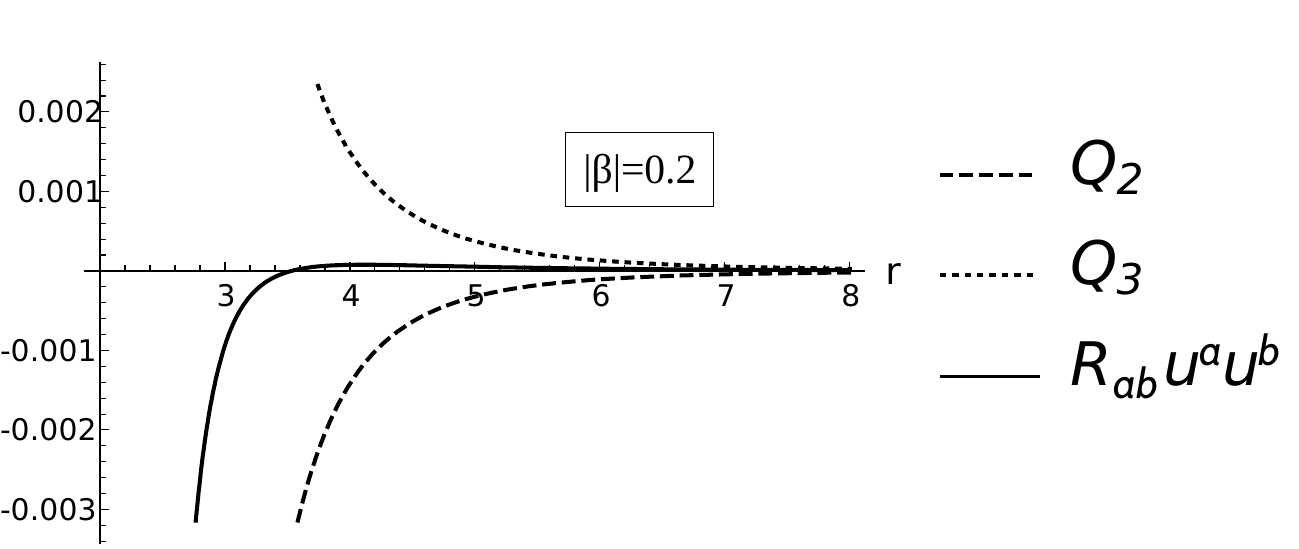}}\\[2ex]
 \frame{\includegraphics[width=0.7\linewidth]{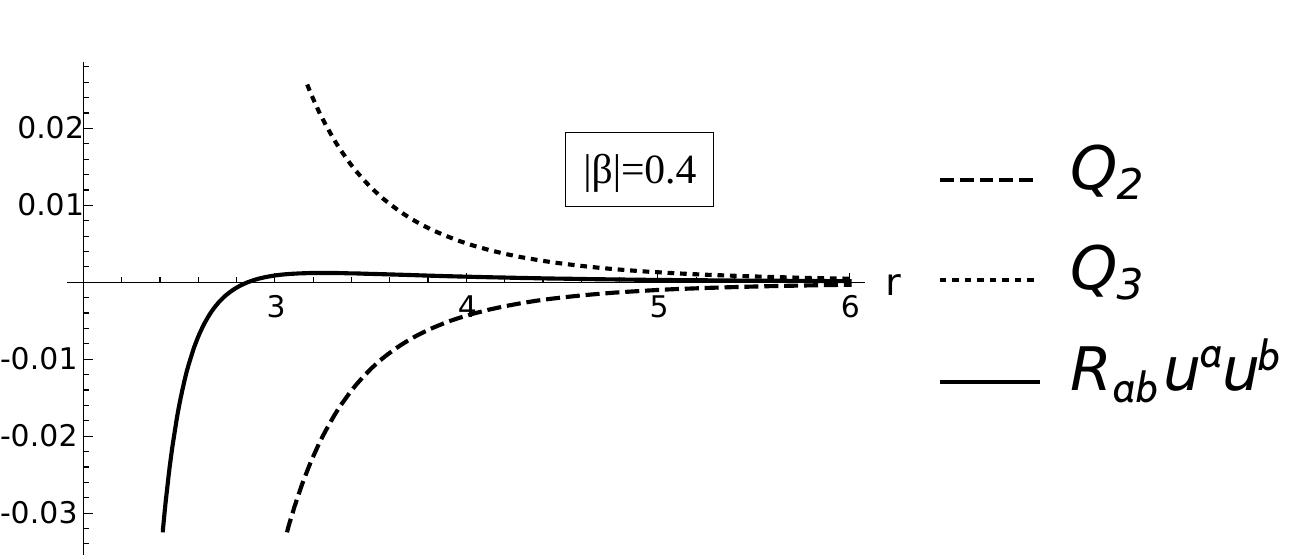}}
 \caption{{Radial variation of different terms present in the expression \eqref{rmunubb} of $R_{ab}u^a u^b$ for $|\beta|<\frac{1}{2}$. Here we have used $M=1$.}
 \label{figbkad1}}
\end{figure}

\begin{figure}[h]
\centering
 \frame{\includegraphics[width=0.7\linewidth]{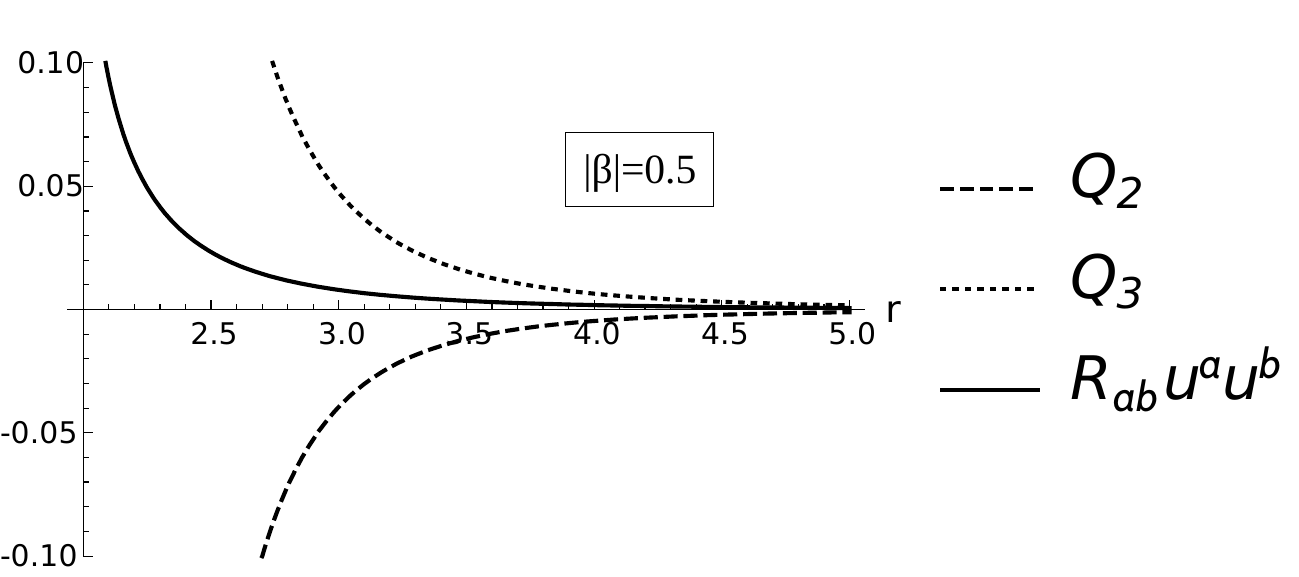}}
\caption{{Radial variation of different terms present in the expression \eqref{rmunubb} of $R_{ab}u^a u^b$ for $|\beta|=\frac{1}{2}$. Here we have used $M=1$.}
 \label{figbk3}} 
\end{figure}

\begin{figure}[h]
\centering
 \frame{\includegraphics[width=0.7\linewidth]{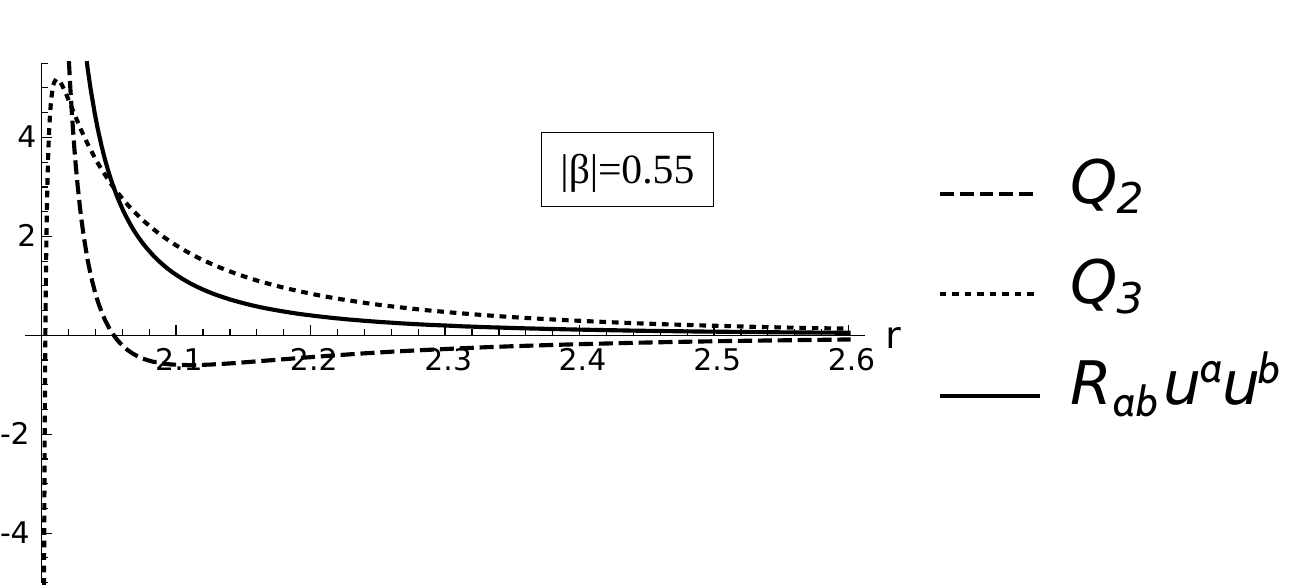}}\\[2ex]
 \frame{\includegraphics[width=0.7\linewidth]{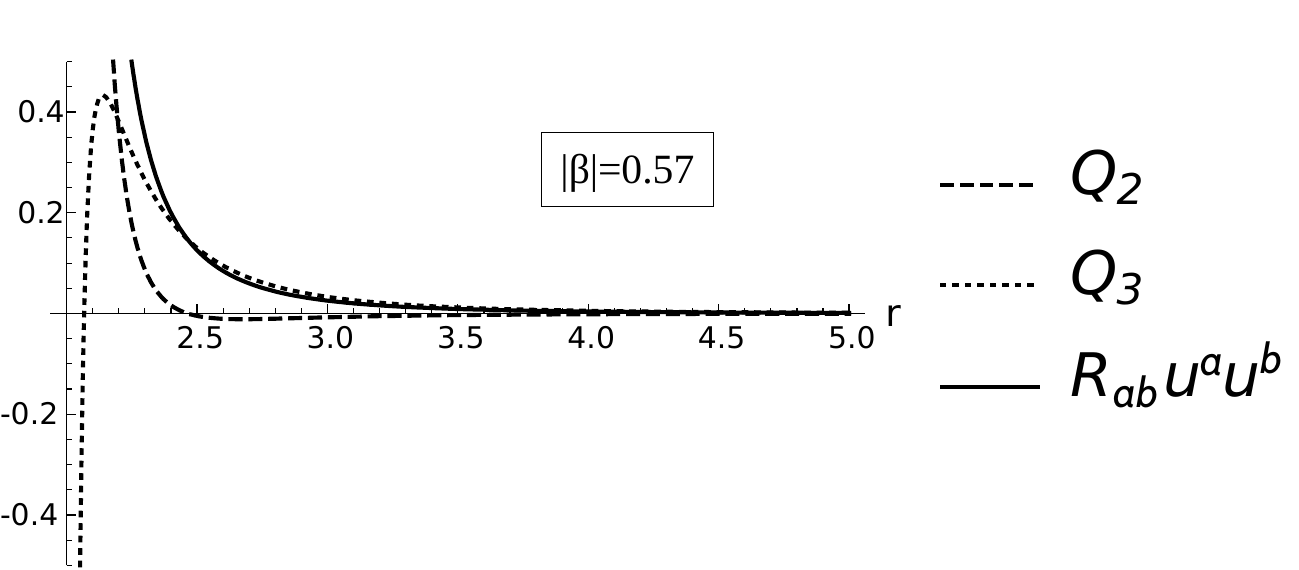}}
 \caption{{Radial variation of different terms present in the expression \eqref{rmunubb} of $R_{ab}u^a u^b$ for $\frac{1}{2}<|\beta|\leq\frac{1}{\sqrt{3}}$. Here we have used $M=1$.}
 \label{figbkad2}} 
\end{figure}

Using equations \eqref{abcp}, \eqref{conphi} and \eqref{wr}, we can explore the behaviour of $Q_2$, $Q_3$ and $R_{ab}u^a u^b$ with the help of plots. These plots are presented in figures \ref{figbkad1}--\ref{figbkad2} for different values of $\beta$. We can observe that qualitatively there exists three distinctive behaviour. This is expected as the solution \eqref{abcp} represents only three different physical scenarios depending on the value of $\beta$.

For the case of $|\beta|<\frac{1}{2}$ (i.e. from figure \ref{figbkad1}) it is evident that $R_{ab}u^a u^b$ is negative in a region close to $r=2M$. As $r$ increases, it becomes positive and approaches zero for large $r$. $R_{ab}u^a u^b$ always remains non-negative for $\frac{1}{2}\leq|\beta|\leq\frac{1}{\sqrt{3}}$ (figures \ref{figbk3} and \ref{figbkad2}).  Therefore, the TCC can be violated only when $|\beta|<\frac{1}{2}$. But, this is restricted to a small range of $r$. This result agrees with the findings reported in \cite{Bekenstein:1974sf}.

\section{Spatially homogeneous and isotropic case}
We now turn to cosmology and consider a spatially homogeneous and isotropic universe. The universe is represented by the Friedmann--Robertson--Walker (FRW) metric,
\begin{equation}
 \mathrm{d}s^2=-\mathrm{d}t^2+a^2(t)\left(\frac{\mathrm{d}r^2}{1-kr^2}+r^2\mathrm{d}\theta^2+r^2\sin^2\theta \mathrm{d}\varphi^2\right),
\end{equation}
where $a(t)$ is the scale factor and $k$ is the curvature index. The cases corresponding to $k=0$, $1$ and $-1$ represent spatially flat, closed and open universes, respectively.
The universe is assumed to be filled with a perfect fluid whose energy-momentum tensor is given by,
\begin{equation}
 T_{ab}=(\rho+p)u_a u_b+pg_{ab}.
\end{equation}
Here $u^a=(1,0,0,0)$ is the velocity of the fluid in the comoving frame; $\rho$, $p$ are the energy density and the pressure of the fluid.

\subsection{The Brans-Dicke theory}
In the spatially homogeneous and isotropic case, equation \eqref{maineq} yields,
\begin{equation}\label{frbd1}
 R_{ab}u^a u^b=\frac{\rho+3p}{2\phi}+\frac{3}{2}\frac{\ddot{\phi}}{\phi}+\frac{3}{2}\frac{\dot{a}}{a}\frac{\dot{\phi}}{\phi}+\omega\frac{\dot{\phi}^2}{\phi^2},
\end{equation}
under the framework of the BD theory.
Upon further mathematical manipulations, we have,
\begin{equation}\label{frbd}
 R_{ab}u^a u^b=\frac{\omega}{(2\omega+3)\phi}(3p-\rho)+3\frac{\dot{a}^2}{a^2}+\frac{3k}{a^2}+\frac{\omega}{2}\frac{\dot{\phi}^2}{\phi^2}.
\end{equation}
In this expression, the second term in the right-hand side is always positive. The next term is negative only for $k=-1$, whereas the last term is negative when $\omega<0$. If the fluid satisfies the condition, $\rho>3p$, the first term is negative for $\omega>0$ or $\omega< -\frac{3}{2}$. On the other hand, if $\rho<3p$, we need $-\frac{3}{2}<\omega<0$ for the first term to be negative.

It is evident from equation \eqref{frbd1} that signs of the derivatives of $\phi$ and that of $\omega$ have a significant role in determining the fate of the TCC. In the absence of exact solutions, making any specific comment on these signs is difficult. This is why we had to use the field equations to arrive at \eqref{frbd}. This equation indicates the role of the curvature $k$ in this context. In this theory, an open universe may actually lead to non-singular models.

We will use an exact solution to explicitly check whether the TCC can be violated in some situations. This solution corresponds to a spatially flat geometry with the fluid in the form of a pressure-less $(p=0)$ dust. The solution is given by,
\begin{equation}
 a=a_0 t^{\frac{2(\omega+1)}{3\omega+4}},
\end{equation}

\begin{equation}
 \phi=\frac{(3\omega+4)\rho_0 t^{\frac{2}{3\omega+4}}}{2a_0^3(2\omega+3)}.
\end{equation}
Brans and Dicke proposed this solution in their seminal paper \cite{Brans:1961sx}. Using this solution, we have,
\begin{equation}
 R_{ab}u^a u^b=\frac{6 (\omega +1) (\omega +2)}{t^2 (3 \omega +4)^2}.
\end{equation}
It is clear from the above expression that the TCC will be violated only when $-2<\omega< -\frac{3}{2}$. Banerjee and Pavon\cite{Banerjee:2000mj} found that this same condition on $\omega$ is needed to explain the accelerated expansion of the Universe characterized by this solution. But, we should note that this range of $\omega$ is observationally unfavourable (see section \ref{bdtheory}).

Here we have considered a very simple example. It is still sufficiently useful to show that the SEC alone cannot guarantee the TCC in the BD theory. The value and sign of $\omega$ play a crucial role in this context.

\subsection{Bekenstein's scalar-tensor theory}
For this case, equation \eqref{maineq} gives us,
\begin{equation}\label{bksfrw}
 R_{ab}u^a u^b=\frac{1}{2\left(1-\frac{\phi^2}{6}\right)}\left[(\rho+3p)+\dot{\phi}^2-\phi\ddot{\phi}-\frac{\dot{a}}{a}\phi\dot{\phi}\right].
\end{equation}
Using \eqref{fsce1} and \eqref{fsce2} in the above equation \eqref{bksfrw} leads to
\begin{equation}\label{frbk}
 R_{ab}u^a u^b=\frac{3p-\rho}{2}+3\frac{\dot{a}^2}{a^2}+\frac{3k}{a^2}.
\end{equation}
The first term in the right-hand side of this equation is negative if the fluid obeys $\rho>3p$. The second term is always positive. The third term is negative only when $k=-1$. Therefore, for a universe consisting of radiation $(p=\frac{1}{3}\rho)$, $R_{ab}u^a u^b$ can be negative only when the universe is open.

As explicit examples, we will now refer to exact solutions corresponding to the $k=0$, $1$ and $-1$ cases. All of them represent radiation-filled universes and were reported by Bekenstein in \cite{Bekenstein:1974sf}.

\begin{itemize}
 \item Solution for the scale factor in the spatially flat case is,
 \begin{equation}
 a(t)=C (t-t_0)^\frac{1}{2},
\end{equation}
where $C$ and $t_0$ are arbitrary constants. Using this solution, we obtain,
\begin{equation}
  R_{ab}u^a u^b=\frac{3}{4}\frac{1}{(t-t_0)^2}.
\end{equation}
This is always positive.

\item For the $k=1$ case we have,
\begin{equation}
 a(t)=\sqrt{C^2-(t-t_0)^2}.
\end{equation}
Then the quantity $R_{ab}u^a u^b$ given by,
\begin{equation}
 R_{ab}u^a u^b=\frac{3C^2}{\left[C^2-(t-t_0)^2\right]^2},
\end{equation}
is always positive definite.
Thus, for both the spatially flat and closed cases, the TCC holds and focusing of geodesics is inevitable.

\item Finally, for the $k=-1$ case,
\begin{equation}\label{keqm1}
 a(t)=\sqrt{C^2+(t-t_0)^2}.
\end{equation}
Here we have,
\begin{equation}
 R_{ab}u^a u^b=-\frac{3C^2}{\left[C^2+(t-t_0)^2\right]^2},
\end{equation}
which is negative. This gives rise to the possibility of avoiding focusing and the formation of singularities.
\end{itemize}

The above conclusions are in excellent agreement with those which follow from equation \eqref{frbk}. The TCC can be violated only when $k=-1$, and the solution for this case, given by equation \eqref{keqm1}, is indeed non-singular, as shown by Bekenstein\cite{Bekenstein:1974sf}. Here we should mention the following point. It is well-known that our Universe appears to be spatially flat. However, the present exercise aims to explore what spatial geometry could have led to non-singular models in BCCSTT.
\section{Summary and discussion}
As mentioned in section \ref{phst}, the assumption of the convergence conditions is necessary to prove the inevitable existence of singularities. The SEC and the TCC are equivalent in GTR. In other relativistic theories of gravity, this TCC may be modified due to the characteristics of those theories. Therefore, violation of the TCC is possible even if we continue with the assumption of the SEC. In this chapter, we have investigated this modified convergence condition in two NMCSTTs, namely the BD theory and the BCCSTT. We have selected a few exact solutions for two representative situations in both the theories to obtain specific conclusions. The first one is the static spherically symmetric case which represents the analogue of the Schwarzchild geometry. The second represents Friedmann cosmologies.

We have found that for static spherical distributions, violation of the TCC is possible depending on the value of $\omega$ in the BD theory and on $|\beta|$ in Bekenstein's theory. Here it should be mentioned that $\omega$ is a parameter of the BD theory, whereas $\beta$ is a constant of integration in Bekenstein's theory.

In the cosmological setting, an important finding is that the TCC can indeed be violated at least in an open universe. Using a limited number of examples, we have seen that while such violation is possible only for an open universe in Bekenstein's theory, this may also be achieved for a spatially flat geometry in the BD theory.

\cleardoublepage
\chapter[Reconstruction of $f(R)$-gravity models for an accelerated universe]{Reconstruction of $f(R)$-gravity models for an accelerated universe}
\chaptermark{Reconstruction of $f(R)$-gravity models}
\label{chapter5}

We have discussed in section \ref{accex} that the phenomenon of the late time accelerated expansion of the Universe still lacks complete theoretical understanding. We have also discussed a few possible ways considered in the literature to explain this dark energy phenomenology. In this chapter, we attempt to find an explanation of this phenomenon. We consider the approach of using a modified theory of gravity, namely $f(R)$-gravity, rather than introducing an exotic matter component.

There exists a plethora of models that can account for the late-time acceleration of the Universe, but all of them lack substantial support from other related branches of physics. In the absence of a widely accepted theoretical model for dark energy phenomenology, reconstructing models using observational data proves to be a good option. This uses the idea of starting from a suitable evolution history of the Universe and finding the required distribution of matter\cite{Ellis:1990wsa}.

Here our endeavour is the reconstruction of $f(R)$-models, not by employing the observational data but rather by following the idea put forward by Ellis and Madsen\cite{Ellis:1990wsa}. We start from a particular form of evolution resulting in an accelerated expansion characterized by a kinematical quantity. After that, we try to find the forms of $f(R)$ that can cause such an evolution. For a discussion regarding the significance of the kinematical quantities in the context of reconstruction of models of an accelerating universe, we refer to \cite{Visser:2003vq, Zhai:2013fxa, Mukherjee:2016trt}.

While various attempts addressing this kind of reverse engineering technique have already been reported in the literature\cite{Song:2007da, Capozziello:2008qc, Pogosian:2007sw, Nojiri:2009kx, Dunsby:2010wg, Carloni:2010ph, Lombriser:2010mp, He:2012rf}, we shall follow a different strategy. We use the Raychaudhuri Equation (RE) to carry out the process of reconstruction.
The RE has been applied within the framework of $f(R)$-gravity to explore effective energy conditions and investigate whether repulsive gravity can be obtained out of the geometry\cite{Santos:2007bs, Albareti:2012va, Santos:2016vjg}.
\section{The Raychaudhuri equation in $f(R)$ gravity}
Using the field equations \eqref{fieldeqfr} for $f(R)$-gravity we can write,
\begin{equation}
 R_{ab}u^a u^b=\frac{1}{f^\prime}\left[T_{ab}+\frac{f}{2} g_{ab}+(\nabla_a\nabla_b-g_{ab}\square)f^\prime\right]u^a u^b.
\end{equation}
If we use the above expression to replace the last term in the right-hand side of the equation \eqref{rc eq ng}, we get the timelike RE for $f(R)$-gravity.

The Universe is described by the spatially flat Friedmann--Robertson--Walker (FRW) metric,
\begin{equation}
 \mathrm{d}s^2 = -\mathrm{d}t^2 + a^2 (t) (\mathrm{d}r^2 + r^2 \mathrm{d}{\theta}^2 + r^2 \sin^2 \theta \mathrm{d}{\phi}^2),
\end{equation}
where $a(t)$ is the scale factor.

We assume that the Universe is filled with a perfect fluid. The corresponding energy-momentum tensor is thus given by,
\begin{equation}\label{emtpf}
T^{ab} = (\rho + p)u^{a}u^{b} + p g^{ab}.
\end{equation}
Using, $u^a=(1,0,0,0)$, the RE in this setting can be written as\cite{Guarnizo:2011bnf},
\begin{equation}\label{rceq1}
 \frac{\ddot{a}}{a}=\frac{1}{f^\prime}\left(\frac{f}{6}+H f^{\prime\prime}\dot{R}-\frac{\rho}{3}\right),
 \end{equation}
 where $dot$ denotes derivative with respect to $t$.
 It should be clarified that we have used the field equations \eqref{fieldeqfr} to arrive at the above expression \eqref{rceq1}. Up to this point, we have not assumed any equation of state for the fluid.

 \section{Reconstruction of $f(R)$-models}
 In this section, we will attempt to reconstruct models of $f(R)$-gravity for a specified mode of accelerated expansion. We will employ equation \eqref{rceq1} for this purpose. The mode of acceleration is characterized by the kinematical quantities, e.g., the deceleration parameter $q$ or the jerk parameter $j$. We consider two examples in this study. The first one is specified by a constant deceleration parameter $q$. This gives rise to an ever-accelerating universe. For the other, we take $j=1$, which indicates an evolution mimicking that due to the $\Lambda$CDM model in standard General Theory of Relativity (GTR).

 \subsection{Model I: deceleration parameter being a constant}
The well-known Hubble parameter, defined by $H=\dfrac{\dot{a}}{a}$, is the earliest observable quantity in physical cosmology. Though this is often called Hubble's constant, observations indicate that this parameter evolves with time. Therefore the deceleration parameter, $q=-\dfrac{a\ddot{a}}{{\dot{a}}^2}$, related with the next order derivative of the scale factor finds a lot of interest.

 We begin with a toy model of a universe which assumes,
 \begin{equation}\label{q}
 q=-\frac{1}{H^2}\frac{\ddot a}{a}=\mbox{constant}=-m,
\end{equation}
with $m$ being restricted as $0<m<1$. When $m<0$, the universe decelerates during the expansion. On the other hand, for $m>1$, we have accelerated contraction. Also, we have not included $m=1$ in our discussion. This case describes an exponential expansion signifying a pure deSitter universe without any matter evolving with time.

We can obtain the expression for the scale factor by solving equation \eqref{q}. This is given by,
\begin{equation}\label{sc-fact}
 a(t)=C (t-t_0)^{\frac{1}{1-m}},
\end{equation}
where $C$ and $t_0$ are integration constants.

At first, we shall discuss the \emph{effective equation of state}, $\omega_\mathrm{eff}$ for this kind of a model. For that, we will use the field equations written in terms of the effective energy density $\rho_\mathrm{eff}$ and the effective pressure $p_\mathrm{eff}$\cite{Sotiriou:2008rp}. These equations are,
\begin{equation}\label{hubble}
 H^2=\frac{\rho_\mathrm{eff}}{3},
\end{equation}
and
\begin{equation}\label{dec}
 \frac{\ddot{a}}{a}=-\frac{3 p_\mathrm{eff}+\rho_\mathrm{eff}}{6}.
\end{equation}
The above two equations, along with the solution for the scale factor (equation \eqref{sc-fact}) yield,
\begin{equation}
 w_\mathrm{eff} = \frac{p_\mathrm{eff}}{\rho_\mathrm{eff}}=-\frac{1+2m}{3}.
\end{equation}
The upper and lower limits of $m$, namely $1$ and $0$, lead to $w_\mathrm{eff}=-1$ and $-\frac{1}{3}$ respectively.

For a perfect fluid with the energy-momentum tensor given by equation \eqref{emtpf}, the Strong Energy Condition (SEC) is, $(\rho+3p)\geq 0$. In a similar way,
the \emph{effective energy condition} in the current context is characterised by the quantity, $\rho_\mathrm{eff}+3p_\mathrm{eff}$. For the present model, this quantity is given by,
\begin{equation}
 \rho_\mathrm{eff}+3p_\mathrm{eff}=\frac{-6m}{(1-m)^2(t-t_0)^2},
\end{equation}
which is always negative. Hence the effective energy condition will always be violated. One should expect this for an eternally accelerated universe.\\

In this case, the Ricci scalar can be expressed in terms of the scale factor as,
\begin{equation}
R=6\dfrac{\dot{a}^2}{a^2}+6\dfrac{\ddot{a}}{a}.
\end{equation}
We can write the derivatives of $a$ as functions of $a$ using the solution \eqref{sc-fact}. Hence the RE \eqref{rceq1} and the Ricci scalar are respectively given by,
\begin{equation}\label{Rceq2}
\begin{split}
\frac{12(1+m)}{(1-m)^3}{\left(\frac{C}{a}\right)}^{4(1-m)}f^{\prime\prime}(R)+\frac{m}{(1-m)^2}\left(\frac{C}{a}\right)^{2(1-m)}f^{\prime}(R)
-\frac{f(R)}{6}=-\frac{\rho}{3}.
\end{split}
\end{equation}
and
\begin{equation}
R=6\frac{(1+m)}{(1-m)^2}\left(\frac{C}{a}\right)^{2(1-m)}.
\end{equation}
As the Ricci scalar and the scale factor are functionally related, we can recast equation \eqref{Rceq2} as,
\begin{equation}\label{Rceq3}
\frac{1-m}{3(1+m)} R^2 f^{\prime\prime}(R)+\frac{m}{6(1+m)} R f^{\prime}(R)-\frac{f(R)}{6}=-\frac{\rho}{3}.
\end{equation}
From the energy-momentum conservation equation for the fluid distribution, i.e. $\nabla_a T^{ab}=0$, we obtain, $\dot{\rho} + 3H (\rho + p)=0$. Thus, if we assume that the universe is dominated by pressure-less dust, we have,
\begin{equation}
 \rho=\dfrac{\rho_0}{a^3}.
\end{equation}
Replacing this in equation \eqref{Rceq3} we get,
\begin{equation}\label{const-acc-fdprime}
 R^2f^{\prime\prime}(R)+\frac{m}{2(1-m)} R f^{\prime}(R)-\frac{1+m}{2(1-m)}f(R)=ER^{\frac{3}{2(1-m)}},
\end{equation}
with $E=-\dfrac{\rho_0}{C^3}\dfrac{(1+m)}{(1-m)}
\left[\dfrac{(1-m)^2}{6(1+m)}\right]^\frac{3}{2(1-m)}.$\\

The analytical solution for $f(R)$ which follows from this equation is,
\begin{equation}\label{solconacc}
f(R)=C_1 R^\alpha+C_2 R^\beta+F R^\gamma,
\end{equation}
where $C_1$, $C_2$ are integration constants and \\[2ex]
$F=\dfrac{4E(1-m)^2}{1+m(9+2m)}$,\\[2ex]
$\alpha=\dfrac{1}{4(1-m)}\left[(2-3m)+\sqrt{m^2-12m+12}\right]$,\\[2ex]
$\beta=\dfrac{1}{4(1-m)}\left[(2-3m)-\sqrt{m^2-12m+12}\right]$,\\[2ex]
$\gamma=\dfrac{3}{2(1-m)}$.\\

\begin{figure}[h]
\centering
 \frame{\includegraphics[width=0.7\linewidth]{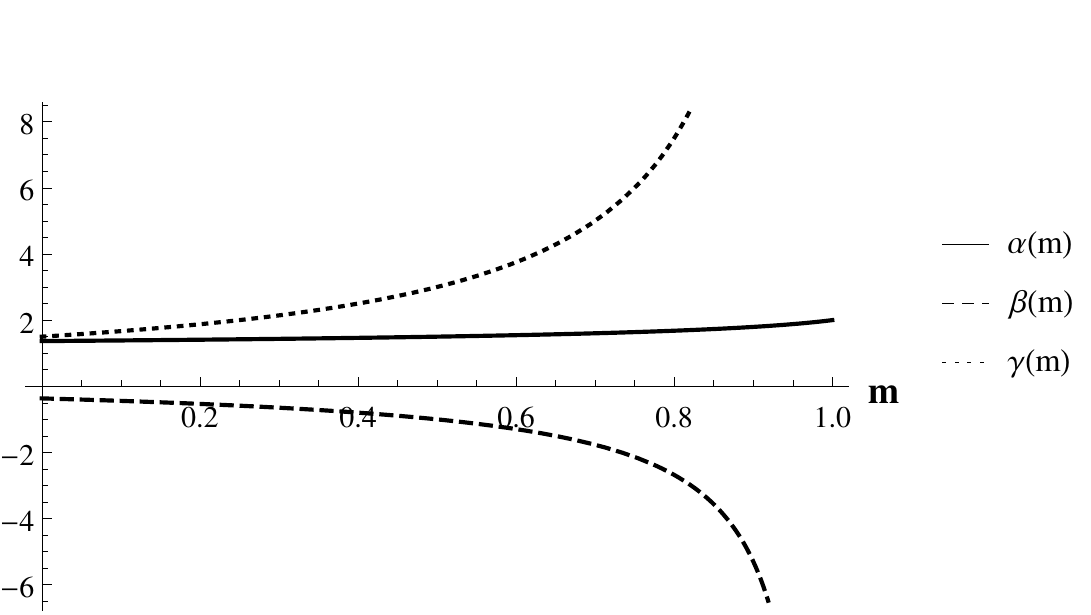}}
   \caption{Plot of $\alpha$, $\beta$, $\gamma$ with $m$.}
 \label{figpow}
\end{figure}

In the solution \eqref{solconacc} for $f(R)$, there are three different powers of the Ricci scalar. The variations of these three powers with $m$ are shown in the figure \ref{figpow}.
Within the range $0<m<1$, we have $\alpha, \gamma>0$ while $\beta<0$. Note that choosing $C_2=0$ gives an $f(R)$, which leads to accelerated expansion despite having only positive powers of $R$.
This has already been observed by Capozziello et al.\cite{Capozziello:2003gx}. If we want to recover GTR as a special case from this kind of a model, we need one of the two positive powers to be equal to $1$. This is not possible for the given range of $m$.

\subsubsection*{Viability analysis}
\begin{figure}[h]
\centering
 \frame{\includegraphics[width=0.7\linewidth]{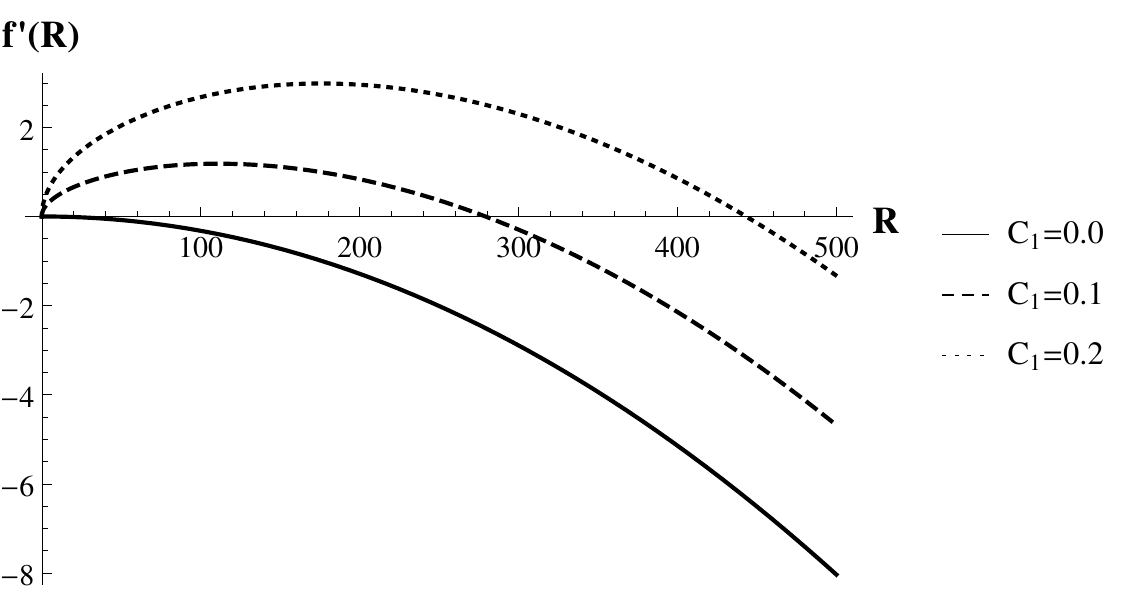}}\\[2ex]
 \frame{\includegraphics[width=0.7\linewidth]{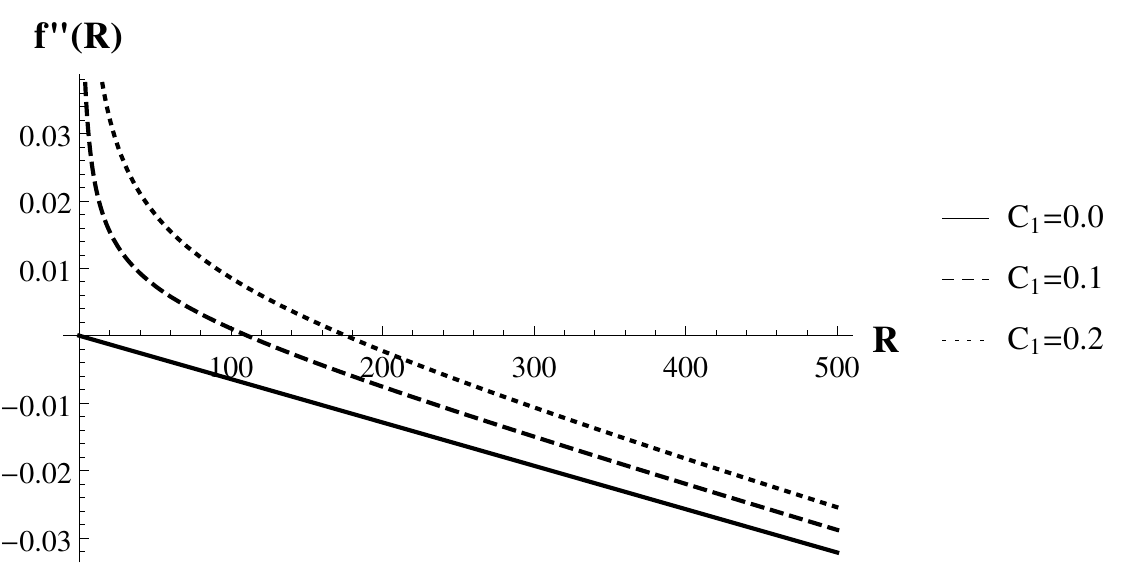}}
   \caption{{Plot of $f'$ and $f''$ (corresponding to the solution \ref{solconacc}) with $R$ for different values of $C_1$, where we have chosen $C_2=0$, $m=0.5$ and $\dfrac{\rho_0}{C^3}=1$.}}
 \label{sv1}
\end{figure}

The model under consideration definitely does not fit with the observations. Leaving this aside, we should also check if this is theoretically consistent. We have discussed the viability criteria for $f(R)$ models in section \ref{vcfr}. We need $f'(R)$, $f''(R)>0$ for an $f(R)$ model to be viable. In the solution for $f(R)$ (equation \eqref{solconacc}), the second term dominates in the low curvature (i.e. $R\rightarrow 0$) limit because $\beta<0$.   Therefore, in this limit, we must have $C_1$ to be positive and $C_2=0$ for viability. In the high curvature limit, the third term is dominant as $\gamma>\alpha$. But, the coefficient of this term, namely $F$, is negative. So, the model fails to be viable at high curvature. For illustration of the above mentioned points we have plotted $f'(R)$ and $f''(R)$ with $R$ in figure \ref{sv1}.

\subsection{Model II: a constant jerk parameter}
It has been found from observations that the deceleration parameter also evolves. Therefore there should be a natural shift of attention towards its evolution which is specified by the jerk parameter. This is related to the third-order derivative of the scale factor. The jerk parameter is defined as,
\begin{equation}\label{const_j}
 j=\frac{1}{H^3}\frac{\dddot{a}}{a}.
\end{equation}
Reconstruction of accelerated models of the Universe employing the jerk parameter as the starting point is gaining increasing interest. For motivation regarding this, we refer to \cite{Zhai:2013fxa, Mukherjee:2016trt}.

Although there is a huge discrepancy between the theoretically predicted and the cosmologically required values of the cosmological constant $\Lambda$, the $\Lambda$CDM model fits very well with the observed accelerated expansion of the Universe. In this section our reconstruction of $f(R)$-models starts with the ansatz,
\begin{equation}\label{j1}
 j=\frac{1}{H^3}\frac{\dddot{a}}{a}=1.
\end{equation}
In this case, the resulting evolution mimics the same due to the $\Lambda$CDM model.

We can obtain the general solution of the equation \eqref{j1} as,
\begin{equation}\label{scale}
 a(t)=\left[A\exp(\lambda t)+B\exp(-\lambda t)\right]^{\frac{2}{3}},
\end{equation}
where $A, B,$ and  $\lambda$ are integration constants.
Here it is to be noted that when $AB<0$, the solution for the scale factor can be rewritten as, $a(t) = a_0 \left[\sinh(\lambda(t-t_0))\right]^{\frac{2}{3}}$. We shall refer to this case as type I evolution in the following. In the case of $AB>0$, $a(t)$ can be expressed as $a(t) = a_0 \left[\cosh(\lambda(t-t_0))\right]^{\frac{2}{3}}$. We shall call this type II evolution.

Let us now discuss the effective equation of state parameter and the effective energy condition for this model. Following a process similar to that used in the previous case, we have,
\begin{equation}
 w_\mathrm{eff} = \frac{p_\mathrm{eff}}{\rho_\mathrm{eff}}=
 -\frac{\left[A\exp(\lambda t)+B\exp(-\lambda t)\right]^2}{\left[A\exp(\lambda t)-B\exp(-\lambda t)\right]^2}.
\end{equation}
For type I evolution,
\begin{equation}
 w_\mathrm{eff}= -\tanh[\lambda(t-t_0)]^2.
\end{equation}
In the limit $t\rightarrow\infty$, $w_\mathrm{eff}$ tends to $-1$, while for $t\rightarrow t_0$, it tends to zero. This is quite promising as an accelerated expansion follows after a long matter-dominated era. This is expected to happen for a $\Lambda$CDM model. For this kind of evolution,
\begin{equation}\label{encond1}
 \rho_\mathrm{eff}+3p_\mathrm{eff}=\frac{4}{3} \lambda ^2 \left(\coth ^2[\lambda  (t-{t_0})]-3\right),
\end{equation}
which tells that the effective energy condition is obeyed or violated depending on the epoch $t$ one is looking at.

In the case of type II evolution,
\begin{equation}
 w_\mathrm{eff} = \frac{p_\mathrm{eff}}{\rho_\mathrm{eff}}=
 -\coth[\lambda(t-t_0)]^2.
 \end{equation}
 This also tends to $-1$ in the limit $t\rightarrow \infty$. However, as $t\rightarrow t_0$, it takes very large values. Therefore, we cannot have a matter era in the past as required for the observed Universe. This is a major drawback of this type of evolution. In this case,
 \begin{equation}\label{encond2}
 \rho_\mathrm{eff}+3p_\mathrm{eff}=\frac{4}{3} \lambda ^2 \left(\tanh ^2[\lambda  (t-{t_0})]-3\right),
\end{equation}
which indicates that the effective energy condition is violated throughout the evolution.\\

For the type I evolution, the Ricci scalar is given by,
\begin{equation}\label{ricciI}
 R=\frac{16\lambda^2}{3}\left(1+\frac{a_0^3}{a^3}\right).
\end{equation}
The same for the type II evolution is,
\begin{equation}\label{ricciII}
 R=\frac{16\lambda^2}{3}\left(1-\frac{a_0^3}{a^3}\right).
\end{equation}
The requirement of the scale factor being real and positive leads to the following conditions on $R$: we have $R>\dfrac{16\lambda^2}{3}$ for the type I evolution, whereas $R<\dfrac{16\lambda^2}{3}$ for the type II.\\

For both the type I and the type II evolutions, the RE \eqref{rceq1} can be expressed as,
\begin{equation}\label{mastereq}
\begin{split}
 \left(R-{4}\lambda^2\right)\left(R-\frac{16}{3}\lambda^2\right)f^{\prime\prime}(R)-\frac{1}{6}\left(R-8\lambda^2\right) f^\prime(R)
 -\frac{1}{6}f(R)=-\frac{\rho}{3}.
 \end{split}
\end{equation}
If we substitute $z=\dfrac{3(R-4\lambda^2)}{4\lambda^2}$,  we can transform the corresponding homogeneous equation (when $\rho=0$) into the standard hypergeometric equation:
\begin{equation}\label{mastereqz}
 z(1-z)\frac{\mathrm{d}^2 f}{\mathrm{d}z^2}+[c-(a+b+1)z]\frac{\mathrm{d}f}{\mathrm{d}z}-ab f=0,
 \end{equation}
with $a=\dfrac{-7+\sqrt{73}}{12}$, $b=\dfrac{-7-\sqrt{73}}{12}$ and
$c=-\dfrac{1}{2}$.
This equation has three different singular points at $z=0$, $1$ and $\infty$. These correspond to $R=4\lambda^2$, $\dfrac{16}{3}\lambda^2$, $\infty$, respectively. As we are interested in real solutions, the points $+ \infty$  and $- \infty$ must be treated as distinct. The homogeneous equation then has different real solutions around four different singular points at $R=-\infty, 4\lambda^2, \dfrac{16}{3}\lambda^2, \infty$. Table \ref{tab} summarizes these solutions and the region of their validity. Detailed discussion regarding the hypergeometric functions can be found in \cite{maier2006generalization}.

We have listed four different solutions of the equation \eqref{mastereqz} around four different singular points. We have to find out which of these solutions are actually relevant as complementary functions of equation \eqref{mastereq}. This is determined by the boundary conditions, and the range of $R$ one is interested in. We shall elaborate on this when we write down the complete general solution for $f(R)$.\\

\begin{landscape}
\begin{table}
 \centering
 \begin{tabular}{|c|c|c|}
\hline
Singular Point & Solution & Range of applicability  \\
\hline
$z=0$ $\left(R=4\lambda^2\right)$ &  $f_{h}(z)=W_1\hspace{0.2cm} {}_2F_1\left( a,b; c; z \right)
 $ & $-1<z<1$ $\left(\dfrac{8\lambda^2}{3}<R<\dfrac{16\lambda^2}{3}\right)$ \\
 & $+W_2 z^{1-c} {}_2F_1\left(1+a-c,1+b-c; 2-c; z \right)$ & \\[2ex]
\hline
$z=1$ $\left(R=\dfrac{16\lambda^2}{3}\right)$ &  $f_{h}(z)=W_1\hspace{0.2cm} {}_2F_1\left( a,b; 1+a+b-c; 1-z \right)
 $ & $0<z<2$ $\left(4\lambda^2<R<\dfrac{20\lambda^2}{3}\right)$ \\
 & $+W_2 \left(z-1\right)^{c-a-b} {}_2F_1\left(c-a,c-b; 1+c-a-b; 1-z \right).$ &  \\[2ex]
\hline
$z=\infty$ $\left(R=\infty\right)$ &  $f_{h}(z)=W_1 \left(z-1\right)^{-a}
 {}_2F_1\left( a,c-b; a-b+1; \left(1-z\right)^{-1}  \right)
 $ & $1<z<\infty$ $\left(\dfrac{16\lambda^2}{3}<R<\infty\right)$ \\
 & $+W_2 \left(z-1\right)^{-b}
 {}_2F_1\left( b,c-a; b-a+1; \left(1-z\right)^{-1}  \right)$ &   \\ [2ex]
\hline
$z=-\infty$ $\left(R=-\infty\right)$ &  $f_{h}(z)=W_1 (-z)^{-a}
 {}_2F_1\left( a,a-c+1; a-b+1; {z}^{-1}  \right)
 $ & $-\infty<z<0$ $\left(-\infty<R<4\lambda^2\right)$ \\
 & $+W_2 (-z)^{-b}
 {}_2F_1\left( b,b-c+1; b-a+1; {z}^{-1}  \right)$ & \\[2ex]
\hline
\end{tabular}
\caption{The solutions of the equation \eqref{mastereqz} around different singular points and their region of validity. Subscript $h$ is used to denote the homogeneous part. $W_1$ and $W_2$ are integration constants.
\label{tab}}
\end{table}
\end{landscape}

Our next job is to find the particular integral for equation \eqref{mastereq}. We again assume the Universe to be dust dominated, i.e. $\rho=\dfrac{\rho_0}{a^3}$. Now, the inhomogeneous term on the right-hand side, namely $\frac{\rho}{3}$, can be expressed as a function of the Ricci scalar. Equation \eqref{mastereq} thus takes the following form -
\begin{equation}
\label{rceq-j}
 \begin{split}
 \left(R-{4}\lambda^2\right)\left(R-\frac{16}{3}\lambda^2\right)f^{\prime\prime}(R)-\frac{1}{6}\left(R-8\lambda^2\right)
 f^\prime(R) -\frac{1}{6}f(R) \\= k \left(R-\frac{16\lambda^2}{3}\right),
 \end{split}
\end{equation}
where $ k =\dfrac{\rho_0}{16\lambda^2 a_0^3}$ for type II evolution and $k =-\dfrac{\rho_0}{16\lambda^2 a_0^3}$ type I evolution. The particular integral for this equation is,
\begin{equation}
 f_{p}(R)=-3k\left(R-\frac{8\lambda^2}{3}\right),
\end{equation}
where $p$ is used to denote the particular integral.

If we want to find the relevant forms of $f(R)$ which can account for a late time $\Lambda$CDM behaviour, we have to choose proper conditions. We shall illustrate this using two examples, one each for the type I and the type II mode of evolution.

\subsubsection*{General solution for the type I evolution: an example}
For type I evolution, we have two possible choices for the complementary function from table \ref{tab}. These are the second one and the third one. The former can be used in the range $\dfrac{16\lambda^2}{3}<R<\dfrac{20\lambda^2}{3}$ while the latter is valid for
$\dfrac{16\lambda^2}{3}<R<\infty$. It is natural to choose the latter as it serves the purpose for the entire region $R>\dfrac{16\lambda^2}{3}$.

Now, we can write down a complete solution for $f(R)$ for the type I evolution. This works for the entire range $R>\dfrac{16\lambda^2}{3}$. The solution in terms of $z=\dfrac{3(R-4\lambda^2)}{4\lambda^2}$ is given by,
\begin{equation}\label{finalfr}
 \begin{split}
  f(z)=W_1 \left(z-1\right)^{-a}
 {}_2F_1\left( a,c-b; a-b+1; \left(1-z\right)^{-1}  \right)\\ +W_2 \left(z-1\right)^{-b}
 {}_2F_1\left( b,c-a; b-a+1; \left(1-z\right)^{-1}  \right)-4k\lambda^2\left(z+1\right).
 \end{split}
\end{equation}
As the curvature is anticipated to be high at the early stages and to decrease during the evolution, the above solution is consistent with a distant past.
Also, we can make this work at present times by choosing $\lambda$ in such a way that the present value of the Ricci scalar $R_0>\dfrac{16}{3}\lambda^2$.\\

\begin{figure}[h]
\centering
  \frame{\includegraphics[width=0.7\linewidth]{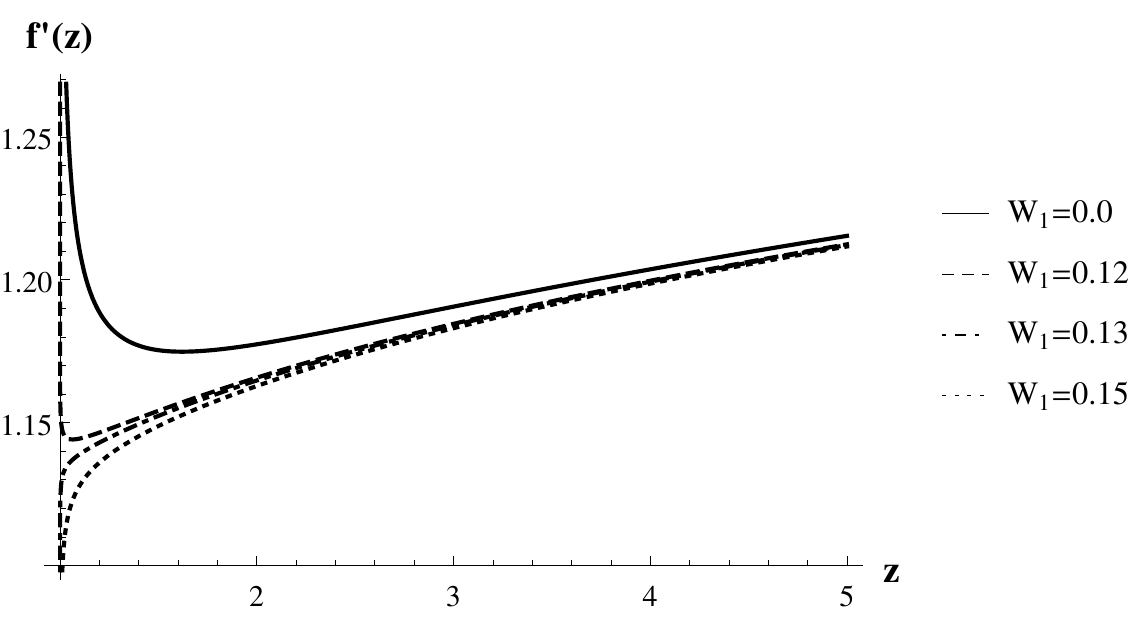}}\\[2ex]
  \frame{\includegraphics[width=0.7\linewidth]{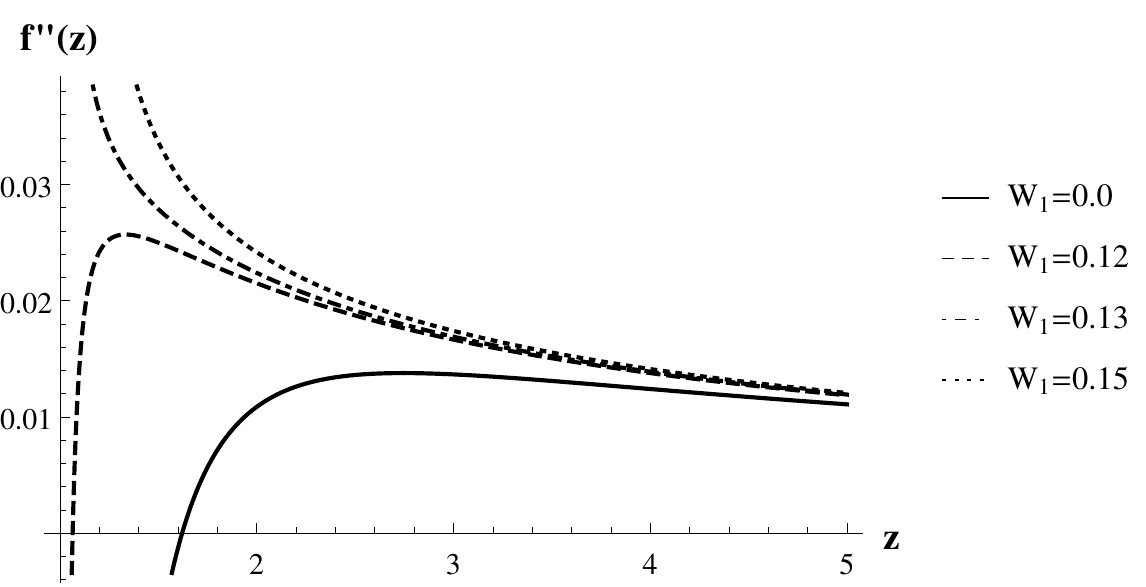}}
 \caption{{Plot of $f'$ and $f''$(corresponding to the solution \ref{finalfr}) with $z$ for different values of $W_1$, where we have chosen $W_2=0.1$, and $4k\lambda^2=-\dfrac{\rho_0}{a_0^3}=-1$.}}
 \label{sv3}
\end{figure}

\textbf{Viability check:} The second term in the solution of $f(z)$ (equation \ref{finalfr}) dominates as $z\rightarrow \infty$. Therefore, for the solution to be viable at high curvatures, we need $W_2>0$. When the curvature is low, $z$ approaches unity and the first two terms containing hypergeometric functions dominate. Then the relative sizes of $W_1$ and $W_2$ determine the behaviour of the solution. It can be shown using properties of the hypergeometric functions that $f'$ and $f''$ are of opposite signs when $z\rightarrow 1$. This is illustrated in figure \ref{sv3} where for example we set $W_2=0.1$ and $\dfrac{\rho_0}{a_0^3}=1$ (we have chosen $\dfrac{\rho_0}{a_0^3}=1$ so that $f'_p(z)=1$). The qualitative inferences do not depend on this choice. Therefore, we see that both $f'$ and $f''$ cannot be positive at the same time, and either of them takes high negative values as $z\rightarrow 1$. Therefore, this model cannot serve as a viable option.

\subsubsection*{General solution for the type II evolution: an example}
In the case of the type II evolution, the first solution in the table \ref{tab} can be used for the range $\dfrac{8\lambda^2}{3}<R<\dfrac{16\lambda^2}{3}$. The second and the fourth can do the job in the ranges  $4\lambda^2<R<\dfrac{16\lambda^2}{3}$ and $-\infty<R<4\lambda^2$, respectively. But none of them is suitable for the whole range $0<R<\dfrac{16\lambda^2}{3}$. Here as a complementary function, we choose the first one. Hence the complete solution for $f(z)$ is,
\begin{equation}\label{finalfr2}
 \begin{split}
  f(z)=W_1\hspace{0.2cm} {}_2F_1\left( a,b; c; z \right)
 +W_2 z^{1-c} {}_2F_1\left(1+a-c,1+b-c; 2-c; z \right)\\-4k\lambda^2\left(z+1\right).
 \end{split}
\end{equation}
This solution is valid for the range $0<z<1$ when $W_2\neq 0$ and $-1<z<1$ otherwise. Here also, we can make the solution work at the present epoch by adjusting the value of $\lambda$, but an extension to a distant past is not possible.\\

\begin{figure}[h]
\centering
  \frame{\includegraphics[width=0.7\linewidth]{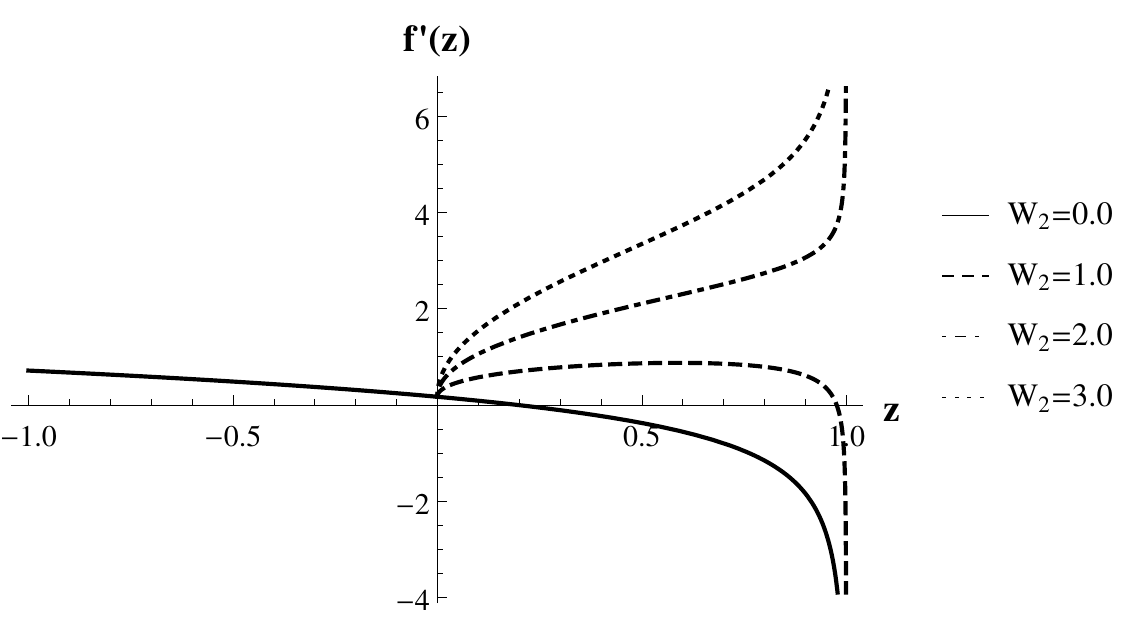}}\\[2ex]
  \frame{\includegraphics[width=0.7\linewidth]{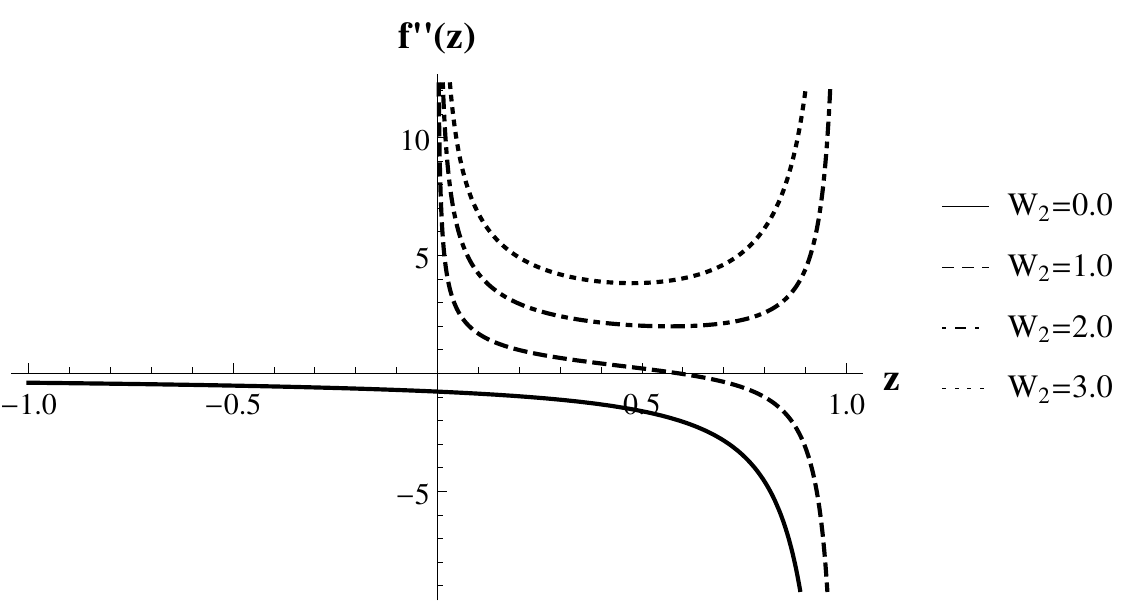}}
  \caption{{Plot of $f'$ and $f''$  with $z$ for different values of $W_2$, where we have chosen $W_1=3.5$, and $4k\lambda^2=\dfrac{\rho_0}{a_0^3}=1$.}}
 \label{sv5}
\end{figure}

{\bf Viability of the solution:} When $W_2\neq 0$, the contribution in $f'$ coming from the term with coefficient $W_2$ (the second term in the solution \ref{finalfr2}) is negligible in the limit $z\rightarrow 0$. For the choice $4k\lambda^2=\dfrac{\rho_0}{a_0^3}=1$, we need $W_1>3$ for having $f'>0$ as $z\rightarrow 0$. This follows from the properties of the hypergeometric series. As the contribution coming from this second term is dominant in $f''$ as $z\rightarrow 0$, we must have $W_2>0$ for $f''>0$. On the other hand, for $W_2=0$, we need $W_1<0$ for $f''$ to be positive. It is clear from the above discussion that having both $f'$ and $f''$ to be positive as $z\rightarrow 0$ is not possible when $W_2=0$. A non-zero $W_2$ is needed for the viability of this example. For a more detailed illustration $f'$ and $f''$ are plotted in figure \ref{sv5}. We have used $4k\lambda^2=\dfrac{\rho_0}{a_0^3}=1$
and $W_1=3.5$ in these plots. One can see from the plots that for higher values of $W_2$ (e.g. $W_2=2$ or $3$ in the plots), both $f'$ and $f''$ are positive. However, the type II model suffers from the shortcoming of leading to a completely unfavourable effective equation of state for low values of $t$ (i.e. high values of $z$). \\

We have discussed the viability of two particular examples in detail. We can also carry out the viability analysis for the remaining solutions in the table. The solution around $z=1$ (i.e. the second one in the table) leads to $f'$ and $f''$ which have opposite signs when $z\rightarrow 1$. The solution around $z=-\infty$ (the last one in the table) has the same problem as $z\rightarrow 0$.\\

An important feature of the $j=1$ model is that the corresponding particular integral contains a term proportional to $R$. So, GTR can always be recovered as a special case from this kind of a model. However, it is clear from the above discussion that the only viable choice in the present case is given by $f(R)=R-2\Lambda$, where $\Lambda$ is a constant equivalent to the cosmological constant.\\

Here it is worthwhile to mention a few earlier works. Dunsby et al.\cite{Dunsby:2010wg} showed that, for a universe filled with dust, the only real-valued $f(R)$ which can mimic the behaviour of the $\Lambda$CDM model is the Einstein-Hilbert term ($R$) along with a positive cosmological constant. Their starting point was to consider an evolution ansatz for the Hubble parameter. They extracted the above inference from the Friedmann equations including a cosmological constant in the context of $f(R)$-gravity. Later, He and Wang\cite{He:2012rf} used a slightly different approach and found another real-valued analytical $f(R)$ which leads to an evolution mimicking the $\Lambda$CDM model. This can be expressed in terms of hypergeometric functions. The solution for $f(R)$ obtained by them matches with one of the present cases, namely equation \eqref{finalfr}, after identifying the corresponding variables. Also, they have chosen $W_2=0$. This choice follows from the requirement of the chameleon property, i.e., $f_h$ and $f_h^\prime$ are convergent in the limit $R\rightarrow \infty$.

\section{Summary and discussion}
In this chapter, we have discussed a new strategy for the reconstruction of models of $f(R)$-gravity for a specified evolution history of the Universe. We have used the RE for this purpose. We have successfully carried out the investigation for two examples. It should be mentioned that the examples considered here are simple; therefore, even without using the RE, we can obtain the same results with a few additional steps. For more involved scenarios, this technique may lead to information that is difficult to obtain otherwise.

The first example is a simple model describing an eternally accelerating universe. The reconstructed $f(R)$ is a simple combination of powers of the Ricci scalar. One can obtain a variety of models in this case because these powers are not uniquely determined. An important conclusion in this context is that an $f(R)$ gravity model for which GTR can be recovered as a special case cannot yield an ever-accelerating model of universe.

The second example corresponds to an evolution which is equivalent to the celebrated $\Lambda$CDM mode of evolution in GTR. We have discussed two possible types of evolution in this case. For all the models, the reconstructed $f(R)$ contains a combination of hypergeometric functions of $R$. The example considered for the type I case is valid for the early Universe to present times subject to a tuning of a free parameter. The example discussed for the type II mode works only for a limited portion of the evolution. This is applicable for the current state of the Universe but is unable to account for a distant past.

The major conclusions of this work can be summarised as follows. The second example, which reproduces the most favoured evolution history of the Universe, i.e. the $\Lambda$CDM model, leads to the trivial choice $f(R)=R-2\Lambda$ as the only viable option. All other non-trivial possibilities corresponding to $j=1$ suffer from pathologies like instability or a negative effective gravitational coupling or not accounting for a sufficient matter-dominated era in the past or some combination of the above. The first example of the toy model having a constant negative deceleration parameter fails the viability test for moderately high curvatures, apart from being inconsistent with observational data.

\cleardoublepage
\chapter{Conclusions}
\label{chapter6}



This thesis addresses two essential themes, each focused on a critical problem in classical gravitational physics. The first one is the problem of spacetime singularities. Laws of physics break down and physical quantities diverge at a singularity. The other one is the late time accelerated expansion of the Universe. Currently known forms of matter cannot explain this phenomenon within the scope of GTR. We have explored the utility of the Raychaudhuri Equation (RE) in investigating these two crucial issues. The motivation for using the RE comes from the fact that we can obtain valuable inferences about different kinds of gravitational systems without having to refer to solutions of field equations. This enables us to perform analytical treatment of systems for which finding exact solutions is difficult.

A major part of this thesis studies the subject of singularities. Here we aim to find escape routes from these troublesome predictions of GTR. This part extends from chapter \ref{chapter2} to \ref{chapter4}.\\

We have studied two different gravitational collapse scenarios in chapters \ref{chapter2} and \ref{chapter3}, respectively, within the scope of General Theory of Relativity (GTR). Unhindered gravitational collapse of matter distributions is believed to result in singularities. Hence, analyzing collapsing configurations is undoubtedly one of the best approaches to a careful investigation of this issue. We need to define a relevant congruence to study gravitational collapse if we wish to use an approach based on the RE. To serve this purpose, we have considered the collection of worldlines of particles inside the collapsing distribution as a congruence.

In chapter \ref{chapter2}, we have studied the dynamics of a spherically symmetric self-similar gravitational collapse. The matter distribution for the system contains a fluid and a scalar field. Although we have imposed a stringent symmetry requirement on spacetime in this work, the matter content has been kept quite general. We have first determined the general Focusing Condition (FC) for the system. We have discussed that this condition is helpful in the absence of exact solutions. From this condition, we have derived important constraints on the metric and matter variables for some special cases characterized by specific choices of the matter content. Using these constraints, we have discovered situations where the formation of singularities due to collapse is either inevitable or can be prevented. In addition, we have shown that the inferences drawn from the FC are consistent with those from exact solutions wherever available. This study also reveals a connection between the FC and the Critical Phenomena (CP) in gravitational collapse. Collapse to a singularity occurs when the FC is obeyed, whereas a dispersal is possible only when the FC is violated. We have found this conclusion consistent by referring to a few exact solutions. We have shown this explicitly with a particular example. We have also determined the critical parameter and its critical value for this example. Using both approaches (i.e. using the RE and the exact solutions), we have found that non-singular evolution, namely dispersal, is possible for the system under consideration.

In chapter \ref{chapter3}, we have explored how non-gravitational agents, specifically a magnetic field in our case, affect the dynamics of a gravitational collapse. It has been found in some previous and more recent studies that magnetic fields can give rise to repulsive stresses. These stresses can prevent the final singularity formation if they overpower the gravitational attraction. In our study we have examined a scenario where an inhomogeneous distribution of a charged fluid collapses in the presence of a magnetic field within the ideal--MHD limit. We assume that the distribution is cylindrically symmetric, and the magnetic field points along the axis of symmetry. This kind of magnetized collapse with no restriction on the magnetic field, to begin with, has not been studied previously. We have considered two representative situations for our investigation. In our first example, we have assumed the metric coefficients to be separable into radial and time dependent parts. For the second one, we have considered a self-similar evolution. Using a few simplifying assumptions for both examples, we have found that repulsive stresses arise from the magnetic field's effect. We have obtained constraints on the magnetic field's strength for averting the formation of a singularity due to collapse. This study has led to important conclusions about magnetized inhomogeneous collapse under cylindrical symmetry. Our investigation acquires importance in this regard as there do not exist exact solutions for such systems in the literature. We should note that although we have not referred to any solutions of the field equations in this study, these solutions are necessary in order to reach a definite conclusion.

For future investigations of the problem of singularities within the context of gravitational collapse, one may consider more general and realistic models incorporating pressure gradient, scalar fields, non-gravitational agents or a combination of them. Then it will be clearer whether these agents can prevent the formation of singularities. For this purpose, one can use either the metric-based approach, where the field equations have to be solved explicitly, or the RE-based method we have adopted in this thesis.\\

In chapter \ref{chapter4}, we have explored whether avoiding singularities is possible in Non-Minimally Coupled Scalar-Tensor Theories (NMCSTT). The Penrose-Hawking singularity theorem establishes the inevitable existence of singularities for purely gravitational systems within the framework of GTR. The convergence conditions are essential for proving the singularity theorem. For example, Timelike Convergence Condition (TCC) is necessary to prove that timelike geodesics are incomplete. In GTR, the Strong Energy Condition (SEC) implies the TCC. The SEC is physically well-motivated. Ordinary or Newtonian matter satisfies this condition. But, in NMCSTT, the assumption of the SEC on the matter distribution does not necessarily lead to the TCC as the field equations in these theories contain additional terms. So, the TCC will be violated when some of these terms give rise to repulsive effects and dominate. Singularities may also be avoided in such scenarios. We have investigated whether a violation of the TCC is possible in two well-known candidate theories in the class of NMCSTT - the Brans-Dicke (BD) theory and the Bekenstein Conformally Coupled Scalar-Tensor Theory (BCCSTT). We have considered two specific situations in both of these theories. These correspond to the static spherically symmetric case and the spatially homogeneous and isotropic case. The first one represents a case analogous to Schwarzschild geometry. The second case helps in examining cosmological big bang type singularities. The field equations of NMCSTT are complex. Hence, it is difficult to extract useful inferences even for specific theories and symmetry assumptions. As a result, we have used examples of exact solutions to reach a definite conclusion. For the static spherically symmetric case, we have found that violation of the TCC is possible in both the BD theory and the BCCSTT. This depends on the parameters involved in these theories (e.g. $\omega$ in the BD theory and $\beta$ in the BCCSTT). An important conclusion in the context of cosmology is that violation of the TCC can indeed occur, at least for open universes in both theories. In addition, such a violation can happen for a spatially flat universe in the BD theory.

We should note that though most of the exact solutions we considered in the settings of NMCSTT are indeed singular, they give us an idea that these theories allow violation of the TCC even with conventional matter distribution. So, as a possible future avenue, one can consider finding exact non-singular solutions which are physically and observationally viable in different candidate theories in this class. In principle, many different theories of gravity can emerge from the general action of the NMCSTT (equation \eqref{nmcstta}) for different choices of $f(\phi)$ and $\omega(\phi)$. One can explore in more detail whether other such specific theories lead to escape routes from singularities. One can also attempt to develop the corresponding singularity theorem for these theories. However, it seems very difficult to generalize this kind of treatment for a generic NMCSTT represented by the action \eqref{nmcstta}.\\


In chapter \ref{chapter5}, we have attempted to find a possible explanation of the late time accelerated expansion of the Universe. Instead of assuming the presence of exotic matter, we have considered $f(R)$-gravity for this purpose. In $f(R)$-gravity theories, the geometry itself can give rise to repulsive effects needed to drive the late-time acceleration. We have reconstructed models of $f(R)$-gravity which can account for the accelerated expansion of the Universe. We have studied two particular examples in this context. The starting point for the first one is an ansatz that the Universe's deceleration parameter is constant. This gives rise to an ever-accelerating model. For the second one, we have started with the assumption that the jerk parameter is equal to $1$ ($j=1$). In this case, the evolution of the Universe mimics the well-known $\Lambda$ Cold Dark Matter ($\Lambda$CDM) expansion history, which is arguably the most favoured evolution of the Universe according to observations. We have obtained the relevant forms of $f(R)$ by solving the RE.

For the ever-accelerating model, the expression for $f(R)$ is a combination of three different powers of $R$. This kind of an $f(R)$ can lead to a number of models as these powers are not uniquely determined. However, we cannot recover GTR as a special case from this kind of model. The second model gives rise to two types of evolution, which we have referred to as type I and type II. The reconstructed forms of $f(R)$, for all the models leading to $j=1$, are given by combinations of hypergeometric functions of the Ricci scalar. The example discussed for the type I evolution can be made to work for the early Universe to the recent times by tuning a free parameter of the model. The example considered for the type II evolution can account for the current Universe. However, we cannot extend it to a distant past.

The viability analysis for the second example leads only to the trivial choice $f(R)=R-2\Lambda$, which corresponds to the Einstein-Hilbert action along with a cosmological constant. All other non-trivial options suffer from pathologies like instabilities or negative effective gravitational coupling or not having a sufficient matter-dominated era in past or a combination of them. The first example also fails to be viable at moderately high curvature regimes.

As non-trivial $f(R)$ models consistent with the $\Lambda$CDM mode of evolution fails the viability test, one may search for $f(R)$-theories which are theoretically viable and, at the same time, more or less consistent with observational data. One may also consider other modified theories to investigate this issue. A significant amount of study has been carried out and is still being done in this direction. The method of using the RE for the theoretical reconstruction of cosmological models appears to be quite useful in this context. Reconstructions of cosmological models from the theoretical perspective or observational data are gaining increasing interest. This exercise is crucial when a reasonable description of dark energy phenomenology is still absent.


\section*{Outlook of the research and further avenues}
In conclusion, we have explored several aspects of gravitational systems using the RE as the major tool in this thesis. Specifically, we have focused on the problems of figuring out escape routes from spacetime singularities and understanding the mechanism behind the late-time accelerated expansion of the Universe. These problems are challenging and we may have to shift the paradigm entirely for this purpose. However, we have obtained valuable insights regarding these issues by analyzing different gravitational systems within the purview of GR and modified theories of gravity. While there are a lot of things that can still be done in this vein, we hope that we have been able to contribute to this in our own small way.

We have already mentioned some possible future directions earlier in this chapter. Furthermore, it is essential to address these issues in quantum settings. This will further strengthen our understanding of these two demanding issues. It is worth mentioning in this context that recently, we have reported a work based on the quantization of geodesic congruences in \cite{Choudhury:2021huy}. We have obtained a quantum analogue of the RE in this work and found that the probability of focusing of a timelike geodesic congruence vanishes in the quantum regime. In future, we want to pursue further research to understand the nature of gravitational interaction in the quantum regime. There are several avenues that can be pursued: considering theories of gravity where quantum effects are manifested through specific higher-order correction terms, terms related to dilaton and form fields, non-local terms etc., in the gravitational action; studying gravitational collapse in quantum settings;  exploring dynamical Chern-Simons gravity, which can be treated as an effective quantum field theory; investigating different quantum field theory phenomena in gravitational backgrounds etc.

A crucial job in investigating the problem of dark energy phenomenology is to test the theoretical and observational viability of existing dark energy models. Also, we want to extensively pursue the subject of reconstruction of cosmological models in the future. Another interesting topic in this context is the chameleon and the symmetron mechanisms\cite{Khoury:2003rn, Khoury:2003aq, Hinterbichler:2011ca}. These are potential candidates for explaining dark energy phenomenology while keeping their effect screened at local scales. It will be interesting to see if the RE can provide some new insights into these topics.

\cleardoublepage

\appendix


\appendix 

\chapter{A brief review of some essential concepts of General Theory of Relativity}\label{appenpre}\chaptermark{A brief review of some essential concepts of GTR}
Here we shall very briefly discuss a few essential and fundamental concepts of General Theory of Relation (GTR). For details we refer the reader to \cite{Misner:1974qy, Wald:1984rg, Carroll:1997ar, Carroll:2004st, JayantVNarlikar:2010vvf, d1992introducing, hooft2001introduction, book:Schutz, hartle2003gravity}.

In Special Theory of Relativity (STR), the speed of light has a special status. No signal can travel faster than the speed of light, which ensures causality in spacetime. But in Newtonian gravity, the gravitational interaction is instantaneous. Therefore, Newtonian gravity needs to be modified to fit with the concepts of STR. On the other hand, we should also note that STR needs modifications when gravity is present. The reason is the following. STR deals with inertial observers, but it is not possible to have global inertial observers when gravity is present.

If we compare electromagnetism and gravitation, we see that it is possible to have a closed space completely sealed from outside electrical effects. However, an analogous scenario is impossible in gravitation. Gravity is ever-present. This remarkable feature of gravity plays a vital role in the discovery of Einstein's GTR.

\section{The principle of equivalence}\label{peq}
We have discussed that the effect of gravity cannot be screened. Now, one may argue that an observer inside a freely falling lift in the Earth's gravitational field feels weightless. Therefore, the effect of gravity is absent inside the lift. Motivated by this idea, Einstein earlier thought that gravity might be interpreted as an effect of being in an accelerated frame which creates a pseudo-gravitational field. However, later he recognized that this is true only if we have a uniform gravitational field. Even inside this freely falling lift, one can detect the effect of gravity if it falls from a sufficiently large height and for a sufficiently long time. Suppose an observer inside the lift releases two objects at rest with respect to the instantaneous rest frame of the lift, one initially near the ceiling of the lift and the other near the floor. If gravity had really been a pseudo force, the separation between the objects would not have changed with time. However, the acceleration of the object nearer to the Earth will be slightly greater, which will cause a change in the separation of these objects. This is an example of geodesic deviation (refer to section \ref{geoddevdis} for a brief discussion on geodesic deviation), which occurs in the presence of gravity. So, an extended object in a gravitational field experiences different forces on its different parts. This gives rise to the tidal effects.

The discussion above leads to the conclusion that it is possible to transform away the effects of gravity only locally and for a short time by going to a suitable accelerated frame. This is the statement of the \emph{weak equivalence principle}. Such frames of reference where gravity is absent locally are called locally inertial frames. The more robust version of the equivalence principle, known as the \emph{strong equivalence principle} or the \emph{Einstein's equivalence principle} states that experiments in locally inertial frames lead to results which are indistinguishable from those obtained from the same experiments in the absence of gravity. The equivalence principle played an important part in Einstein's ideas about a theory of gravity.

Therefore, the effect of gravity always prevails. This led Einstein to argue that gravity must be associated with some intrinsic characteristic of spacetime. He proposed that the effect of gravity is attributed to the geometry of spacetime. If a spacetime is flat, the situation is the same as in STR. Therefore, we must have a non-Euclidean (or curved) geometry of spacetime when gravity is present. This led Einstein to his general theory.

\section{Spacetime metric}
The ingenious idea by Einstein that theory of gravity (now known as GTR) should be described in terms of spacetime geometry is now well established. A (pseudo)-Riemannian geometry is defined using a line element, aka a metric which gives the elementary notion of distance. We will begin with a familiar result in STR. The invariant interval between two events occurring at two nearby spacetime points $(x, y, z, t)$ and $(x+\mathrm{d}x, y+\mathrm{d}y, z+\mathrm{d}z, t+\mathrm{d}t)$ (in Cartesian coordinates) is given by,
\begin{equation}
 \mathrm{d}s^2=-\mathrm{d}t^2+\mathrm{d}x^2+\mathrm{d}y^2+\mathrm{d}z^2.
\end{equation}
When we wish to implement this idea of Einstein and make a transition from STR to GTR, we need a more general version of this distance measure. This is written as,
\begin{equation}\label{spmetric}
 ds^2=g_{ab} \mathrm{d}x^a \mathrm{d}x^b,
\end{equation}
where $a=1,2,3$ gives three spatial coordinates and $a=0$ gives the temporal coordinate. Here we have used the Einstein summation convention. The coefficients $g_{ab}$ are now functions of $x^a$, i.e. they depend on the spacetime points. The above expression \eqref{spmetric} for the invariant distance is known as the \emph{spacetime metric}. This describes the geometry of a spacetime. The determinant of the matrix $||g_{ab}||$ is denoted by $g$, which is always negative.

Even if two sets of metric coefficients $g_{ab}$ and $\tilde{g}_{ab}$ depend differently on the coordinates $x^a$, they may not represent different geometries. This may happen due to the choice of coordinate system. For example, if we express three-dimensional Euclidean geometry in spherical polar coordinates, we have coordinate-dependent $g_{ab}$,
\begin{equation}
 \mathrm{d}s^2=\mathrm{d}r^2+r^2\mathrm{d}\theta^2+r^2\sin^2\theta \mathrm{d}\phi^2,
\end{equation}
but the geometry is, of course, still flat. Therefore, the property of coordinate dependence of metric coefficients may not convey the real physical picture. We need a mathematical structure which can differentiate between physical and coordinate effects. This structure should distinguish the quantities which do not change under coordinate transformations. Such structure can be built using the well-known concepts of scalars, vectors and tensors. In such a structure, $g_{ab}$ is known as the metric tensor, which is a symmetric covariant tensor of rank $2$.

We can define vectors and tensors not only in inertial frames but in any coordinate frame. This is quite useful because we should, in principle, be able to describe physics in any general reference frame. This is one of the goals of GTR.

\section{Covariant differentiation}\label{covdiff}
GTR is based on the principle of general covariance, which says {\it laws of physics are generally covariant} i.e. independent of coordinate choices. Physical laws are actually written in terms of differential equations involving vectors and tensors. These laws will be generally covariant by construction if we can express them in the language of tensors.

\subsection{Parallel transport}
We need to consider a derivative of tensors which also transforms as a tensor under coordinate transformations. Vectors, tensors etc., are elements of locally defined vector spaces. We can compare them if they are defined at the same point. Partial derivatives of scalars do transform as vectors. However, this is not the case when we consider the partial derivative of a vector field. Partial derivative of a {\it covariant} vector $B_a$ is defined as,
\begin{equation}
 \frac{\partial B_a}{\partial x^b}=\lim_{\delta x^b \rightarrow 0} \left(\frac{B_a(x^d+\delta x^d)-B_a(x^d)}{\delta x^b}\right),
\end{equation}
where the two vectors in the numerator are compared at two different points. The difference between vectors will be a vector if we define them at the same point. This is achieved by defining the concept of parallel transport. We have to move the vector from one point to another, keeping it parallel to itself all the time. The difference between $B_a(x^d+\delta x^d)$ and the vector resulting from parallel propagation will give us the real physical change. Change in $B_a$ due to the parallel transport is expected to be proportional to the components of $B$ and the displacement $\delta x^d$. Therefore, as a measure of this change, we can write,
\begin{equation}\label{delb}
 \delta B_a= \Gamma^c_{ad}B_c \delta x^d.
\end{equation}
The coefficients $\Gamma^c_{ad}$ are functions of spacetime points and are known as the {\it Christoffel symbols}. At this point, we should note that we have introduced an additional concept along with the metric. The metric gives us distance between neighbouring points, and the Christoffel symbols enable us to consider parallel vectors at these points. The newly introduced property of connecting vectors at neighbouring points is known as the {\it affine connection} of spacetime.

\subsection{The covariant derivative}
Using equation \eqref{delb} we can write the difference between $B_a(x^d+\delta x^d)$ and the vector due to parallel transport as,
\begin{equation}
 B_a(x^d+\delta x^d)-\left(B_a(x^d)+\delta B_a\right)=\left(\frac{\partial B_a}{\partial x^d}- \Gamma^c_{ad}B_c\right)\delta x^d.
\end{equation}
Now, we can define the physically meaningful derivative of a vector as,
\begin{equation}
 \nabla_d B_a\equiv \frac{\partial B_a}{\partial x^d}- \Gamma^c_{ad}B_c,
\end{equation}
which transforms as a tensor of rank $(0,2)$. This is known as the {\it covariant derivative}.

A scalar does not change under parallel transport. So, the covariant derivative of a scalar is the same as the partial derivative. The covariant derivative of a {\it contravariant} vector is given by,
\begin{equation}\label{covdcontra}
 \nabla_d A^b\equiv \frac{\partial A^b}{\partial x^d}+ \Gamma^b_{cd}A^c.
\end{equation}
This ensures that the scalar $A^c B_c$ does not change under parallel transport, and therefore covariant derivative of this quantity is the same as its partial derivative.
We can extend this definition to include tensors of arbitrary rank. For example,
\begin{equation}
 \nabla_d {T^a}_{bc}=\frac{\partial {T^a}_{bc}}{\partial x^d}+\Gamma^a_{kd} {T^k}_{bc}- \Gamma^k_{bd} {T^a}_{kc}-\Gamma^k_{cd} {T^a}_{bk}.
\end{equation}

\subsection{Riemannian geometry}
Einstein chose the non-Euclidean Riemannian geometry to describe spacetime in GTR. In Riemannian geometry, there are two additional assumptions about the connection -
\begin{equation}
 \Gamma^c_{ad}=\Gamma^c_{da}; \hspace{0.2cm} \nabla_d g_{ab}=0.
\end{equation}
This leads to the following expression for the Christoffel symbols,
\begin{equation}
 \Gamma^a_{bc}=\frac{1}{2}g^{ad} \left(\frac{\partial g_{db}}{\partial x^c}+\frac{\partial g_{cd}}{\partial x^b}-\frac{\partial g_{bc}}{\partial x^d}\right).
\end{equation}
which means that once the metric is known, the connection is also fixed.

\section{Spacetime curvature}\label{rieric}
Let us consider parallel transport on a curved space. The outcome of this transport between two different points depends on the path followed. This is not the case for a flat surface. On the other hand, on a flat surface, parallel transport does not depend on the path. Path dependence of parallel transport is a property which distinguishes a curved space from a flat one.

\subsection{Riemann tensor}
Let us now try to find a condition for which parallel transport becomes independent of path. The change in a vector $B_a$ due to parallel transport is given by equation \eqref{delb}. So, this transport will be path independent if a non-trivial solution of the following set of differential equations exist,
\begin{equation}
 \frac{\partial B_a}{\partial x^b}=\Gamma^c_{ab} B_c.
\end{equation}
The necessary condition for the existence of a solution can be found from the identity,
\begin{equation}
  \frac{\partial^2 B_a}{\partial x^b \partial x^c}=\frac{\partial^2 B_a}{\partial x^c \partial x^b}.
\end{equation}
This condition is given by,
\begin{equation}
 {{R_a}^k}_{cd} B_k=0,
\end{equation}
where
\begin{equation}
 {{R_a}^k}_{cd}=\frac{\partial \Gamma^k_{ac}}{\partial x^d}-\frac{\partial \Gamma^k_{ad}}{\partial x^c}+\Gamma^l_{ac}\Gamma^k_{ld}-\Gamma^l_{ad}\Gamma^k_{lc}.
\end{equation}
Therefore the necessary condition translates into,
\begin{equation}\label{rie}
 {{R_a}^k}_{cd}=0.
\end{equation}
The quantity ${{R_a}^k}_{cd}$ is a tensor. This follows from a simple calculation which leads to,
\begin{equation}\label{defriemann}
 \nabla_c \nabla_d B_a-\nabla_d \nabla_c B_a\equiv {{R_a}^k}_{cd} B_k.
\end{equation}
This tensor is commonly known as the {\it Riemann tensor} (aka {\it Riemann-Christoffel tensor} or {\it curvature tensor}). The condition \eqref{rie} is also the sufficient condition for parallel transport to be path-independent.

A spacetime is {\it flat} if the Riemann tensor vanishes everywhere and {\it curved} otherwise.

 \subsection{Symmetry properties of Riemann tensor}
 We will now discuss the symmetry properties of $R_{abcd}$, which can be obtained by lowering the second index of ${{R_a}^k}_{cd}$. We have,
 \begin{equation}
  R_{abcd}=-R_{bacd}=-R_{abdc}=R_{cdab}.
 \end{equation}
Riemann tensor also satisfies the following relation,
\begin{equation}
 R_{abcd}+R_{adbc}+R_{acdb}=0.
\end{equation}
It follows that the Riemann tensor has at most $20$ independent components.

\subsection{Ricci and Einstein tensors}\label{ret}
Using the well-known process of contraction in tensor algebra, we can construct a low rank tensor from the Riemann tensor as,
\begin{equation}
 R_{ab}\equiv g^{cd}R_{acbd}.
\end{equation}
This is known as the {\it Ricci tensor}, which is a symmetric tensor. $R_{abcd}$ cannot lead to any other independent tensor of rank $2$ by contraction because of the symmetries of the Riemann tensor.

Using further contraction, we have,
\begin{equation}
 R=g^{ab} R_{ab},
\end{equation}
which is called the {\it Ricci scalar} or the {\it scalar curvature}. We can now construct a tensor as,
\begin{equation}
 G_{ab}=R_{ab}-\frac{1}{2}g_{ab}R.
\end{equation}
This is known as the {\it Einstein tensor}.

\subsection{Bianchi identities}
It follows from the definition and properties of the Riemann tensor that
\begin{equation}
 \nabla_a R_{bcdk}+\nabla_k R_{bcad}+\nabla_d R_{bcka}\equiv 0.
\end{equation}
These identities are called the {\it Bianchi identities}. If we multiply this identity with $g^{bk} g^{ca}$ and use the expressions for the Ricci tensor and the Ricci scalar, we get,
\begin{equation}
 \nabla_a \left(R^{ab}-\frac{1}{2}g^{ab}R\right)=\nabla_a G^{ab}\equiv 0.
\end{equation}
Therefore, {\it the Einstein tensor has identically zero divergence}.

\subsection{Geodesics}\label{geodesics}
Geodesics correspond to force-free motion. These happen to be straight lines when the geometry is flat. In Euclidean geometry, the concept of straight lines is well-known. Now we will discuss how to develop an equivalent notion in Riemannian geometry.

The concept of a straight line incorporates two essential properties - straightness and shortest distance. Straightness implies that the direction of the line will not change as one moves along the line. Let us consider a curve in spacetime in parametric representation $x^a(\lambda)$. The tangent vector to this curve is given by,
\begin{equation}
 u^a=\frac{\mathrm{d} x^a}{\mathrm{d}\lambda}.
\end{equation}
According to the concept of straightness, $u^a$ should not change as it follows the curve. When $\lambda$ changes to $\lambda+\delta \lambda$, we have,
\begin{equation}
 \Delta u^a=\frac{\mathrm{d}u^a}{\mathrm{d}\lambda}\delta \lambda+\Gamma^a_{bc} u^b \delta x^c,
 \end{equation}
as the change in $u^a$. The second term in the right-hand side comes into the picture as a result of parallel transport. Now, $\delta x^c=u^c \delta \lambda$. So, the condition for no change in the direction of $u^a$ is,
\begin{equation}\label{geod1}
 \Delta u^a=0 \implies \frac{\mathrm{d}u^a}{\mathrm{d}\lambda}+\Gamma^a_{bc} u^b u^c=0.
\end{equation}
This is the condition for the curve to remain straight.

Now, we will discuss the second property, namely shortest distance. The distance between two points with parameters $\lambda_1$ and $\lambda_2$ is defined as,
\begin{equation}\label{eqdis}
 s=\int_{\lambda_1}^{\lambda_2} \left(g_{ab}\frac{\mathrm{d}x^a}{\mathrm{d}\lambda}\frac{\mathrm{d}x^a}{\mathrm{d}\lambda}\right)^\frac{1}{2}\mathrm{d}\lambda\equiv\int_{\lambda_1}^{\lambda_2} L \mathrm{d}\lambda.
\end{equation}
According to the calculus of variations, this distance will be stationary if we have,
\begin{equation}
 \frac{\mathrm{d}}{\mathrm{d}\lambda}\left(\frac{\partial L}{\partial \dot{x}^a}\right)-\frac{\partial L}{\partial x^a}=0,
\end{equation}
where $\dot{x}^a\equiv \frac{\mathrm{d}x^a}{\mathrm{d}\lambda}$.
The above equation leads to,
\begin{equation}\label{geodeq2}
 \frac{\mathrm{d}^2 x^a}{\mathrm{d}s^2}+\Gamma^a_{bc} \frac{\mathrm{d}x^a}{\mathrm{d}s}\frac{\mathrm{d}x^b}{\mathrm{d}s}=0,
\end{equation}
where
\begin{equation}
 \mathrm{d}s=L \mathrm{d}\lambda.
\end{equation}
There are a few important points to note. $L$ is real for spacelike curves only. For having a real parameter along the curve in the timelike case, we must use,
\begin{equation}
 \mathrm{d}\sigma=i \mathrm{d}s.
\end{equation}
For the null case, $L=0$. In this case we have to choose a new parameter $\tilde{\lambda}=\tilde{\lambda}(\lambda)$ and write the corresponding integral as in \eqref{eqdis} so that we get the similar equation as \eqref{geodeq2} with $s$ replaced by $\tilde{\lambda}$.

Now, equations \eqref{geod1} and \eqref{geodeq2} are the same. Here, $s$ can be interpreted as length along the curve, whereas $\lambda$ is more general. However, if equation \eqref{geod1} is satisfied, $\lambda=k s$, where $k$ is a constant. This follows from the first integral of \eqref{geod1} -
\begin{equation}
 g_{ab}\frac{\mathrm{d}x^a}{\mathrm{d}\lambda}\frac{\mathrm{d}x^b}{\mathrm{d}\lambda}=C,
\end{equation}
where $C$ is a constant.
Curves which satisfy the above condition of stationary distance are known as geodesics. $C<0$, $C=0$ and $C>0$ for timelike, null and spacelike curves, respectively. If we consider $u^a=\frac{\mathrm{d}x^a}{\mathrm{d}\lambda}$ to be normalized then we have $u^a u_a=-1$,  $u^a u_a=0$ and $u^a u_a=1$ for the timelike, null and spacelike cases respectively. $\lambda$ is known as an {\it affine parameter}, which can always be chosen as the proper time $\tau$ for the timelike case.  We should note that the geodesic equation \eqref{geod1}  can be expressed as,
\begin{equation}\label{geodeqv}
 u^a\nabla_a u^b=0.
\end{equation}

\subsection{Geodesic deviation}\label{geoddevdis}
We shall discuss another essential feature which differentiates a curved geometry from a flat one. If we consider two neighbouring geodesics in Euclidean geometry, the separation between these geodesics increases at a uniform rate. However, this does not happen in the case of curved geometries. Therefore, how deviation between neighbouring geodesics changes must give information about the curvature of spacetime.

Let us consider the figure \ref{fig0} (chapter \ref{chapter1}, section \ref{tmlkc})  of two neighbouring geodesics in spacetime. We consider $\lambda$ as the affine parameter (this is denoted by $\tau$ in the figure as we were concerned with timelike geodesics). A specific point on any of these two geodesics can be represented by the coordinates $x^a(\lambda, s)$, where $s$ is the parameter specifying the particular geodesic on which this point lies. The velocity vector $u^a$ is given by, $u^a=\frac{\partial x^a}{\partial \lambda}$. The vector $\xi^a=\frac{\partial x^a}{\partial s}$ gives the rate of deviation between the neighbouring geodesics. This is called the deviation vector. Now, employing equation \eqref{covdcontra}, we can easily show that,
\begin{equation}\label{equxi}
  u^b\nabla_b\xi^a=\xi^b\nabla_b u^a,
\end{equation}
where we have to use the facts that partial derivatives commute and the connection is symmetric.

We have to find the evolution of the deviation vector, which gives the idea of the underlying geometry. Now,
\begin{equation}
 u^c \nabla_c \left(u^b\nabla_b\xi^a\right)=  u^c \nabla_c\left(\xi^b\nabla_b u^a\right)=u^c\xi^b\nabla_c \nabla_b u^a+ u^c \left(\nabla_c \xi^b\right)\left(\nabla_b u^a\right).
\end{equation}
Using equation \eqref{defriemann}, we can write,
\begin{equation}
 u^c \nabla_c \left(u^b\nabla_b\xi^a\right)+{R^a}_{bcd} u^b \xi^c u^d=u^c\xi^b\nabla_b \nabla_c u^a+u^c \left(\nabla_c \xi^b\right)\left(\nabla_b u^a\right),
\end{equation}
which implies,
\begin{equation}\label{geoddev}
 \frac{\mathrm{d}^2 \xi^a}{\mathrm{d}\lambda^2}+{R^a}_{bcd}u^b u^d \xi^c=0.
\end{equation}
Here we have used,
\begin{equation}
 u^c\xi^b\nabla_b \nabla_c u^a=\xi^b \nabla_b\left(u^c \nabla_c u^a\right)-\xi^b \left(\nabla_b u^c\right) \left(\nabla_c u^a\right)
\end{equation}
and equations \eqref{geodeqv} and \eqref{equxi}.
The equation \eqref{geoddev} is called the equation of \emph{geodesic deviation}. The presence of the Riemann curvature tensor indicates the effect of curvature in the evolution of the geodesic deviation. For the flat case, we have ${R^a}_{bcd}=0$. So, the rate of change of the deviation vector is uniform, as expected.
One can now understand the intrinsic relationship between gravity and geometry more clearly by referring to the example of two objects released in a freely falling lift discussed in section \ref{peq}.

\section{Einstein's equations}
In GTR, gravity is described in terms of spacetime geometry. As the source of gravity is known to be the presence of matter, Einstein conjectured that the latter must be responsible for the curvature of spacetime. The matter distribution is characterized by the energy-momentum tensor $T_{ab}$. To obtain a complete description of how matter creates spacetime curvature, one must determine an equation that connects $g_{ab}$ to $T_{ab}$. Einstein obtained these fundamental equations, famously known as the Einstein field equations, which relate the geometry part with the matter part.
We will now briefly discuss the reasoning that leads to the Einstein equations.

\subsection{Relation between curvature and matter}
We need a second-rank symmetric tensor which is associated with spacetime curvature. This should have similar properties as $T_{ab}$ and involves second derivatives of $g_{ab}$. So, this tensor must be related to the Riemann curvature tensor. The natural first choice is $R_{ab}$ and we can attempt to write dynamical equations like,
\begin{equation}\label{eneqt}
 R_{ab}=\text{constant} \times T_{ab}.
\end{equation}
Now, conservation of energy-momentum implies,
\begin{equation}\label{emc}
 \nabla_a T^{ab}=0.
\end{equation}
Since $\nabla_a R^{ab}$ does not vanish in general, equation \eqref{eneqt} is not consistent. To remedy this,  we also need a divergence-less quantity in the left-hand side. This leads to the choice introduced in section \ref{ret}  - the Einstein tensor $G_{ab}$. Thus, the field equations of GTR (aka the Einstein equations) are given by,
\begin{equation}
 G_{ab}\equiv R_{ab}-\frac{1}{2}g_{ab} R=\kappa T_{ab}.
\end{equation}
If we impose the condition that in the limit of a weak gravitational field, the theory should reduce to that of Newton's, $\kappa$ takes the value ${8\pi G}$ (in natural units $c=1=\hbar$).

There are $10$ Einstein equations corresponding to $10$ unknown metric coefficients $g_{ab}$. However, the condition \ref{emc} reduces this to $6$ independent equations. Therefore, the problem is underdetermined. The underlying reason is the general covariance of GTR -- let us consider that a solution for the Einstein equations is $g_{ab}$. Then any other $\tilde{g}_{ab}$ resulting from a coordinate transformation $g_{ab}$ is also a solution. Therefore, several solutions which look different may represent the same spacetime.

\subsection{The action for general theory of relativity: the Einstein-Hilbert action}\label{ehac}
In this section, we will discuss an action which leads to the Einstein field equations. Hilbert found this action soon after Einstein reported the field equations of GTR. To formulate an action principle, we need a suitable scalar related to geometry. Such scalars containing up to first derivatives of $g_{ab}$ do not exist. However, if we include second-order derivatives, the simplest choice comes out to be the Ricci scalar $R$. This is taken as the scalar quantity, which sits in the action of gravity. This action is called the Einstein-Hilbert action, which is given by,
\begin{equation}\label{ehc}
 S=\frac{1}{2\kappa}\int R \sqrt{-g} d^4x+ S_m.
\end{equation}
Here,
\begin{equation}
 S_m=\int L_m \sqrt{-g} d^4x,
\end{equation}
 is the corresponding action for the matter distribution, which acts as the source of gravity;  $\sqrt{-g} d^4x$ is the invariant volume element which remains unchanged under coordinate transformations. If we vary this action $S$ with respect to $g^{ab}$ $\left(g^{ab}\rightarrow g^{ab}+\delta g^{ab}\right)$ with the restriction that variation of the metric and variation of its derivatives vanish on the boundary, the action principle leads to the Einstein field equations,
\begin{equation}
 R_{ab}-\frac{1}{2}g_{ab} R=\kappa T_{ab}.
\end{equation}
Here the energy-momentum tensor $T_{ab}$ is defined as,
\begin{equation}
 T_{ab}\equiv-\frac{2}{\sqrt{-g}}\frac{\delta S_m}{\delta g^{ab}}.
\end{equation}

\section{Extended theories of gravity via modifying gravitational action}\label{ehtmod}
If we replace the Ricci scalar in the Einstein-Hilbert action \eqref{ehc} by an analytic function of the Ricci scalar $f(R)$ such that,
\begin{equation}
 S=\frac{1}{2\kappa}\int f(R) \sqrt{-g} d^4x+ S_m,
\end{equation}
we will have a different theory of gravity. The resulting theory is called the $f(R)$-theory of gravity. Many such extended theories of gravity can result from different modifications of gravitational action. For example, we can introduce a scalar field non-minimal coupled with the curvature --
\begin{equation}
 S=\frac{1}{2}\int \sqrt{-g} \mathrm{d}^4 x \left( \phi R-\frac{\omega}{\phi}\partial_a \phi \partial^a \phi\right)+S_m.
\end{equation}
This is the action for the well-known \emph{Brans-Dicke theory}. We have discussed such extended theories of gravity in more detail in section \ref{modg}.

\cleardoublepage
\chapter{Congruence of null geodesics and the corresponding Raychaudhuri equation}\label{appen1}\chaptermark{Null geodesic congruences and the corresponding RE}
We have discussed timelike geodesic congruences in section \ref{tmlkc}. Null geodesics are another type of causal geodesics. Though we have not considered this kind of geodesics in the thesis, we shall briefly review the topic of null geodesic congruences for completeness. Here the tangent vector field is denoted by $k^a$, which is a null vector. We choose the geodesics to be affinely parametrized by the parameter
$\lambda$. So, $dx^a = k^a d\lambda$. The deviation vector is denoted by  $\eta^a$. Therefore, we have (similar to the timelike case discussed in section \ref{tmlkc}),
\[k^ak_a=0, \hspace{0.2cm} k^b\nabla_b k^a=0, \hspace{0.2cm} \eta^b\nabla_b k^a=k^b\nabla_b{\eta^a}, \hspace{0.2cm} k^a\eta_a=0.\]

\section{Derivation of the null Raychaudhuri equation}
The study of transverse properties of null congruences is a bit involved in this case. As $k^a$ is null, $h_{ab}=g_{ab}+k_{a}k_{b}$ does not work as the transverse metric. One can note that $h_{ab}k^b=k^a\neq 0$. To construct the correct transverse metric, it is necessary to introduce another null vector $N_{a}$ such that $k^a N_a\neq 0$.
Using the arbitrariness of normalization of a null vector, we may always impose the condition $k^a N_{a}=-1$.
Now we can introduce a purely transverse and effectively two-dimensional metric,
\begin{equation}\label{2.28}
 h_{ab}=g_{ab}+k_aN_b+N_ak_b,
\end{equation}
and we have,
\begin{equation}\label{2.29}
 h_{ab}k^b=h_{ab}N^b=0, \hspace{0.2cm} {h^a}_a=2, \hspace{0.2cm} {h^a}_c {h^c}_b={h^a}_b.
\end{equation}

In this case also, the  spatial tensor field
\begin{equation}
 B_{ab}=\nabla_b k_a,
\end{equation}
is introduced. This measures the failure of $\eta^a$ to be parallel transported if we move along the congruence,
\begin{equation}
k^b\nabla_b \eta^a={B^a}_b \eta^b.
\end{equation}
Here $B_{ab}$ is orthogonal to the tangent vector field but not to $N^a$. Thus $\eta^a$ contains a non-transverse component which should be removed. The purely transverse part of the
deviation vector is given by,
\begin{equation}
 \tilde{\eta}^a\equiv {h^a}_c \eta^c = \eta^a+(N_c \eta^c)k^a.
\end{equation}
Now,
\begin{equation}
 k^b\nabla_b\tilde{\eta}^c={h^c}_d {B^d}_b \eta^b+k^b\nabla_b{h^c}_d\eta^d ={h^c}_d {B^d}_b \eta^b+(k^b\nabla_b N_d\eta^d)k^c.
\end{equation}
So, $k^b\nabla_b\tilde{\eta}^c$ has a component along $k^c$. We have to remove it again by projecting with ${h^a}_c$. Using equation \eqref{2.29} we can write,
\begin{equation}
{\left(k^b\nabla_b\tilde{\eta}^a \right)}^\sim\equiv {h^a}_c (k^b\nabla_b\tilde{\eta}^c )={h^a}_c {B^c}_d \eta^d={h^a}_c {B^c}_d \tilde{\eta}^d={h^a}_c {h^d}_{b} {B^c}_d \tilde{\eta}^b.
\end{equation}
Hence
\begin{equation}\label{2.33}
 {\left(k^b\nabla_b\tilde{\eta}^a\right)}^\sim={\tilde{B}^a}_{\hspace{0.1cm}b}\tilde{\eta}^b,
\end{equation}
where $\tilde{B}_{ab}$ denotes the purely transverse part of $B_{ab}$,
\begin{equation}
 \tilde{B}_{ab}={h^c}_a {h^d}_b B_{cd}.
 \end{equation}
 The purely transverse behaviour of the congruence is governed by equation \eqref{2.33}.

Now using equation \eqref{2.28}, the explicit form of $\tilde{B}_{ab}$ can be found out as,
\begin{equation}\label{2.35}
 \tilde{B}_{ab}=B_{ab}+k_a N^c B_{cb}+k_b B_{ac} N^c+k_a k_b B_{cd}N^cN^d.
\end{equation}

Following the same line as for the timelike case (section \ref{tmlkc}), $\tilde{B}_{ab}$ is decomposed  as,
\begin{equation}
 \tilde{B}_{ab}=\frac{1}{2} \theta h_{ab}+\sigma_{ab}+\omega_{ab},
\end{equation}
with $\theta={\tilde{B}^a}_{\hspace{0.1cm}a}$ being the \textit{expansion scalar}, $\sigma_{ab}=\tilde{B}_{(ab)}-\frac{1}{2}\theta h_{ab}$ being the
\textit{shear tensor} and $\omega_{ab}=\tilde{B}_{[ab]}$ being the  \textit{rotation tensor}. It follows from equation \eqref{2.35} that,
\[\theta=g^{ab}\tilde{B}_{ab}=g^{ab}B_{ab}.\]
Therefore we have,
\begin{equation}
 \theta=\nabla_a {k^a}.
\end{equation}
We see though the auxiliary null vector, $N^a$ and hence the transverse metric is not unique,
the expansion is indeed unique.

The next job is to write down the evolution equations for Expansion, Shear and Rotation (ESR). Following the same derivation as in
the timelike case (see equation \eqref{eq 9.2.10}) , we have,
\begin{equation}\label{eq 9.2.30}
 k^c\nabla_cB_{ab}+{B^c}_bB_{ac}={R_{cba}}^d k_d k^c,
\end{equation}
which implies
\begin{equation}\label{nullmain}
 k^c\nabla_c\tilde{B}_{ab}+{\tilde{B}^c}_{\hspace{0.1cm}b}\tilde{B}_{ac}=({R_{cba}}^d k_d k^c)^\sim.
\end{equation}
Equation \eqref{nullmain} yields the corresponding evolution equations for ESR,
\begin{equation}\label{rc eq n}
 \frac{d\theta}{d\lambda}=-\frac{1}{2}\theta^2-{\sigma}_{ab}{\sigma}^{ab}+{\omega}_{ab}{\omega}^{ab}-R_{cd}k^ck^d,
\end{equation}
\begin{equation}
 k^c\nabla_c\sigma_{ab}=-\theta {\sigma}_{ab}+(C_{cbad}k^ck^d)^\sim,
\end{equation}
\begin{equation}
 k^c\nabla_c{\omega}_{ab}=-\theta {\omega}_{ab},
 \end{equation}
 when we take the trace, symmetric trace-free and antisymmetric parts, respectively. The first (equation \eqref{rc eq n}) of these three equations is the Raychaudhuri Equation (RE) for null geodesic congruences. The term \emph{REs} is also used to refer to the set of these three equations.

 \section{Focusing theorem for null geodesic congruences}
Using the Einstein equations, we have,
\begin{equation}
 R_{ab}k^ak^b=8\pi T_{ab} k^a k^b,
\end{equation}
for null geodesic congruences.
Similar to the timelike case (section \ref{tmlkc}), the {\it Null Convergence Condition (NCC)}
\begin{equation}\label{ncc}
 R_{ab}k^ak^b\geq 0,
\end{equation}
 holds if,
\begin{equation}\label{nec}
 T_{ab} k^a k^b \geq 0.
\end{equation}
for all $k^a$.
The above condition \eqref{nec} is known as the \textit{Null Energy Condition (NEC)}. The NEC follows from the assumption of the Strong Energy Condition (SEC) by continuity.
Similarly, the NEC also follows from the Weak Energy Condition (WEC) by continuity.

Therefore, following the same sequence of arguments as in the timelike case, for hypersurface orthogonal null geodesic congruences, we have,
\begin{equation}\label{nfc}
\frac{d\theta}{d\lambda}\leq -\frac{1}{2}\theta^2.
\end{equation}
 This yields,
\[\theta^{-1}(\lambda)\geq {\theta_0}^{-1}+\frac{\lambda}{2},\]
where $\theta_0\equiv\theta(0)$.
Therefore, if this congruence is initially converging, it will focus
 within an affine parameter value $\lambda \leq 2 /|{\theta_0}|$. This is the null version of the Focusing Theorem (FT).

\cleardoublepage
\chapter{Chapter 2: A Few Exact Solutions}\label{appen}
In chapter \ref{chapter2}, we have referred to a few exact solutions corresponding to different special cases. We have used them to discuss the consistency of the findings from the focusing condition with the exact solutions. The derivation and features of these solutions are discussed in detail in the following.
\section{Massless scalar field}\label{massless}
 At first, we consider the simplest possible scenario where only a massless scalar field is present as the matter distribution. Then from equation \eqref{eqfu}, we have,
\begin{equation}\label{mlu}
 \frac{1}{B}\frac{dB}{dz}=\frac{1}{z}.
\end{equation}
The solution for B is,
\begin{equation}\label{solB}
 B=m z,
\end{equation}
where $m$ is a constant of integration. This implies,
\begin{equation}\label{solA}
 A=rB=m t.
\end{equation}
But it should be noted that if we use equation \eqref{mlu}, equations \eqref{fe1s}, \eqref{fe2s} and \eqref{fe3s} all yield,
\begin{equation}
\frac{3}{z^2(z^2-1)}=0.
\end{equation}
 This is not possible for finite $z$ values. Thus, a consistent solution of the field equations does not exist in this case.

\section{Scalar field with a non-zero potential}\label{massive}
Even if a potential is included in the energy-momentum tensor of the scalar field,
solutions for $B$ and $A$ remain the same as in the previous case (equations \eqref{solB} and \eqref{solA} respectively).
Here equations \eqref{fe1s}, \eqref{fe2s} and \eqref{fe3s} all lead to,
\begin{equation}
 V(\phi)=3m^2,
\end{equation}
after putting the solution for $B$ (equation \eqref{solB})
Therefore, the potential must be constant.

Again, from equation \eqref{fe4s} we have,
\begin{equation}
 \phi=\text{ constant}.
\end{equation}
This is also consistent with the equation \eqref{wave3}. Thus, this case is equivalent to the case where only a cosmological constant is present.

In this case, from equation \eqref{solA}, we can conclude that a singularity ($A\rightarrow \infty$) occurs only in the limit $t\rightarrow\infty$. This corresponds to an ever
collapsing solution.
At $t=0$, $A=0$ which means the scale factor (i.e. inverse of $A$) tends to infinity. This signifies a dispersal.

\section{Scalar field along with a perfect fluid}\label{perfect}
In this example, we consider a matter distribution consisting of a perfect isotropic fluid ($p_r=p_t$ and $q=0$) along with the scalar field. As one can see from the equation \eqref{eqfu}, the solution for $B$ will not change even in this case.
Here from equation \eqref{rhopr} we have,
\begin{equation}
 \rho=-p.
\end{equation}
The consistency with the field equations demands $\rho$, $p$, $\phi$ and $V(\phi)$ have to remain constant. Thus the matter again turns out to be
the cosmological constant. Clearly, the conclusions, in this case, will not change as compared to the previous case.

 It should be noted that in the examples discussed so far, the formation of singularities within finite time has not been observed. The possible reason is that the matter behaves like a cosmological constant in these cases.

\section{Fluid with isotropic pressure and radial heat flux}\label{fourth}
This example corresponds to the case where only a fluid with isotropic pressure and non-zero heat flux is present. This example has already been considered by Chan, Silva and Rocha \cite{Chan:2002bn}. The authors found that a singularity appears as $t\rightarrow 0$. In this case, equations \eqref{eqfu} and \eqref{fe4s} yield,
\begin{equation}
 \frac{z}{B}\frac{d^2B}{dz^2}+\frac{1}{B}\frac{dB}{dz}-\frac{1}{z}=0.
\end{equation}
The solution for $B$ is thus given by,
\begin{equation}\label{s1B}
 B=\frac{C z^2+2D}{2z},
\end{equation}
where $C$ and $D$ are constants of integration.
 So, the expression for $A$ is,
\begin{equation}
 A=\frac{Ct^2+2Dr^2}{2t}.
\end{equation}
This confirms the presence of a zero proper volume singularity at $t=0$. The scale factor also vanishes in the limits $t\rightarrow \infty$ and/or $r\rightarrow \infty$. A dispersal may occur
at $z^2=\dfrac{t^2}{r^2}=-\dfrac{2D}{C}$, only if $C$ and $D$ are of opposite signs.

\section{Scalar field along with an anisotropic fluid}\label{fifth}
As our final example, we consider the case where both a scalar field and a fluid are present. The fluid has an anisotropic pressure ($p_r\neq p_t$) and its $q=0$. Now, using equations \eqref{rhopr} and \eqref{eqfu} we can write,
\begin{equation}\label{cc}
 \rho=-p_r,
\end{equation}
and
\begin{equation}\label{Bprop}
 \frac{1}{B}\frac{dB}{dz}=\frac{1}{z}+\frac{p_r-p_t}{2zB^2}.
\end{equation}
Brandt {\it et al}, in \cite{Brandt:2006ai},  obtained the same equation of state as in \eqref{cc}
imposing the requirement of self-similarity. Their system was more general in the sense that the spacetime was not conformally flat.
For the current example, the conservation equation for the fluid (equation \eqref{consfl2}) leads to,
\begin{equation}\label{consprop}
 \frac{dp_r}{dz}=\frac{2}{B}\frac{dB}{dz}(p_r-p_t).
\end{equation}

In this case, finding an exact solution for the conformal factor is difficult without further simplification. Thus, we incorporate an additional assumption that the tangential and the radial pressure are proportional to each other, i.e., $p_t=\beta p_r$.
A similar assumption was used by Brandt {\it et al} in \cite{brandt2003}. Their study involved the evolution of a spherically symmetric spacetime which is not conformally flat. Using this assumption, we get a solution for equation \eqref{consprop} as,
\begin{equation}
 p_r=n B^{2(1-\beta)},
\end{equation}
where $n$ is a constant of integration.
Replacing this in equation \eqref{Bprop}, solution for $B$ comes out to be,
\begin{equation}\label{Bsolpropap}
 B=\left(Fz^{2\beta}-k\right)^\frac{1}{2\beta},
\end{equation}
with $F$ being a constant of integration and $k=\dfrac{n(1-\beta)}{2}$.
 Consequently, the solution for $A$ is,
\begin{equation}
 A=\left(Ft^{2\beta}-kr^{2\beta}\right)^\frac{1}{2\beta}.
\end{equation}\\
In this case, the radial pressure of the fluid comes out to be negative (equation \eqref{cc}). So, we may consider the fluid to be viscous for consistency.

\cleardoublepage



\printbibliography[heading=bibintoc]


\end{document}